\documentclass[a4paper,12pt]{article}
\pdfoutput=1
\usepackage{epsfig}
\usepackage{dsfont}
\usepackage{amssymb}
\usepackage{amsfonts}
\usepackage{amsmath}
\usepackage{blindtext}
\usepackage{hyperref}
\usepackage{esint}
\usepackage{euscript}
\usepackage{verbatim}
\usepackage{latexsym}
\usepackage{graphicx}
\usepackage{caption}
\usepackage{float}
\usepackage{subcaption}
\usepackage{bbm}
\usepackage{circuitikz}
\usepackage{listings}

\usepackage{tikz, pgfplots}
\usetikzlibrary{decorations}
\usepackage{caption}
\usetikzlibrary[positioning]
\usetikzlibrary{snakes}
\usepackage{amsmath,amssymb} 
\usepackage{booktabs}
\usepackage{hyperref}
\hypersetup{
	colorlinks=true,
	linkcolor=black,
	filecolor=red,      
	urlcolor=blue,
	citecolor=blue
}

\DeclareMathOperator{\csch}{csch}

\newif\ifdtup

\catcode`\@=11

\@addtoreset{equation}{section}

\def\@normalsize{\@setsize\normalsize{15pt}\xiipt\@xiipt
\abovedisplayskip 14pt plus3pt minus3pt%
\belowdisplayskip \abovedisplayskip
\abovedisplayshortskip \z@ plus3pt%
\belowdisplayshortskip 7pt plus3.5pt minus0pt}

\def\small{\@setsize\small{13.6pt}\xipt\@xipt
\abovedisplayskip 13pt plus3pt minus3pt%
\belowdisplayskip \abovedisplayskip
\abovedisplayshortskip \z@ plus3pt%
\belowdisplayshortskip 7pt plus3.5pt minus0pt
\def\@listi{\parsep 4.5pt plus 2pt minus 1pt
     \itemsep \parsep
     \topsep 9pt plus 3pt minus 3pt}}

\relax

\catcode`@=12

\catcode`\@=11

\def\section{\@startsection{section}{1}{\z@}{3.5ex plus 1ex minus
   .2ex}{2.3ex plus .2ex}{\large\bf}}

\def\SymBoxes#1#2#3#4{\newdimen\un@t \un@t#3%
\raisebox{#1}{\rule{#2\un@t}{#4}\hskip-#2\un@t
\@tempdimb\un@t \advance\@tempdimb by-#4\@tempcntb#2\relax%
\@whilenum{\@tempcntb>0}\do{
\rule{#4}{\un@t}\hskip\@tempdimb \advance\@tempcntb by\m@ne}%
\hskip-#2\un@t \rule[\un@t]{#2\un@t}{#4}%
\rule[\un@t]{#4}{#4}\hskip-#4
\rule{#4}{\un@t}}\hskip-#4}                

\begin{document}

\newcommand{\beq}{\begin{equation}}
\newcommand{\eeq}{\end{equation}}
\newcommand{\bea}{\begin{eqnarray}}
\newcommand{\eea}{\end{eqnarray}}
\newcommand{\beas}{\begin{eqnarray*}}
\newcommand{\eeas}{\end{eqnarray*}}
\newcommand{\defi}{\stackrel{\rm def}{=}}
\newcommand{\non}{\nonumber}
\newcommand{\bquo}{\begin{quote}}
\newcommand{\enqu}{\end{quote}}
\renewcommand{\(}{\begin{equation}}
\renewcommand{\)}{\end{equation}}
\def \eqn#1#2{\begin{equation}#2\label{#1}\end{equation}}

\def\e{\epsilon}
\def\IZ{{\mathbb Z}}
\def\IR{{\mathbb R}}
\def\IC{{\mathbb C}}
\def\IQ{{\mathbb Q}}
\def\de{\partial}
\def\Tr{ \hbox{\rm Tr}}
\def\H{ \hbox{\rm H}}
\def\HE{ \hbox{$\rm H^{even}$}}
\def\HO{ \hbox{$\rm H^{odd}$}}
\def\K{ \hbox{\rm K}}
\def\Im{ \hbox{\rm Im}}
\def\Ker{ \hbox{\rm Ker}}
\def\const{\hbox {\rm const.}}
\def\o{\over}
\def\im{\hbox{\rm Im}}
\def\re{\hbox{\rm Re}}
\def\bra{\langle}\def\ket{\rangle}
\def\Arg{\hbox {\rm Arg}}
\def\Re{\hbox {\rm Re}}
\def\Im{\hbox {\rm Im}}
\def\exo{\hbox {\rm exp}}
\def\diag{\hbox{\rm diag}}
\def\longvert{{\rule[-2mm]{0.1mm}{7mm}}\,}
\def\a{\alpha}
\def\dag{{}^{\dagger}}
\def\tq{{\widetilde q}}
\def\p{{}^{\prime}}
\def\W{W}
\def\N{{\cal N}}
\def\hsp{,\hspace{.7cm}}

\def\br{\nonumber}
\def\IZ{{\mathbb Z}}
\def\IR{{\mathbb R}}
\def\IC{{\mathbb C}}
\def\IQ{{\mathbb Q}}
\def\IP{{\mathbb P}}
\def \eqn#1#2{\begin{equation}#2\label{#1}\end{equation}}

\newcommand{\C}{\ensuremath{\mathbb C}}
\newcommand{\Z}{\ensuremath{\mathbb Z}}
\newcommand{\R}{\ensuremath{\mathbb R}}
\newcommand{\rp}{\ensuremath{\mathbb {RP}}}
\newcommand{\cp}{\ensuremath{\mathbb {CP}}}
\newcommand{\vac}{\ensuremath{|0\rangle}}
\newcommand{\vact}{\ensuremath{|00\rangle}                    }
\newcommand{\oc}{\ensuremath{\overline{c}}}
\newcommand{\psizero}{\psi_{0}}
\newcommand{\phizero}{\phi_{0}}
\newcommand{\hzero}{h_{0}}
\newcommand{\psiin}{\psi_{\rh}}
\newcommand{\phiin}{\phi_{\rh}}
\newcommand{\hin}{h_{\rh}}
\newcommand{\rh}{r_{h}}
\newcommand{\rb}{r_{b}}
\newcommand{\psibnd}{\psi_{0}^{b}}
\newcommand{\psibndp}{\psi_{1}^{b}}
\newcommand{\phibnd}{\phi_{0}^{b}}
\newcommand{\phibndp}{\phi_{1}^{b}}
\newcommand{\gbnd}{g_{0}^{b}}
\newcommand{\hbnd}{h_{0}^{b}}
\newcommand{\zh}{z_{h}}
\newcommand{\zb}{z_{b}}
\newcommand{\man}{\mathcal{M}}
\newcommand{\hbr}{\bar{h}}
\newcommand{\tbr}{\bar{t}}
\usetikzlibrary{arrows}
\newcommand{\midarrow}{\tikz \draw[-triangle 90] (0,0) -- +(.1,0);}
\begin{titlepage}

\def\thefootnote{\fnsymbol{footnote}}

\begin{center}
{\large
{\bf JT gravity and deformed CFTs}
}
\end{center}

\begin{center}
\ Suchetan Das$^a$\footnote{\texttt{suchetan1993@gmail.com }}, \ Anirban Dinda$^{a,b}$\footnote{\texttt{dindaanirban@gmail.com }} 

\end{center}

\renewcommand{\thefootnote}{\arabic{footnote}}

\begin{center}
{$^a$School of Physical Sciences, Indian Association for the Cultivation of Science,\\
2A and 2B Raja S. C. Mullick Road, Jadavpur Kolkata-700032, India.}\\
{$^b$ Institute of Fundamental Physics and Quantum Technology, \\ 
 Department of Physics, School of Physical Science and Technology,\\
 Ningbo University, Ningbo, Zhejiang 315211, China.} \\   


\end{center}
\vspace{-0.15in}
\noindent
\begin{center} {\bf Abstract}
\end{center}
We propose alternative \textit{UV completion} of pure JT gravity as well as CFT coupled to JT gravity, via a class of \textit{deformed} 2D CFT. In AdS/CFT with a prescribed classical limit, pure JT gravity in \textit{one-sided} AdS$_{2}$ black hole is argued to be described by certain holographic deformed CFT on a strip. Equivalently, these deformed CFTs can be recast as CFTs on one-sided AdS$_{2}$ black hole with  \textit{emergent conformal boundary condition on a stretched horizon}$-$providing a \textit{proper UV frame} of JT gravity. On the other hand, JT gravity coupled to CFT with fixed central charge of $\mathcal{O}(1)$, is also described by deformed CFT on strip satisfying conformal boundary condition, with a different classical limit. The resulting CFT Hilbert spaces in both of the above classical limits yield the black hole entropy as thermal entropy and the high-energy density of states match that of JT gravity with a precise energy scale correspondence. Moreover, the Hilbert space defined for a two-sided black hole factorizes into two one-sided sectors in both limits. Notably in the second limit, degenerate zero modes of the deformed Hamiltonian$-$characterized by conformal primaries localized at the horizon$-$appear as a residual effect of the stretched horizon boundary condition. Exploiting the second limit, we compute entanglement entropy in one-dimensional quantum systems dual to a conformally glued black hole$-$Poincaré geometry in JT gravity, reproducing a `Page curve' via the quantum extremal surface prescription, with `Page time' set by the stretched horizon cutoff.

\vspace{1.6 cm}
\vfill
\end{titlepage}

\setcounter{footnote}{0}
\tableofcontents

\section{Introduction}\label{sec1}
Over the past decade, the renewed interest in studying JT gravity as a solvable quantum gravity toy model has significantly shaped our understanding of the quantum nature of black hole physics\cite{Almheiri:2014cka}-\cite{Penington:2023dql} \footnote{See also \cite{Mertens:2022irh} and references therein for a detailed review on this subject.}. In particular, the universality of JT gravity in the near-extremal limit \cite{Davison:2016ngz}-\cite{Iliesiu:2020qvm} offers a rich framework to explore non-perturbative aspects of quantum black holes beyond the traditional AdS/CFT correspondence. The connection between the quantum JT gravity path integral and certain one-dimensional matrix model descriptions \cite{Saad:2019lba},\cite{Maxfield:2020ale} has opened a new avenue for rethinking holography in lower dimensions, in terms of ensemble-averaged quantum systems\cite{Sachdev:1992fk}-\cite{K}, providing concrete resolutions to the factorization problem \cite{Harlow:2018tqv} and the continuous nature of the density of states \cite{Stanford:2017thb}.

This novel perspective is primarily motivated by the gravitational path integral as a guiding principle, which can be rigorously analyzed in JT gravity through higher-order genus expansions. Notably, JT gravity without its topological component\footnote{This component is responsible for the genus expansion and the connection between JT gravity and matrix integrals.} can also emerge from the s-wave sector of Einstein gravity in three-dimensional asymptotically AdS spacetimes \cite{Achucarro:1993fd}. Therefore, it is crucial to investigate whether pure JT gravity$-$or JT gravity coupled to matter$-$admits a UV completion on genus zero surface.

In this paper, we take an initial step in this direction, building on and motivated by earlier work, which we review next.
\subsection{Large diffeomorphism, deformed CFTs and modular quantization} 
The solutions of vacuum Einstein equation in asymptotically AdS$_{3}$ (AAdS$_{3}$) spacetime are described by a class of Banados metrics \cite{Banados:1994tn}, obtained from large diffeomorphism in the bulk. In AdS$_{3}$/CFT$_{2}$, the large diffeomorphism can be identified with boundary conformal transformation governed by Virasoro algebra with central charge $c=\frac{3}{2G_{3}}$\footnote{We took $l_{AdS}=1$.} \cite{Brown:1986nw}. This class of AAdS$_{3}$ metrics includes solutions like BTZ \cite{Banados:1992gq}, conical defect geometry, which admit a global time-like killing vector. In particular, one can obtain the planar BTZ metric in $(r,s,\theta)$ coordinate by a boundary conformal transformation $z \rightarrow f(\omega) \equiv \sqrt{d}\tan[2\sqrt{d}(s+i\theta)]$, which map complex plane $(z,\bar{z})$ to thermal cylinder $(\omega,\bar{\omega})$, with $\omega=s+i\theta$. AdS/CFT identifies thermal state of the dual CFT, described by a holographic CFT on thermal cylinder as the dual description of planar BTZ. Hence the Hamiltonian on the thermal cylinder should be identified with $H_{th}=\int^{\infty}_{-\infty}d\theta (T(\omega)+\bar{T}(\bar{\omega}))$, which generates the bulk proper time in planar BTZ \cite{Das:2024mlx}. Incorporating the fact that the conformal transformation of the above form $z=f(\omega)$ generates this Hamiltonian on the thermal cylinder, one can immediately see the Hamiltonian in complex plane can not be written as $\tilde{L}_{0}+\bar{\tilde{L}}_{0}$, where $\tilde{L}_{n},\bar{\tilde{L}}_{n}$ are generators of radial quantization on cylinder. In fact, the form of the Hamiltonian can be written in terms of certain linear combinations of SL(2,$\mathbb{R}$) generators $\tilde{L}_{\pm1,0},\bar{\tilde{L}}_{\pm1,0}$. This can be mapped to a \textit{deformed Hamiltonian} on a ring$(x)$ with the form $H=\int dx \tilde{f}(x) T_{00}(x)$ with a deformation function $\tilde{f}(x)$.  Interestingly, this particular type of deformed Hamiltonian can be identified with the vacuum modular Hamiltonian on the ring \cite{Das:2024vqe}.

The vacuum modular Hamiltonian ($\mathcal{K}$) for a subregion ($R, -R$), in two dimensional CFT's on a ring of length $l$, is a global $sl(2,R)$ generator and has a simple expression\cite{Cardy:2016fqc},\cite{Das:2024vqe}: 
\begin{eqnarray}\label{eqn1}
\mathcal{K} &=& \frac{l}{\pi}\int^l_0 \frac{\sin \frac{\pi(x + R)}{l} \sin \frac{\pi(R -x)}{l}}{\frac{\sin 2\pi R}{l}}T_{00}(x) dx  \nonumber \\
&=& \cot{\frac{2\pi R}{l}} \Big(\tilde{L}_0 +\bar{\tilde{L}}_0\Big) -\frac{1}{2\sin{\frac{2\pi R}{l}}}\Big(\tilde{L}_{-1} + \tilde{L}_{1} +\bar{\tilde{L}}_1 +\bar{\tilde{L}}_{-1}\Big)  
\end{eqnarray}
$\mathcal{K}$  generates an automorphism of the algebra of observables ($\mathcal{A}$) associated with the causal diamond anchored at ($R, -R$). The vacuum of the CFT satisfies the KMS condition wrt $\mathcal{K}$ and thus appears to be thermal to an observer evolving in modular time under $\mathcal{K}$\footnote{ Interestingly, there are situations where the modular Hamiltonian is realized as a physical Hamiltonian. One such example is a periodically driven CFT \cite{Wen:2018agb}-\cite{Das:2022pez} in the {\it heating phase} where an effective $sl(2,R)$ Hamiltonian, of the form given in equation \ref{eqn1}, generates its {\it stroboscopic} dynamics. In such situations, the modular evolution corresponds to the real-time evolution of the driven CFT, although at stroboscopic times.}. Under the action of $\mathcal{A}$, one can define a CFT Hilbert space $\mathcal{H}_{B}$,
where $B$ denotes conformal boundary condition imposed at the two fixed points $(R,-R)$. 

Holography implies identification of $\mathcal{K}$, with the bulk modular Hamiltonian $(\mathcal{K}_b)$ \cite{Jafferis:2015del} upto $\mathcal{O}(G_{N})$ corrections, which generates flows inside the associated entanglement wedge, which in this case turns out to be the AdS Rindler wedge\cite{Das:2024mlx},\cite{Das:2024vqe}, described by the following metric. 

\begin{align}
ds^{2} =-(r^{2}-1)ds^{2} + \frac{1}{(r^{2} -1)}dr^{2} + r^{2} d\theta^{2} 
\end{align}

In these coordinates, the conformal boundary is at $r=\infty$ and the horizon is at $r=1$. $s$ is the time generated by $\mathcal{K}_b$ and $\theta$ is the boundary coordinate. 
Its natural to expect that the BCFT associated with $\mathcal{H}_{B}$ should describe the dual gravitational dynamics in the AdS Rindler patch in the presence of a stretched horizon.  Consistent with this expectation, in\cite{Das:2024vqe} it was observed that the ``emergent" Virasoro algebra of BCFT ($\mathcal{H}_{B}$) has a structural similarity to the ``emergent" near-horizon Virasoro algebra, in AdS Rindler\cite{Solodukhin:1998tc}. Moreover, as shown there, one can also match the entanglement entropy of the subregion $(R, -R)$ in the CFT with the AdS Rindler entropy, after identifying the cutoff of the CFT $\epsilon$ with the cutoff of the stretched horizon $\epsilon'$ via the following relation: $\frac{1}{\epsilon'} = \frac{1}{\sqrt{2R}}\log\left(\frac{\sqrt{2R}}{\beta\epsilon}\right) \equiv \Lambda$, with $(\beta =\frac{1}{2}csc\frac{2\pi R}{l})$ . In \cite{Das:2024mlx}, the same entropy is identified with the thermal entropy of modular quantization on a ring. On the other hand, the high energy density of states corresponding to the Hilbert space has the similar form of Cardy density of states  with a different regime of energy scale. In the next section, following \cite{Verheijden:2021yrb}, we will study the AdS Rindler wedge after dimensional reduction to two dimensions to obtain JT gravity on a $AdS_2$ black hole background. 



\subsection{Review of pure JT gravity from dimensional reduction}
In \cite{Achucarro:1993fd}, it was shown that starting with a class of asymptotic Anti-de Sitter metrics,
\begin{align}\label{gauge}
    ds^{2} = h_{ij}(x^{i})dx^{i}dx^{j}+\tilde{\phi}^{2}(x^{i})d\theta^{2}, \; \text{for} \; i,j=0,1,
\end{align}
as solutions of a theory of pure Einstein Gravity in 3 dimensions with negative cosmological constant, with action $S$ given by:
\begin{align}
    S=\frac{1}{16\pi G_{3}} \int d^{3}x \sqrt{-g}(R_{3}+2),
\end{align}
one ends up with the 2d JT gravity action, upon dimensional reduction of $S$ over the coordinate $\theta:(-\Lambda,\Lambda)$ with a reduction parameter $\alpha\in(0,1]$.
\begin{align}\label{pure jt action1}
  S=  \frac{2\Lambda\alpha}{16\pi G_{3}}\int d^{2}x \sqrt{-h}\tilde{\phi}(R_{2}+2)=\frac{1}{16\pi G_{2}}\int d^{2}
x \sqrt{-h}\phi(R_{2}+2)
\end{align}
Here we have identified $G_{3}\equiv 2\pi G_{2}$ and the dilaton identified as $\phi \equiv \frac{\Lambda\tilde{\phi}\alpha}{\pi}$. In \cite{Achucarro:1993fd}, the 3d metrics were taken to be spherically symmetric, with $\theta$ being the angular boundary coordinate. On the other hand, there will be an reduction to Gibbons-Hawking(GH) boundary term as follows \cite{Verheijden:2021yrb}
\begin{align}
    S_{b}=\frac{2\Lambda\alpha}{16\pi G_{3}}\int ds \sqrt{-h_{ss}}\tilde{\phi}_{b}K^{(3)} 
\end{align}
It can be shown for the AdS-Rindler or AdS-Poincare metric, $K^{(3)}=K^{(2)}+1$ \cite{Verheijden:2021yrb}. Hence the boundary action is reduced to
\begin{align}
    S_{b}=\frac{1}{8\pi G_{2}}\int ds \sqrt{-h_{ss}}\phi_{b} (K^{(2)}+1)
\end{align}
Hence the total action of pure JT gravity is
\begin{align}
    S_{total} =S + S_{b} = \frac{1}{16\pi G_{2}}\int d^{2}x \sqrt{-h}\phi (R_{2}+2) + \frac{1}{8\pi G_{2}}\int ds \sqrt{-h_{ss}}\phi_{b} (K^{(2)}+1)
\end{align}
In classical analysis, the equation of motion that comes from varying $\phi$ would set $R=-2$. This fixes the metric everywhere to be locally AdS$_{2}^{b}$ \footnote{We denote the upperscript $b$ in AdS$_{2}^{b}$ as the AdS$_{2}$ blackhole emerged from the dimensional reduction of AdS$_{3}$.}. We work in coordinates, in which the metric takes the standard AdS$_{2}^{b}$-Poincare form: 
\begin{align}\label{1.metric1}
    ds^{2} = \frac{-4 dz^{b} d\bar{z}^{b}}{(z^{b}-\bar{z}^{b})^{2}} = \frac{-dt^{2}+du^{2}}{u^{2}}, \; \text{where} \; z^{b}=t+u, \bar{z}^{b}=t-u, \; \text{with} \; u:(0,\infty) 
\end{align}

Varying $h_{\mu\nu}$ gives us the equation of motion for the dilaton field $\phi$ \cite{Mertens:2022irh}:
\begin{align}\label{dilaton eom}
    \left(\nabla_{\mu}\nabla_{\nu} -h_{\mu\nu}\nabla^{2}+g_{\mu\nu}\right)\phi = 0
\end{align}
After fixing metric and obtaining dilaton equation of motion, we are only left with GH boundary term, which will provide non trivial gravitational dynamics which captures asymptotic symmetries of AdS$_{2}^{b}$ or time reparametrization symmetry at the asymptotic boundary. As  denoted in \cite{Maldacena:2016upp}, we may think of this non trivial dynamics arises due to the motion of dynamical and physical boundary inside the rigid Poincaré AdS$_{2}^{b}$. In other words, we may think that this rigid AdS$_{2}^{b}$ spacetime emerges as a low energy limit of a \textit{UV theory}. The UV theory has an intrinsic notion of time. In our case, we denote $s$ is identified with the UV time,  which is related to the bulk Poincare time as $t=f(s)$. To extract the correct boundary term, following \cite{Maldacena:2016upp} we will first put a near-boundary cutoff and fix a boundary condition on the proper length of boundary curve and dilaton $\phi$. The position of the physical boundary is now determined by choosing a near boundary cut-off. In particular, we fix the proper length of the boundary metric as 
\begin{align}\label{1.metricbc}
    h_{ss}  = \frac{1}{\epsilon^{2}}
\end{align}
This will further fix the locations of wiggly boundaries as:
\begin{align}
    t=f(s), u=\epsilon f'(s)
\end{align}
At the wiggly boundary, we can also compute extrinsic curvatures $K^{(2)}$, which will give the Schwarzian term:
\begin{align}
    K^{(2)} = -1+ \epsilon^{2} Sch\{f(s),s\} +\mathcal{O}(\epsilon^{4})
\end{align}

To find out the finite boundary action, as we mentioned earlier, we also need to impose boundary values of dilaton. We will choose the following values of dilaton on the wiggly boundary \cite{Maldacena:2016upp}:
\begin{align}\label{1.dilatonbc}
    \phi_{b} = \frac{a}{2\epsilon}
\end{align}
Where $a$ is a dimensionful constant. In particular, this boundary condition directly emerges from the asymptotic expansion of the full asymptotically AdS$_{3}$ metric in the usual Fefferman-Graham gauge. Using this Dirichlet boundary conditions we will end up with the following action after taking $\epsilon \rightarrow 0$:
\begin{align}\label{1.Itot2}
    S_{tot} = \frac{a}{16\pi G_{2}}\int ds  Sch\{f(s),s\}
\end{align}
Eventually, the full dynamics of JT gravity is described by a boundary (at $r\rightarrow\infty$) 1D Schwarzian action integrated over $s$. Hence it would be useful to consider the proper time frame. Let us denote the lightcone coordinate in the proper time frame as $\omega^{b},\bar{\omega}^{b}$ which is related to $z^{b},\bar{z}^{b}$ as follows:
\begin{align}
    z^{b}=f(\omega^{b}),\bar{z}^{b}=\bar{f}(\bar{\omega}^{b}), \; \text{with} \; \omega^{b} =s+\theta^{b}, \bar{\omega}^{b}=s-\theta^{b}.
\end{align}
Hence the Poincare metrics (\ref{1.metric1}) will transform as:
\begin{align}\label{1.metric2}
    ds^{2} = -\frac{4\partial_{\omega^{b}}f(\omega^{b})\partial_{\bar{\omega}^{b}}\bar{f}(\bar{\omega}^{b})}{(f(\omega^{b})-\bar{f}(\bar{\omega}^{b}))^{2}}d\omega^{b} d\bar{\omega}^{b} , \; 
\end{align}
If we use the map $z^{b}=\sqrt{d}\tan[2\sqrt{d}(s+\theta^{b})]$, the metric will be AdS$_{2}^{b}$ black hole with temperature $\sqrt{d}$.
\begin{align}\label{heating metric}
    ds^{2} = \frac{4d}{\sinh^{2}(2\sqrt{d}\theta^{b})}(-ds^{2}+(d\theta^{b})^{2})
\end{align}
Notably, the same map in boundary coordinates $z \rightarrow f(\omega)$ also generates the boundary conformal transformation, which corresponds to the large diffeomorphism that maps dual AdS-Poincare to dual AdS-Rindler or planar BTZ as we mentioned earlier \cite{Das:2024mlx}. In this way, \textit{we can identify the proper UV frame($\omega^{b},\bar{\omega}^{b}$) of JT gravity black hole with  proper boundary time$(s)$ frame denoted by $(\omega,\bar{\omega})$ generated by boundary modular Hamiltonian.}

A similar reduction can be done for the AdS Rindler metric which is of the form (\ref{gauge}):
 \begin{align}\label{rindler}
    ds^{2} = -(r^{2}-d)ds^{2}+\frac{dr^{2}}{r^{2}-d}+r^{2}d\theta^{2}
\end{align} 
where  $\theta$ now runs over $(-\Lambda, \Lambda )$, with $\Lambda$ being an IR cut-off. This metric can be approximated by considering a \textit{large BTZ black hole}\cite{Witten:1998zw} in the solution of vacuum Einstein equation with the horizon radius $\frac{\Lambda\sqrt{d}}{\pi}$, having the \textit{same} Bekenstein-Hawking entropy\footnote{See Appendix \ref{app1} for the discussion.}.

In this case, we once again end up with the pure JT Gravity action (\ref{pure jt action1}) in AdS$_{2}^{b}$ black hole. In the metric (\ref{rindler}), we can identify $\tilde{\phi}=r$. The boundary condition on the dilaton should be $\phi_{b} = \frac{\phi_{r}}{\epsilon} \equiv \frac{a}{2\epsilon}$. Here $\phi_{r}$ is defined as $\frac{a}{2}$.  One can consistently take $\tilde{\phi}_{b}=\frac{1}{\epsilon}$. Hence $\phi_{r}=\frac{\Lambda\alpha}{\pi}$ or $a=\frac{\Lambda}{\pi}$.

For this reduction, the dilaton becomes $\phi = \phi_{r}r$. This is matched with the dilaton profile for AdS$_{2}$ black hole in JT gravity\footnote{Note that in AdS$_{2}$, there is no distinction of Rindler and black hole in terms of metric unlike in higher dimension.}. With this identification, one can define the AdS$_2^{b}$ black hole entropy with the AdS rindler entropy in 3 dimensions. 
\begin{align}\label{entropy relation}
     S_{AdS-Rindler}^{(3)} =\frac{\Lambda\sqrt{d}}{2G_{3}} =  \frac{\Lambda\sqrt{d}}{4\pi G_{2}} = \frac{\Lambda\tilde{\phi}}{4\pi G_{2}}|_{\tilde{\phi}=\sqrt{d}} = \frac{2\phi_{h}}{4G_{2}} \equiv 2S_{BH}^{(2)}
\end{align}
Here we choose $\alpha=\frac{1}{2}$ for the half reduction. Also we defined $S_{BH}^{(2)}=\frac{\phi_{h}}{4G_{2}}$, where $\phi_{h}=\frac{\Lambda\sqrt{d}}{2\pi}$, from the standard notion of black hole entropy in JT gravity away from extremality \cite{Verheijden:2021yrb}. Similarly, the full reduction to JT gravity with $\alpha=1$ in two sided AdS$_{2}^{b}$ black hole reproduces the AdS-Rindler entropy as the entropy of two-sided AdS$_{2}^{b}$ black hole.

\subsection{A scaling limit in modular quantization and JT gravity}
In the previous two sections, we reviewed two key facts:

\begin{enumerate}
  \item Pure Einstein gravity in AdS–Rindler with a boundary cutoff reduces$-$via dimensional reduction$-$to pure JT gravity in an AdS$_2^{b}$ black hole.
  \item The entropy of the AdS–Rindler (equivalently the two-sided AdS$_2^{b}$ black hole) can be reproduced from the thermal entropy of the dual holographic CFT, evolved with the modular Hamiltonian on a ring.
\end{enumerate}

This leads to a natural question: can the Hilbert space of pure JT gravity both in one and two-sided black holes be associated with a specific sector of the holographic CFT Hilbert space, via modular quantization?

In modular quantization on a cylinder \cite{Das:2024mlx}, a modular Virasoro algebra emerges with effective central charge $c_{\rm eff} = c\,\Lambda$, where $\Lambda$ is the fixed point-cutoff.
The Hilbert space is constructed from highest-weight representations of this algebra, subject to suitable boundary conditions at the fixed points of the modular flow. In the limit $c_{\rm eff}\to\infty$, the thermal entropy matches the AdS–Rindler entropy or the entanglement entropy of the corresponding boundary subregion\cite{Das:2024mlx}.

There are two inequivalent ways to take the limit $c_{eff}\to\infty$:
\begin{itemize}
  \item Take $c\to\infty$ while keeping $\Lambda = c_{eff}/c$ large but fixed.
  \item Take $\Lambda\to\infty$ while keeping $c = c_{eff}/\Lambda$ finite.
\end{itemize}

Hence, the two inequivalent Hilbert-space descriptions under modular quantization on a cylinder in the $c_{\rm eff}\to\infty$ limit are
\[
  \mathcal{H}^{C}_{c_{eff}\to\infty} = 
  \begin{cases}
    \mathcal{H}^{C}_{c\to\infty,\ \Lambda=\text{fixed}}, \\[6pt]
    \tilde{\mathcal{H}}^{C}_{c=\text{fixed},\ \Lambda\to\infty},
  \end{cases}
\]
where the superscript $C$ denotes construction via modular quantization on a cylinder.


\begin{itemize}
  \item $\displaystyle \mathcal{H}^{C}_{c\to\infty,\ \Lambda=\text{fixed}}$ corresponds$-$in the AdS/CFT framework$-$to the low-energy limit of full quantum gravity theory in AdS–Rindler with boundary cut-off (i.e., Einstein gravity with matter), as arises from approximating a large BTZ black hole. In the leading-order limit $G_3\to0$, this describes the vacuum sector. Assuming the holographic CFT includes only stress-tensor dynamics (i.e., dual to pure AdS$_3$ gravity), its $c\to\infty$ limit corresponds to pure Einstein gravity in AdS–Rindler, which further reduces to pure JT gravity in an AdS$_2^{b}$ black hole, with $\Lambda$ sourcing the boundary dilaton.
  \item $\displaystyle \tilde{\mathcal{H}}^{C}_{c=\text{fixed},\ \Lambda\to\infty}$ arises from modular quantization of a fixed-$c$ CFT and lacks a standard bulk holographic interpretation. We will revisit this limit shortly.
\end{itemize}
\begin{align}
    \mathcal{H}^{C}_{c_{eff}\rightarrow \infty} =\mathcal{H}^{C}_{c\rightarrow\infty,\Lambda= \text{fixed}} \cong \mathcal{H}^{\text{pure JT in two-sided AdS$_{2}^{b}$}}_{\Lambda=\text{fixed}}  
\end{align}
Hence, pure JT gravity Hilbert space can be defined via a holographic deformed CFT with vacuum sector and certain Virasoro descendants acting on it.
We can recast the modular Hamiltonian in the modular time frame $(s,\theta)$ as CFT Hamiltonian in AdS$_{2}$ black hole background with a conformal boundary condition on a \textit{stretched horizon} cut-off placed at a distance $\theta=\Lambda$. We emphasize that this cut-off is a requirement of defining the CFT Hilbert space under modular quantization which further provides a source to dilaton boundary condition in a natural way as we will show. The full reduction to JT gravity follows from integrating out the bulk Einstein gravity action in boundary coordinate $\theta:[-\Lambda,\Lambda]$. Correspondingly, the boundary CFT action should also be integrated over $\theta:[-\Lambda,\Lambda]$. Hence the CFT Hamiltonian in the dual modular time frame(described by a two-sided AdS$_{2}$ black hole with stretched horizon) can be shown to be proportional to $\int^{\Lambda}_{-\Lambda}d\theta (T(\omega)+\bar{T}(\bar{\omega}))$ with a constant Schwarzian term.
In a similar way, we can identify other class of SL(2,$\mathbb{R}$) deformed holographic CFTs with (only) stress tensor sectors in $c\rightarrow \infty$ limit as CFT Hamiltonian in proper boundary time frame, which describes pure JT gravity in dual UV frame corresponding to Poincare and global AdS$_{2}^{b}$. 

From these observations, we propose: \textit{an alternative UV completion of pure JT gravity could be described by a particular class of holographic deformed CFT made out of stress tensor sector. The low energy limit is defined by taking $c\rightarrow\infty$, which describes classical limit of pure JT gravity. The Hilbert space of JT gravity black hole can be equivalently described by that of the holographic CFT in AdS$_{2}$ black hole background with conformal boundary condition on a stretched horizon.} 

Also, it is natural to expect the half reduction of Einstein gravity in AdS-Rindler to JT gravity in \textit{one-sided} AdS$_{2}^{b}$ black hole, should have a UV completion in terms of modular quantization of dual holographic CFT(made out of stress tensor) on a strip. Naturally the half reduction should accompany with a dual CFT Hamiltonian on one-sided AdS$_{2}$ black hole with a stretched horizon, of the form $\int^{\Lambda}_{0}d\theta (T(\omega)+\bar{T}(\bar{\omega}))$. We can show that this corresponds to the modular Hamiltonian on a strip. With this motivation, in \textbf{section \ref{sec2}} we will study CFT modular Hamiltonian on a strip.
\begin{align}\label{1 isomorph}
    \mathcal{H}^{S}_{c_{eff}\rightarrow\infty} = \mathcal{H}^{S}_{c\rightarrow\infty,\Lambda=\text{fixed}} \cong \mathcal{H}^{\text{pure JT in one-sided AdS$_{2}^{b}$}}_{G\rightarrow 0,\Lambda=\text{fixed}}
\end{align}
Here the superscript $S$ in $\mathcal{H}^{S}$ denotes the Hilbert space constructed out of modular quantization on a strip. This Hilbert space also yields a notion of thermal entropy which matches with the entropy of one-sided AdS$_{2}^{b}$ black hole obtained from the dimensional reduction. Furthermore, we showed the Hilbert space of two-sided black hole in JT gravity factorizes into two one-sided one with the identification of dual holographic deformed CFT Hilbert space i.e. $\mathcal{H}^{C}_{c\rightarrow\infty,\Lambda=\text{fixed}} = \mathcal{H}^{S}_{c\rightarrow\infty,\Lambda=\text{fixed}} \otimes \mathcal{H}^{S}_{c\rightarrow\infty,\Lambda=\text{fixed}}$. 

On the other hand, we show in \textbf{section \ref{sec3}} that, JT gravity coupled to a CFT with central charge $c\sim \mathcal{O}(1)$ is also described by certain class of deformed CFTs on strip satisfying conformal boundary condition with another classical limit. 

For such non-holographic CFTs, one should identify the central charge $c$ with boundary value of dilaton $a$ which should be $\mathcal{O}(1)$, while $\Lambda \propto \frac{1}{G_{2}}$. This provides a new semiclassical limit for JT gravity coupled to non-holographic CFT, by identifying it with a deformed CFT Hilbert space, for which one should take $\Lambda \rightarrow \infty$, with $c$ finite and fixed. This is precisely the other realization of $c_{eff}\rightarrow \infty$ limit, discussed earlier.
\begin{align}
    \mathcal{\tilde{H}}^{C(S)}_{c=\text{fixed},\Lambda\rightarrow\infty} = \mathcal{\tilde{H}}^{\text{JT in two(one)-sided BH + finite $c$ CFT}}_{G \rightarrow 0}.
\end{align}
We can also argue the resulting Hilbert space $\mathcal{\tilde{H}}^{S}_{c=\text{fixed},\Lambda\rightarrow\infty}$ yields a notion of thermal entropy, which matches with the entropy of the one-sided AdS$_{2}$ black hole. Again for this limit, we can show the Hilbert space of CFT in two-sided classical JT black hole factorizes into two one-sided one i.e. $\mathcal{\tilde{H}}^{C}_{\Lambda\rightarrow\infty,c=\text{fixed}} = \mathcal{\tilde{H}}^{S}_{\Lambda\rightarrow\infty,c=\text{fixed}} \otimes \mathcal{\tilde{H}}^{S}_{\Lambda\rightarrow\infty,c=\text{fixed}}$. We note that, this limit reduces to the stretched horizon approaching to real horizon limit, for which the Hilbert space of modular quantization is described by \textit{zero mode sector} of modular Hamiltonian, identified as conformal primaries located at the horizon. 

We also prescribe a gluing of two theories of JT gravities with conformal coupling of matter CFTs in \textbf{section \ref{sec3}}. The resulting on-shell theory becomes a conformal coupling of two CFTs in the UV frame where the boundary condition of the gluing contains the information of Schwarzian degrees of freedom. As an application, we compute holographic entanglement entropy of dual 1D quantum systems in a glued Poincare- black hole spacetimes in JT gravity with matter CFTs in \textbf{section \ref{sec4}}. From the input of semiclassical limit of modular quantization, we have seen the entanglement entropy indeed shows a `Page curve' like behavior, where `Page time' is set by the appearance of stretched horizon in the AdS$_{2}$ black hole background. Details of the modular quantization will be given in \textbf{appendix \ref{app1}}.

The central goal of this paper is to provide evidences, consistency checks and discuss consequences of our proposal on this \textit{alternative UV completion of pure JT gravity as well as JT gravity coupled to finite $c$ CFT, in terms of deformed CFTs}. We end our note with a discussion of interesting future directions in \textbf{section \ref{sec5}}.


\textbf{Notation:} \begin{itemize}
\item Throughout the note, we have used heating phase Hamiltonian and modular Hamiltonian interchangeably.
\item Allover this note, AdS-Rindler denotes the standard metric with fixed boundary IR cut-off $\theta:(-\Lambda,\Lambda)$. This approximates a large BTZ black hole.
    \item  In the rest of the note, we have denoted AdS$_{2}^{b}$ as the metric($s,r$) appears in dimensional reduction of 3D AdS-Rindler and AdS$_{2}$ as the metric($s,\theta$) in the proper frame of modular quantization. Both of these metrics share same notion of time $s$ and same temperature .
\item We have used $\epsilon \rightarrow 0$ and $\Lambda \rightarrow \infty$ in the same footing.
    \item We took $l_{AdS}=c=h=k_{B}=1$ everywhere in the note.
    \item We denote both $G_{N}$ and $G_{2}$ to be the Newton's constant in 2D. $G_{3}$ denotes the same in 3D.
    \item In this note, $\omega$ frame refers to the UV frame of JT gravity.
\end{itemize}

\section{Holographic deformed CFT and pure JT gravity}\label{sec2}
We have argued the CFT Hamiltonian in the $\omega$ frame can be conformally mapped to a modular Hamiltonian of a subregion on a ring\cite{Das:2024mlx}. Similarly, we will show the modular Hamiltonian of a subregion on a strip, can be conformally mapped to a CFT Hamiltonian in $\omega$ frame described by an one-sided AdS$_{2}$ black hole, with the horizon maps to the end points of the subregion. We will use this description to quantize the CFT in the $\omega$ frame. This naturally leads to a stretched-horizon picture in the black-hole description. The resulting Hilbert space for holographic CFT in $c\rightarrow \infty$ limit should describe the dual JT gravity Hilbert space in one-sided AdS$_{2}^{b}$ black hole.
\subsection{sl(2,$\mathbb{R}$) deformed CFTs on the strip}
\begin{figure}
    \centering
   \begin{tikzpicture}[scale=.5]
    \draw[ thin](-15,9)--(-15,8);\draw[ thin](-15,8)--(-14,8); \draw[brown] node at ( -14.5,8.5) {$w$};
\draw [thin,red] (-20,0) -- (-10,0) ;
\draw [ultra thick, ->,blue] (-10,-8) -- (-10,8) ;
\draw [ultra thick, ->] (-20,-8) -- (-20,8) ;
     \draw node at (-20,-9) {$x$=0};
     \draw node at (-10,-9) {$x=\pi$};
     \draw[ thick,->](-21,2)--(-21,5);
     \draw node at (-21.5,3.5) {$\sigma$};
         \draw[ thick,->](0,0)--(0,8);
        \draw[ thick,blue,->](0,0)--(8,0);
         \draw[ thick,black,<-](-8,0)--(0,0);

        \draw[color=red] (5,0) arc (0:90:5);
         \draw[color=red] (0,5) arc (90:180:5);
         \draw[->,color=green] (-3,4.8) to [bend left=-15] (-4,4);
          \draw[->,color=purple] (3,4.8) to [bend left=15] (4,4);
        \draw[ thin](5,8)--(5,9);\draw[ thin](5,8)--(6,8); \draw[brown] node at ( 5.5,8.5) {z};

   \end{tikzpicture}
    \caption{Strip to UHP mapping: the constant $\tilde{t}$ line maps to a semi-circle on UHP. Two opposite arrows along the semi-circle indicate two contours $C$ and $\bar{C}$ corresponding to holomorphic and anti-holomorphic Virasoro modes on UHP.}
    \label{strip UHP map}
\end{figure}
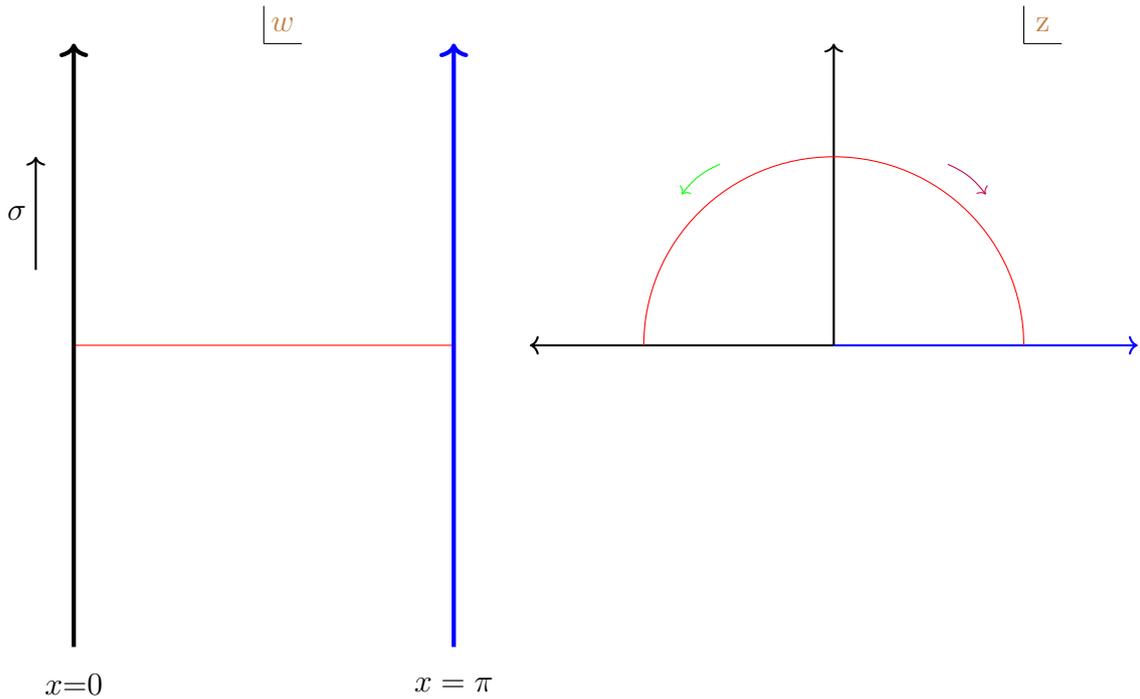

 In more detail, consider a CFT on a strip of length $\pi$, with the ${\it deformed}$ Hamiltonian\cite{Wen:2020wee}. 

\begin{align}
    H = \frac{1}{\pi}\int^{\pi}_{0} \tilde{f}(x) T_{00}(x), 
\end{align}

where $\tilde{f}(x)$ is the deformation function, with $\tilde{f}(x)=1$, corresponding to the undeformed CFT and the (anti)holomorphic strip coordinates are given by $(\bar{w}=\sigma-ix)w=\sigma+ix$, where $\sigma$ is the Euclidean time coordinate and $x$ is the spatial direction on strip.  Without loss of generality, we can expand $\tilde{f}(x)$ as follows.
\begin{align}\label{f(x)}
    \tilde{f}(x) = \sum_{j=0}^{N}a_{j}\cos(jx) 
\end{align}
where $\{a_{j}\}$ are $N$ real non vanishing constants. Upon mapping to Upper Half Plane (UHP) using the map ($w \rightarrow z=e^{w}=\tilde{t}+i\tilde{u}$), we end up with the Hamiltonian on the UHP:
\begin{align}
    H= \sum_{n=-N}^{N}a_{n}L_{n}+ \sum_{n=-N}^{N}a_{n}\bar{L}_{n}-\frac{c}{12},
\end{align}
which is expressed in terms of the modes of $T(z)$ and $\bar{T}(\bar{z})$ on the strip: $L_n$ and $\bar{L}_n$ which are defined in the usual manner:
\begin{align}\label{Fourier mode}
    &L_{n}=\frac{c}{24}\delta_{n,0} + \frac{1}{2}\int^{\pi}_{0}\frac{dx}{\pi} e^{i n x}T(x) = \frac{c}{24}\delta_{n,0} + \frac{1}{2\pi i} \int_{C}dz z^{n+1}T(z)  \\
    &\bar{L}_{n}=\frac{c}{24}\delta_{n,0} + \frac{1}{2}\int^{\pi}_{0}\frac{dx}{\pi} e^{-i n x}\bar{T}(x) = \frac{c}{24}\delta_{n,0} + \frac{1}{2\pi i} \int_{\bar{C}}d\bar{z} \bar{z}^{n+1}\bar{T}(\bar{z}).
\end{align}
Here, the contours $C$ and $\bar{C}$ refer to a half circle around the origin of UHP as shown in fig(\ref{strip UHP map}). 

A natural boundary condition on the strip would be to impose that no momentum flows across the boundaries of the strip, that is, $T(x)=\bar{T}(x)$ at $x=0,\pi$, which translates, in the $UHP$ geometry, to the boundary condition $T(z)=\bar{T}(\bar{z})$ at $z=\bar{z}$. The metric on the strip is the flat metric, so that after the conformal map the metric on the UHP becomes AdS Poincare. 
\begin{align}\label{poincare metric}
    ds^{2}_{UHP} \equiv ds^{2}_{Poincare} = \frac{-4 dz d\bar{z}}{(z-\bar{z})^{2}}
\end{align}
The map $z =f(\omega) = \sqrt{d}\tan(\mu\omega)$, now leads us to the $\omega$ frame metric given by
\begin{align}\label{heating metric1}
    ds^{2} = \frac{4\mu^{2}}{\sinh^{2}(2\mu\theta)}(ds^{2}+d\theta^{2})
\end{align}

As a concrete illustration, we will now discuss the case of sl(2,$\mathbb{R}$) deformed Hamiltonian on the strip, with  $\tilde{f}(x)=\alpha+\beta\cos(x)$ in equation (\ref{f(x)}). 
For this choice, the Hamiltonian on the UHP takes the form:
\begin{align}\label{deformed Ham}
    H = \alpha (L_{0}+\bar{L}_{0})+ \beta (L_{1}+L_{-1}+\bar{L}_{1}+\bar{L}_{-1}) 
\end{align}
It is known\cite{Wen:2020wee}, that depending on the choice of parameters ($\alpha,\beta$), three qualitatively different dynamics are obtained, referred to in the Floquet literature as distinct ${\it dynamical\;  phases}$, which are characterized by qualitatively distinct temporal behavior of the stress energy and unequal time correlators. More specifically, depending upon the sign of $(D\equiv \alpha^2 -4\beta^2)$, the dynamics of the Hamiltonian can be classified into three classes: ${\it heating\; phase\; dynamics}$ ($D<0$), ${\it non\; heating\;  phase\; dynamics}$ ($D>0$) and the ${\it phase\; boundary\; dynamics}$ $(D=0)$. Note that, the undeformed Hamiltonian (with $\alpha=1, \beta=0$) belongs to the class of non-heating phases and in fact they are related by Unitary transformation \cite{Das:2024vqe}.
We will restrict our discussion to the case of interest which is the ${\it heating\; phase}$. As a first step, we will make a conformal transformation to the proper Euclidean time ($s$) frame under this Hamiltonian\cite{Das:2022pez}. This involves solving for the integral curves ($z(s),\; \bar{z}(s)$)of $H$:

\begin{align}\label{deformed Ham1}
\alpha (l_{0}+\bar{l}_{0})+ \beta (l_{1}+l_{-1}+\bar{l}_{1}+\bar{l}_{-1}) = \frac{dz}{ds}\frac{\partial}{\partial z} + \frac{d\bar{z}}{ds}\frac{\partial}{\partial \bar{z}}
\end{align}

Where $l_n\equiv z^{n+1}\frac{\partial}{\partial z}$ are the space time representations of $L_n$. The solution of the curve equation corresponding to (\ref{deformed Ham1}) has been explicitly written down in \cite{Das:2022pez},\cite{Das:2024vqe} for the different phases. For the heating phase in the Euclidean time $s$, this is $z = \sqrt{d}\tan[\mu (s+i\theta)]$ and $\bar{z}= \sqrt{d} \tan[\mu (s-i\theta)]$. Here $\mu = \beta \sqrt{d}$ and $d \equiv \frac{D}{4\beta^{2}}$. This is equivalent to a Mobius transformation once we write $z(s)$ in terms of $z(0)=i\tanh\theta$\footnote{Note that, we could in principle choose $z(s)$ with any $s$=constant to get a similar Mobius transformation.}. It gives
\begin{align}\label{heating mobius}
    z(s) = \frac{\cos(\mu s) z(0) + \sqrt{d}\sin(\mu s)}{-\frac{1}{\sqrt{d}}\sin(\mu s) z(0)+\cos(\mu s)}
\end{align}
Similarly, it's complex conjugate will give $\bar{z}(s)$. One can easily check that for $s \in (-\infty,\infty)$ and $\theta\in(0,\infty)$, $Im(z)\in (0,\infty)$ or the other way round. Note that, there exists solution to the equation $z(s) = z(0)$ which is $z(0) = i\sqrt{d}$. This special point is the fixed point of the Hamiltonian evolution\cite{Fan:2019upv}. The metric in $\omega$ plane is given by (\ref{heating metric1}).
As mentioned earlier, after analytic continuation to real-time, this metric describes a one-sided AdS$_{2}$ black hole, with the horizon at $(\theta = \infty)$, which coincides with the position of the fixed point, since $z(0)|_{\theta =  \infty} =  i\sqrt{d}$. As we will see, the fixed point will play a crucial role in the quantization of the heating phase Hamiltonian as described in appendix (\ref{app1}). 

One can rewrite the Hamiltonian (\ref{deformed Ham}) as 
\begin{align}
    H= \frac{\beta}{2\pi i} \int_{C} dz (z-z_{+})(z-z_{-})T(z) +\frac{\beta}{2\pi i}\int_{\bar{C}} d\bar{z} (\bar{z}-z_{-})(\bar{z}-z_{+})\bar{T}(\bar{z}); \; \text{with} \; T(z)=\bar{T}(\bar{z})|_{z=\bar{z}}
\end{align}
Here, $z_{\pm}=\pm i\sqrt{d}$. To write down the Hamiltonian in $\omega$ coordinates, we need to use $T(\omega) = (\frac{\partial z}{\partial \omega})^{2}T(z)+\frac{c}{12}Sch\{z,\omega\}$. This yields:
\begin{align}\label{defomred Ham2}
    H= &-\frac{i}{2\pi}\int_{\theta=0}^{\theta=\infty}d\omega T(\omega)-\frac{i}{2\pi}\int_{\theta=\infty}^{\theta=0}d\bar{\omega} \bar{T}(\bar{\omega}) +\frac{ic}{24\pi}\left[\int^{\theta=\infty}_{\theta=0} d\omega Sch\{z,\omega\}+\int^{\theta=0}_{\theta=\infty} d\bar{\omega}Sch\{\bar{z},\bar{\omega}\}\right] \nonumber \\
    &= \frac{1}{2\pi}\int^{\infty}_{0}d\theta (T(\omega)+\bar{T}(\bar{\omega}))-\frac{c}{12\pi} Sch\{f(s),s\}\int^{\infty}_{\theta=0}d\theta
\end{align}
Here in the second line, we used the constant $s$ contour in the first term to extract the conserved Hamiltonian associated to time translation $s$. Also in the second term, we have used constant Schwarzian derivative for such mapping $z \rightarrow \omega$, such that $Sch\{z,\omega\}=Sch\{\bar{z},\bar{\omega}\} =Sch\{f(s),s\}$. Here $f(\omega)=\sqrt{d}\tan\mu\omega$. However note that, the second term is diverging, which we need to regulate by some IR cut-off $\theta=\Lambda$. Hence the total Hamiltonian can be written as
\begin{align}\label{strip Ham}
    H= \frac{1}{2\pi}\int^{\Lambda}_{0}d\theta (T(\omega)+\bar{T}(\bar{\omega}))-\frac{c\Lambda}{12\pi} Sch\{f(s),s\}
\end{align}
Here we can identify the first term as CFT Hamiltonian in AdS$_{2}$ black hole background with a stretched horizon at $\theta=\Lambda$. Whereas the second term is identical to the ADM Hamiltonian of JT gravity\footnote{Note that there is a relative sign in the definition of ADM Hamiltonian appeared in (\ref{EC}), which purely arises from taking Euclidean or Lorentzian signature.}, where one can identify $a=\frac{\Lambda}{\pi}$ with $c=\frac{3}{4\pi G_{N}}$ for the holographic CFT, which we obtained from \textit{half} dimensional reduction of 3D Einstein gravity in AdS-Rindler gauge, as we described in the introduction.
 Immediately, we can have
\begin{align}
    \partial_{s}H = \frac{1}{2\pi}(T(\Lambda)-\bar{T}(\Lambda)) 
\end{align}
Here we have used $T(\omega)=\bar{T}(\bar{\omega})$ at $\theta=0$. Since the Hamiltonian $H$ is made out of local stress tensor and it generates a geometric modular flow in a subregion of the strip, it should be conserved as $\partial_{s}H=0$. Hence, we need to show
\begin{align}
    T(\Lambda)-\bar{T}(\Lambda)=0.
\end{align}
From the conformal transformation of the stress tensor, we have the following:
\begin{align}
    T(\omega)-\bar{T}(\bar{\omega}) = \left(\frac{\partial z}{\partial\omega}\right)^{2}T(z)- \left(\frac{\partial \bar{z}}{\partial\bar{\omega}}\right)^{2}\bar{T}(\bar{z})+\frac{c}{12}\left( Sch\{z,\omega\}-Sch\{\bar{z},\bar{\omega}\}\right)
\end{align}
For the map $z=\sqrt{d}\tan\mu\omega$, we have
\begin{align}
     T(\omega)-\bar{T}(\bar{\omega}) = \mu^{2}d\left(1+\frac{z^{2}}{d}\right)T(z) - \mu^{2}d\left(1+\frac{\bar{z}^{2}}{d}\right)\bar{T}(\bar{z})
\end{align}
Here we used the fact $Sch\{z,\omega\}=Sch\{\bar{z},\bar{\omega}\} = \text{constant}$. We also observed, $\theta = \infty$ corresponds to $z=i\sqrt{d}$ and $\bar{z}=-i\sqrt{d}$, where $T(z_{+}),\bar{T}(z_{-})$ are finite. This yields
\begin{align}\label{cutoff1}
 (T(\omega)-\bar{T}(\bar{\omega}))|_{\theta=\infty}  = 0
\end{align}
In this way, we can see
\begin{align}\label{cutoff2}
 \lim_{\Lambda\rightarrow \infty}(T(\Lambda)-\bar{T}(\Lambda)) = \lim_{\Lambda \rightarrow \infty}\mu^{2}de^{-4\mu\Lambda}(T(z_{c})-\bar{T}(\bar{z}_{c})) = 0
\end{align}
Here $z_{c},\bar{z}_{c}$ corresponds to a cut-off around the fixed point $z_{\pm}$ where also the stress tensor is non-vanishing. We will see later how $z_{c}$ is related to $\Lambda$ explicitly. From (\ref{cutoff1}) and (\ref{cutoff2}), without loss of generality we can write the following for $\Lambda$ is finite yet large
\begin{align}
    T(\Lambda)-\bar{T}(\Lambda) = 0
\end{align}

In other words, defining the heating phase Hamiltonian $H$ in $\omega$ frame, enforces a cut-off at stretched horizon $\theta= \Lambda$, where one need to impose conformal boundary condition. We note a similar observation has been made in \cite{Burman:2023kko} in the context of quantizing massless scalars in BTZ background with a stretched horizon cut-off, where the \textit{unique} Dirichlet condition is emergent to define a Klein-Gordon norm in the same background. 

One can repeat a similar exercise for the case of the phase boundary dynamics ($D=0$). solving the integral curve equation for this case, gives us $z'$(here we denote the $UHP$ for phase boundary as $z'$) of the following form:
\begin{align}\label{phase bdy curve}
    z(s) = \frac{-1}{\beta}\frac{1}{s+i\theta} \equiv \frac{-1}{\beta\omega}, \; \bar{z}(s) = \frac{-1}{\beta} \frac{1}{s-i\theta} \equiv \frac{-1}{\beta\bar{\omega}}, \; \text{with} \; z(0) = \frac{i}{\beta\theta}
\end{align}
with the analogue of equation\ref{heating mobius}:
\begin{align}\label{phase bdy mobius}
z(s) = \frac{z'(0)}{1-\beta s z(0)}
\end{align}
As in the heating phase case, there also exists a fixed point corresponding to $z(0)=z(s) = 0$. Using the same procedure as before, we will end up with the following effective metric for the phase boundary:
\begin{align}\label{phase bdy metric}
    ds^{2}_{\text{phase boundary}} = \frac{ds^{2}+d\theta'^{2}}{\theta'^{2}}
\end{align}
This is exactly the same form of a AdS$_{2}$ Poincare metric in Lorentzian time (after doing wick rotation $s\rightarrow is$) with coordinate ranges $s \in [-\infty,\infty]$ and $\theta \in [0,\infty]$. Since (\ref{phase bdy curve}) is a global or Mobius transformation, the metric remains Poincare(from $z\rightarrow\omega$), with the Poincare horizon located at $\theta = \infty$, which, as before, coincides exactly with the fixed point $z(\theta) = 0$.

One can again perform the similar analysis for non-heating phase($D>0$), for which one end up with CFT in global AdS$_{2}$ metrics with a constant Schwarzian term with an overall negative sign.

\subsection{Hilbert space of modular quantization} 
In this section, we will be interested in quantizing the heating phase holographic CFT Hamiltonian on the strip geometry, which is equivalent to quantizing the holographic CFT Hamiltonian with constant Schwarzian in the $\omega$ frame.  A similar quantization has previously been done of the deformed CFT on the ring, which was dubbed as `modular quantization' in \cite{Das:2024mlx},\cite{Das:2024vqe}. Our discussion and results of this section will be similar to those obtained there. In fact, we can identify the same Hamiltonian as the holographic CFT modular Hamiltonian in the half-plane. The details of the quantization are described in Appendix \ref{app1}. In the following, we summarize the key points.
\begin{itemize}
    \item The dynamics of the heating phase Hamiltonian on a strip $H=\frac{1}{2}(L_{1}+L_{-1}+\bar{L}_{1}+\bar{L}_{-1})$, is described by a half or semi-infinite (thermal) cylinder. In this set-up, the angular direction is the Euclidean time ($s$) direction generated by the Hamiltonian or, $H=\mathcal{L}_{0}+\bar{\mathcal{L}}_{0}$ and the flow along spatial coordinate ($\theta$) of the cylinder, extended from the location of the boundary to the fixed point, is generated by $i(\mathcal{L}_{0}-\bar{\mathcal{L}}_{0})$. Here $\mathcal{L}_{0}=\frac{1}{2}(L_{1}+\bar{L}_{-1})$ defines a scaling generator in $(s,\theta)$ plane.

    \item To construct a well-defined Hilbert space of the quantization and to obtain a \textit{modular Virasoro algebra} associated to the conserved charges of the eigenmodes of the Hamiltonian, a cut-off ($\theta=\Lambda$) with conformal boundary condition is needed. Translating to the UHP, the location of the cut-off is related to $z=z_{+}+\epsilon e^{i\theta}$, where $z_{+}$ denotes the fixed point corresponds to $\theta=\infty$. The \textit{modular Virasoro algebra} has the following form:
    \begin{align}
        [\hat{\mathcal{L}^{\epsilon}_{k}},\hat{\mathcal{L}^{\epsilon}_{k'}}] = (k-k')\hat{\mathcal{L}}^{\epsilon}_{k+k'} + \frac{c_{eff}}{12}(k^{3}+k ) \delta_{k+k'} ; \; c_{eff} \equiv c\Lambda.
    \end{align}
    Here the modes $\hat{\mathcal{L}}_{k}$s are defined in (\ref{redefined vir gen}). This structure of \textit{modular Virasoro algebra} with near-fixed point cut-off is similar to those obtained \cite{Solodukhin:1998tc} in higher-dimensional spherically symmetric black holes, where the Virasoro charges generates large diffeomorphism at the horizon. Since the fixed point describes both the future and past horizon of the one-sided AdS$_{2}$ blackhole $\theta=\infty$\footnote{This is slightly different from the modular quantization in a cylinder, where two different fixed points correspond to future and past horizons of AdS$_{3}$ Rindler.}, we should identify the cut-off $\theta=\Lambda$ as the stretched horizon with conformal boundary condition. The relation between $\Lambda$ and $\epsilon$ is $\Lambda=\log(\frac{2}{\epsilon})$. Hence $\Lambda \rightarrow \infty$ corresponds to $\epsilon \rightarrow 0$ limit.
\item We have studied the Hilbert space corresponding to highest weight representation of the modular Virasoro algebra following \cite{Das:2024mlx}\footnote{In that reference, the effect of boundary condition and construction of vacuum has not been studied properly. Also some comments on density of states are misleading.}, with the boundary condition imposed by the vanishing of momentum flux at the cut-off $\theta=\Lambda$. In the limit of $\epsilon \rightarrow 0$, the boundary states $|B\rangle_{\epsilon}$ shrink to primary operators placed at the fixed point as in (\ref{shrinking cond}). Notably, this primaries are \textit{zero modes} of the Hamiltonian\footnote{Physically this is the result of inequivalent Hilbert space construction corresponding to deformed and undeformed Hamiltonian which results a different vacuum sector corresponding to each Hamiltonians. Hence those zero modes could be thought of as Goldstone bosons associated to the breaking of time-reparametrization symmetry. The non-trivial point in modular quantization is that those charges can be explicitly constructed at the horizon, instead of at the asymptotic boundary.}. With this boundary condition, we can construct eigenstates of the modular Hamiltonian $\{|B\rangle_{\epsilon}\}$ as in (\ref{basis}), which either shrinks to primary operators(including the identity operator) located at the horizon or becomes vanishing in $\epsilon \rightarrow 0$ limit. Hence the Hilbert space constructed with this boundary condition, detects the existence of \textit{soft modes} living at the horizon\cite{Hawking:2016msc},\cite{Haco:2018ske} (or fixed point) in the $\epsilon \rightarrow 0$ limit with finite and fixed $c$. However, in our prescribed limit, for which the Hilbert space should isomorphic to JT Hilbert space, we \textit{can not} take $\epsilon \rightarrow 0$ limit. Hence in the dual JT Hilbert space description, the states are defined by eigenstates $\{|B\rangle_{\epsilon}\}$ of cut-off Hamiltonian with a finite stretched horizon cut-off $\epsilon$. 

    \item By exploiting the equivalence of annulus partition function in two different channels \cite{Cardy:2016fqc}, we obtain a partition function of the quantized Hilbert space which in turn gives the expression for thermal entropy  $S \sim \frac{c}{6}\ln\frac{2}{\epsilon}$ in $c_{eff} \rightarrow \infty$ limit as in (\ref{entropy1}). This exactly matches with the leading term of entanglement entropy of an interval which ends on the boundary of the half plane. However this matching should be meaningful while taking $\epsilon\rightarrow 0$ limit. This also shows why the heating phase Hamiltonian is the \textit{modular Hamiltonian} of a CFT on a boundary interval $Im(z):(0,1)$ in the presence of a boundary $Im(z)=0$. 

    \item We can interpret the above thermal entropy as the AdS$_{2}^{b}$ black hole entropy from the point of view of dimensional reduction of AdS$_{3}$ Rindler while taking $c\rightarrow\infty$ with fixed $\epsilon$. Here we can construct a TFD state by considering another copy of the Hamiltonian as in (\ref{TFD1}). This state is by definition constructed out of the eigenstates of the two copies of the Hamiltonians   $\{|\tilde{B}\rangle_{\epsilon}\}$ as modular Virasoro descendants on vacuum in the presence of conformal fixed point cut-off.  
    This further provides a description of \textit{full} CFT Hilbert space on the strip $\mathcal{H}^{S(L,R)}_{c_{eff}\rightarrow\infty}$ constructed out of modular quantization on two strips $L$ and $R$:   $\mathcal{H}^{S(L,R)}_{c_{eff}\rightarrow\infty} = \mathcal{H}^{S(L)}_{c_{eff}\rightarrow\infty} \otimes \mathcal{H}^{S(R)}_{c_{eff}\rightarrow\infty} $ Consequently we can argue $\mathcal{H}^{S(L,R)}_{c_{eff}\rightarrow\infty}$ is isomorphic to $\mathcal{H}^{C(L\cup R)}_{c_{eff}\rightarrow\infty}$,  the Hilbert of modular quantization on cylinder with same Hamiltonian. By the virtue of dimensional reduction in dual picture, this will describe JT gravity Hilbert space in \textit{two-sided} AdS$_{2}^{b}$ black hole and hence provides a factorization of two-sided Hilbert space.
\item We obtain the asymptotic high energy density of state as $\rho(E) =e^{2\pi\sqrt{cE/6}}$. This energy dependence also matches with the asymptotically high energy density of state $\rho_{JT}(E')$ of \textit{quantum} JT gravity \cite{Stanford:2017thb}, obtained from \textit{half} dimensional reduction of Einstein gravity in AdS-Rindler, where $aE'=E$ as obtained in (\ref{jt deform equiv}). From CFT Hilbert space description, we have seen the continuous density of state emerges \textit{universally} in $c_{eff} \rightarrow \infty$ limit. For finite $c$ CFT, this is related to the emergent \textit{continuous Virasoro algebra} in $\Lambda \rightarrow \infty$ limit, in which the spatial region becomes non-compact. On the other hand, for a fixed $\Lambda$, this continuous density is expected to be related to the \textit{emergent type-III} factor constructed out of single trace sector of holographic CFT (in a compact manifold) in $c\rightarrow \infty$ limit \cite{Leutheusser:2021frk}. The corresponding microcanonical entropy of this quantization also agrees with thermal entropy as well as with the black hole entropy for $E=\frac{c\Lambda^{2}}{24\pi^{2}}$ .
\end{itemize}

From all of these observations in this section, we can conclude \textit{the dual JT gravity Hilbert space of one-sided AdS$_{2}^{b}$ black hole is constructed by quantizing a modular Hamiltonian on a strip, of a holographic CFT made out of only stress tensor sector. The dual CFT Hilbert space can be well-defined by imposing conformal boundary condition at the stretched horizon placed at $\epsilon$ distance away from the dual one-sided AdS$_{2}$ black hole horizon. The action of the boundary condition remains non-trivial in $\epsilon \rightarrow 0$ limit, which consists of degenerate vacuum sector of the Hamiltonian. However this vacuum sector is not part of the dual JT gravity Hilbert space which is defined only at fixed yet small $\epsilon$. By the virtue of AdS/CFT, this is a candidate UV Hamiltonian, which describes pure JT gravity in the low energy or $c_{eff}\rightarrow\infty$ limit. In any of the above description, we will not take $\epsilon \rightarrow 0$ limit while describing the dual JT gravity Hilbert space.}

\section{JT gravity coupled to CFT and deformed CFT}\label{sec3}
\subsection{Classical JT gravity with CFT as deformed CFT }We begin the section with a summary of JT gravity coupled to a 2D matter CFT, as reviewed in \cite{Mertens:2022irh}. It's action is given by:
\begin{align}\label{1.Itot1}
    I_{tot} = I_{JT}  + I_{CFT} 
\end{align}
The first term in the above action is the JT gravity action, which has the following form:
\begin{align}
    I_{JT} = -\frac{1}{16\pi G_{N}}\int_{M}\sqrt{-g}\phi(R+2) - \frac{1}{8\pi G_{N}}\int_{\partial M} \sqrt{-h}\phi_{b}(K-1)
\end{align}
The final term in $I_{tot}$ describes the CFT action. Again we reviewed in the introduction, the equation of motion that comes from varying $\phi$ would set $R=-2$. The corresponding metric will have the form (\ref{1.metric1}). 
Similarly, the equation of motion for the dilaton field will be \cite{Mertens:2022irh}:
\begin{align}\label{1.dilaton eom}
    \left(\nabla_{\mu}\nabla_{\nu} -g_{\mu\nu}\nabla^{2}+g_{\mu\nu}\right)\phi = -8\pi G_{N}T_{\mu\nu}
\end{align}
Here $T_{\mu\nu}$ is CFT stress energy tensor in the background manifold $M$.
The general solution of the dilaton can be written as\footnote{Note that, in general the solution of three equations (by the virtue of conservation of energy momentum tensor) for $\phi$ consists of three terms with undetermined constants. Among them, using pSL$(2,\mathbb{R})$ symmetry we can set one to be zero and the other dimensionful constant can be absorbed in the energy tensor for constant energy sources.}:
\begin{align}\label{1.dilaton solution}
    &\phi = \frac{a}{z-\bar{z}}\left(1-\frac{8\pi G_{N}}{a}(I_{+}+I_{-})\right); \; \text{where} \; \\
  &I_{+} = \int^{\infty}_{z} dx (x-z)(x-\bar{z}) T_{zz}(x), \;  I_{-} = \int^{\bar{z}}_{-\infty} dx (x-z)(x-\bar{z}) \bar{T}_{\bar{z}\bar{z}}(x),
\end{align}
In a similar way, after fixing the boundary condition on dilaton and the proper length of boundary curve, we will finally obtain
\begin{align}\label{11.Itot2}
    I_{tot} = I_{CFT}- \frac{a}{16\pi G_{N}}\int ds  Sch\{f(s),s\}
\end{align}

Note that the dilaton boundary condition explicitly breaks the time reparametrization symmetry of the boundary theory, as we can see from (\ref{1.dilaton solution}) after imposing (\ref{1.dilatonbc}). From this we have:
\begin{align}\label{Sch}
    \frac{a}{16\pi G_{N}}\frac{d(Sch\{f(s),s\})}{ds} = (f'(s))^{2}[T(f(s))-\bar{T}(f(s))] + \mathcal{O}(\epsilon^{2}) 
\end{align}
This is also the equation of motion of the Schwarzian term in the AdS spacetime(at finite $\epsilon$ cut-off) once we identify Schwarzian as the ADM mass of the cut-off AdS. 
In the limit $\epsilon \rightarrow 0$,  (\ref{Sch}) reduces to the following, in the proper boundary time frame $(\omega,\omega')$:
\begin{align}\label{1.totalenergy cons}
    & \frac{a}{16\pi G_{N}}\frac{d(Sch\{f(s),s\})}{ds} - (T'(\omega)-\bar{T'}(\bar{\omega}))|_{\theta\rightarrow 0} = 0
\end{align}
Where $T'(\omega)$ and $\bar{T'}(\bar{\omega})$ are the stress tensor components in the $\omega$ conformal frame and are related to $T(z)$ and $\bar{T}(\bar{z})$ by the usual conformal transformations. The boundary condition (\ref{1.totalenergy cons}), can be interpreted as the conservation of total energy in the $\omega$ frame at the asymptotic boundary. To see this, consider the CFT Hamiltonian in the $\omega$ frame, given as:
\begin{align}\label{1.AdS ham}
    H_{M}= \int^{\infty}_{0}d\theta (T'(s+\theta)+\bar{T'}(s-\theta))
\end{align}
then, taking derivative with respect to proper time $s$ yields 
\begin{align}\label{1.diff of Ham}
    \partial_{s}H_{M} = (T'(s+\theta)-\bar{T'}(s-\theta) )\Big{|}^{\theta =\infty}_{\theta =0}
\end{align}
Combining (\ref{1.diff of Ham}) and (\ref{1.totalenergy cons}), 
we immediately have the following relation:
\begin{align}\label{EC}
    \frac{a}{16\pi G_{N}}\frac{d(Sch\{f(s),s\})}{ds} + \frac{d H_{M}}{d s} = T'(\infty)-\bar{T}'(\infty) 
\end{align}
For \textit{pure} JT gravity, we can identify $\frac{a}{16\pi G_{N}}Sch\{f(s),s\}$ as the ADM mass in the $\omega$ frame with Lorentzian time $s$ \cite{Maldacena:2016upp}. 

Thus, for a constant Schwarzian term along with $T'(\infty)-\bar{T}'(\infty)=0$, 
equation \ref{EC} implies that the dynamics of the CFT coupled to the JT gravity system simplifies in the $\omega$ frame to that of an isolated BCFT. Now, the constant Schwarzian term could arise in two situations without choosing $f(s)$:
\begin{itemize}
    \item If the matter CFT satisfies conformal boundary condition  at $\theta=0$, i.e. when $T'(\omega)-\bar{T}'(\bar{\omega})|_{\theta=0}=0$, the equation (\ref{1.totalenergy cons}) implies the Schwarzian will be constant. This is true for any fixed $G_{N}$.
    \item Also if we consider $G_{N}\rightarrow 0$ and $c\sim \mathcal{O}(1)$, the dilaton equation of motion (\ref{1.dilaton eom}) simplifies to vacuum solution and hence the Schwarzian term will be constant. This is regardless of the boundary condition of matter CFT at $\theta=0$.
\end{itemize}
An interesting example of a constant Schwarzian is  $z =f(\omega) = \sqrt{d}\tan(\mu\omega)$, with $(\sqrt{d}, \mu >0)$, which described AdS$_{2}$ black hole upon analytic continuation $(s\rightarrow is)$ as in (\ref{heating metric1}). With this observation, we can explicitly see \textit{deformed CFTs on strip} in the $\omega$ frame or the proper time frame of the Hamiltonian, describes a CFT coupled to JT gravity in black hole background. In such cases, $T'(\infty)-\bar{T}'(\infty) = 0$ and $T(0)-\bar{T}(0)=0$ by construction. As we have shown in (\ref{strip Ham}), the Hamiltonian can be written in $\omega$ frame as
\begin{align}\label{UV Ham}
    H= \frac{1}{2\pi}\int^{\Lambda}_{0}d\theta (T(\omega)+\bar{T}(\bar{\omega}))-\frac{c\Lambda}{12\pi} Sch\{f(s),s\}
\end{align}
Hence we can immediately identify the first term with the matter CFT Hamiltonian $H_{M}$ in AdS$_{2}$ black hole as in (\ref{1.AdS ham}). 
On the other hand, for non-holographic CFTs with \textit{finite} $c$, one should identify 
\begin{align}\label{other effective c}
    c \propto a \sim \mathcal{O}(1) \; \text{and}\; \Lambda\propto \frac{1}{G_{N}}; \; \text{such that} \; \frac{a}{4G_{N}} = \frac{c\Lambda}{3} .
\end{align}
From this observation and identification, we conclude that \textit{JT gravity coupled to CFT with $c\sim \mathcal{O}(1)$ can be described by deformed CFT on strip with a classical limit. Here we identify the proper UV time $s$ in the black hole with stretched horizon background as the time generated by heating phase Hamiltonian on strip. The classical limit is defined via the identification of $G_{N} \rightarrow 0$ to $\Lambda \rightarrow \infty$ of modular quantization which describes approaching the stretched horizon to real horizon. This is also same as taking the other $c_{eff} \rightarrow \infty$ limit with $\Lambda \rightarrow \infty$ by keeping $c$ fixed and finite.}

\subsection{Stress tensor and vacua}
 In the previous section, we showed that the CFT in classical JT gravity  background is described by a deformed CFT Hamiltonian in another $c_{eff}\rightarrow \infty$ limit. In two dimensions, the different AdS patches are descriptions of the spacetime by different observers, with their different notions of proper time, who could thus detect different nature of stress energy in the same vacuum. Our goal in this section, is to identify the different AdS$_{2}$ vacuum with CFT vacuum on strip. To begin, we will first review the different descriptions of the AdS$_{2}$ vacuum. 

The metric of AdS$_{2}$ Poincare \ref{phase bdy metric} and AdS$_{2}$ black hole \ref{heating metric} are related by following coordinate transformation:
\begin{align}\label{coord change}
    &ds^{2} = \frac{4 d\omega' d\bar{\omega'}}{(\omega'-\bar{\omega'})^{2}} = \frac{(2\mu)^{2}d\omega d\bar{\omega}}{\sinh^{2}[\mu(\omega-\bar{\omega})]} , \; \text{where} \nonumber \\
   &  2\mu \omega \equiv 2\mu (\theta + s) = \ln(\theta' + s) \equiv \ln \omega'
\end{align}
Similarly $\bar{\omega} \equiv (\theta - s)$ and $\bar{\omega'} \equiv (\theta' - s)$ are also related by the same transformation. If we consider massless scalar field in AdS$_{2}$ background \footnote{This can be easily generalized to any CFT with central charge $c$ in the same background} then the corresponding stress tensor is related by \cite{Spradlin:1999bn}:
\begin{align}\label{stress tensor ads}
    T(\omega) = (2\mu \omega')^{2}T(\omega') + \frac{\mu^{2}}{12}
\end{align}
In AdS$_{2}$, one can define global AdS vacuum $|0_{\text{global}}\rangle$ which can be defined with respect to global time. Similarly one can also define Poincare vacuum $|0_{\text{poincare}}\rangle$ with respect to Poincare time. In AdS$_{2}$ black hole space time there also exist Hartle-Hawking vacuum and Boulware vacuum depending on the choice of observers. However, observers along the AdS$_{2}$-Schwarzchild time, do not detect any particle production in the Schwarzchild vacuum which coincides with the Boulware vacuum. On the other hand, the same observers(in an eternal black hole) will detect thermal bath in global vacuum which is similar to the fact that Rindler observers detect thermal particle production in Minkowski vacuum. One can also show that a sufficient excitation on top of Boulware vacuum in an one-sided blackhole can also observe the same thermal bath as in Hartle Hawking vacuum\cite{Mukohyama:1998rf},\cite{Soni:2023fke},\cite{Burman:2023kko}. For that, one has to choose Dirichlet boundary condition at the stretched horizon place just outside the horizon($\epsilon$ distance away from the horizon in Schwarzchild coordinate) and taking the limit of approaching the horizon($\epsilon\rightarrow 0$)\cite{Burman:2023kko}. Taking into account all of these observations and by matching Green's function in different patches, it can be shown that in AdS$_{2}$ \cite{Spradlin:1999bn}, 
\begin{align}
 |0_{\text{global}}\rangle =  |0_{\text{poincare}}\rangle = |0_{\text{Hartle-Hawking}}\rangle = |0_{\text{topped up Boulware}}\rangle|_{\epsilon \rightarrow 0}
\end{align}
Hence in the global vacuum (\ref{stress tensor ads}) implies
\begin{align}
    \langle T(\omega) \rangle_{\text{global}} = \frac{\mu^{2}}{12}, \; \langle T(\omega')\rangle_{\text{Poincaré}} = 0
\end{align}
Thus the stress energy of massless scalars in AdS$_{2}$ black hole in global vacuum is same as that of the thermal bath at temperature $\mu$; while the stress tensor of Poincaré AdS$_{2}$ vanishes.

From the deformed CFT point of view, we can demonstrate the same picture in CFT vacuum $|0\rangle_{S}$ on the strip(which respects only half of the conformal symmetry on $UHP$). In heating phase, we have argued that the transformation $z \rightarrow z(s,\theta)$ is equivalent to the CFT in AdS$_{2}$ black hole metric. In other words, doing a conformal transformation $z \rightarrow \sqrt{d}\tan(\mu \omega)$ should generate the same physical picture of an one sided blackhole, where in Lorentzian $(\omega,\bar{\omega}) = (s+\theta, s-\theta)$. Upon mapping to $UHP$, $_{S}\langle 0|T(z)|0\rangle_{S} = 0$ by the conformal invariance in $UHP$. Hence we get the stress tensor in $(\omega,\bar{\omega})$ plane as the following:
\begin{align}
    _{S}\langle 0|T(\omega)|0 \rangle_{S} = \left(\frac{\partial z}{\partial \omega}\right)^{2} \; _{S}\langle 0|T(z)|0\rangle_{S} +\frac{c}{12}Sch(z,\omega) = \frac{c\mu^{2}}{12}
\end{align}
 Hence the vacuum expectation value of the stress tensor exactly matches with the stress tensor of massless scalar fields ($c=1$) in AdS$_{2}$ black hole in the global vacuum. Similarly for the phase boundary, from (\ref{phase bdy curve}) one can define the coordinate transformation $z' \rightarrow -\frac{1}{\beta \omega'}$ for which
\begin{align}
    _{S}\langle 0|T(\omega')|0 \rangle_{S} = \left(\frac{\partial z'}{\partial \omega'}\right)^{2}\; _{S}\langle 0|T(z')|0\rangle_{S} +\frac{c}{12}Sch(z',\omega') = 0
\end{align}
Here the Schwarzian term itself vanishes due to global transformation. Hence again, the vacuum stress tensor in phase boundary coincides with the AdS$_{2}$ Poincare stress tensor of massless scalar fields in global vacuum. Similarly, one can compute boundary correlation functions $_{S}\langle \mathcal{O}(s,0)\mathcal{O}(0,0)\rangle_{S}$ at $\theta,\theta'=0$ for different SL(2,$\mathbb{R}$) deformed CFTs in the vacuum $|0\rangle_{S}$ \footnote{For instance, see section 4 of \cite{Das:2022pez}.} and those can be explicitly matched with the same for different AdS$_{2}$ patches\cite{Spradlin:1999bn}.  From all these observations, we may conclude that,
\begin{align}\label{vacuum equivalence}
    |0\rangle_{S} = |0_{\text{global}}\rangle =  |0_{\text{poincare}}\rangle = |0_{\text{Hartle-Hawking}}\rangle = |0_{\text{topped up Boulware}}\rangle|_{\epsilon \rightarrow 0} 
\end{align}
This observation is consistent with the general correspondence we made on deformed CFT in prescribed classical limit$(\epsilon\rightarrow 0)$ as CFT on AdS$_{2}$ background. The upshot of the discussion is that, \textit{in the semiclassical picture, the thermal particle production in the Hartle Hawking state is described by thermal stress tensor in CFT heating phase with $c_{eff}\rightarrow \infty$ limit. Here the CFT vacuum on the strip looks thermal in the time evolution generated by heating phase Hamiltonian.}

\subsection{Conformal gluing of two AdS spacetimes in JT gravity}
 We will now generalize these results to the case of two JT theories with conformal coupling of matter CFTs. 
 Before returning to the CFT picture, we will first discuss the gluing from the JT Gravity side.

Consider two CFTs on AdS$_{2}$ spacetimes coupled through the gluing of conformal matter within the JT gravity framework. 
The total action $I_{tot}$ of the theory is now given by:
\begin{align}\label{Itot1}
    I_{tot} = \sum_{i=1}^{2} (I_{JT}^{i}+I_{CFT}^{i})
\end{align}
Following the discussion of section\ref{sec4}, we choose AdS$_{2}$-Poincare geometries in the two sides of the boundary with the following metrics:
\begin{align}\label{metric1}
    ds^{2}|_{M^{1}} = \frac{-4 dz d\bar{z}}{(z-\bar{z})^{2}} = \frac{-dt_{1}^{2}+du_{1}^{2}}{u_{1}^{2}}, \; \text{where} \; z=t_{1}+u_{1}, \bar{z}=t_{1}-u_{1}, \; \text{with} \; u_{1}:(-\infty,0) \\
    ds^{2}|_{M^{2}} = \frac{-4 dz' d\bar{z}'}{(z'-\bar{z}')^{2}} = \frac{-dt_{2}^{2}+du_{2}^{2}}{u_{2}^{2}}, \; \text{where} \; z'=t_{2}+u_{2}, \bar{z}'=t_{2}-u_{2}, \; \text{with} \; u_{2}:(0,\infty) 
\end{align}
With the expression for the dilaton given by:
\begin{align}\label{dilaton solution}
    &\phi^{i} = \frac{a_{i}}{z_{i}-\bar{z}_{i}}\left(1-\frac{8\pi G_{N}}{a_{i}}(I_{+}^{i}+I_{-}^{i})\right); \; \text{where} \; \\
  &I_{+}^{i} = \int^{\infty}_{z_{i}} dx (x-z_{i})(x-\bar{z}_{i}) T_{z_{i}z_{i}}(x), \;  I_{-}^{i} = \int^{\bar{z}_{i}}_{-\infty} dx (x-z_{i})(x-\bar{z}_{i}) \bar{T}_{\bar{z}_{i}\bar{z}_{i}}(x),
\end{align}
where $z_{1}=z,z_{2}=z'$ in our convention. 

The boundary metric is fixed as:
\begin{align}\label{metricbc}
    g_{\tau\tau} = g_{\tau'\tau'} = \frac{1}{\epsilon^{2}}
\end{align}
With the locations of wiggly boundaries becoming:
\begin{align}
    t_{1}=f_{1}(\tau), u_{1}=-\epsilon f_{1}'(\tau); \; t_{2}=f_{2}(\tau'), u_{2}=\epsilon f_{2}'(\tau')
\end{align}
We will take $\epsilon \rightarrow 0$ at the end in a way such that $\tau \rightarrow \tau'$ where two spacetimes will glue together. This is a sensible physical requirement which suggests that when we want to couple those two spacetimes at the common boundary, there must be one boundary curve parametrized by an unique boundary time $\tau=\tau'=s$. The extrinsic curvatures $K^{i}$, are given by the Schwarzian terms:
\begin{align}
    K^{1} = 1+ \epsilon^{2} Sch\{f_{1}(\tau),\tau\} +\mathcal{O}(\epsilon^{4}); \; K^{2} = 1+ \epsilon^{2} Sch\{f_{2}(\tau'),\tau'\} +\mathcal{O}(\epsilon^{4})
\end{align}

To find out the finite boundary action, as we mentioned in  earlier section, we also need to impose boundary values of dilatons:
\begin{align}\label{dilatonbc}
    \phi_{b}^{1} = -\frac{a_{1}}{2\epsilon}; \; \phi^{2}_{b} = \frac{a_{2}}{2\epsilon}
\end{align}
Where $a_{1},a_{2}$ are dimensionful constants as described in (\ref{dilaton solution}). Using all these, we will end up with the following final action after taking $\epsilon \rightarrow 0$:
\begin{align}\label{Itot2}
    I_{tot} = \sum_{i=1}^{2}I_{CFT}^{i}+ \frac{1}{16\pi G_{N}}\int ds \left[a_{1} Sch\{f_{1}(s),s\}-a_{2} Sch\{f_{2}(s),s\}\right]
\end{align}
This is the full action where the gravitational boundary term is the difference of two Schwarzian actions. The lightcone coordinates $\omega,\omega'$ in the proper time frame  are related to the same in rigid AdS frame $z,z'$ as the following:
\begin{align}
    z=f_{1}(\omega),\bar{z}=\bar{f}_{1}(\bar{\omega}); \; z'=f_{2}(\omega'),\bar{z'}=\bar{f}_{2}(\bar{\omega'}), \; \text{with} \; \omega =s+\theta, \bar{\omega}=s-\theta; \; \omega' =s+\theta', \bar{\omega'} = s-\theta'
\end{align}
Hence the Poincare metrics (\ref{metric1}) will transform as:
\begin{align}\label{metric2}
    ds^{2}|_{M^{1}} = -\frac{4\partial_{\omega}f_{1}(\omega)\partial_{\bar{\omega}}\bar{f}_{1}(\bar{\omega})}{(f_{1}(\omega)-\bar{f}_{1}(\bar{\omega}))^{2}}d\omega d\bar{\omega} , \; ds^{2}|_{M^{2}} = -\frac{4\partial_{\omega'}f_{2}(\omega')\partial_{\bar{\omega'}}\bar{f}_{2}(\bar{\omega'})}{(f_{2}(\omega')-\bar{f}_{2}(\bar{\omega'}))^{2}}d\omega' d\bar{\omega'}
\end{align}
Note that, by this construction $\theta=\theta'=0$ corresponds to $u_{1}=u_{2}=0$ when we identify the bulk time $t_{1}=t_{2}$ \footnote{Although we should not identify them as a function of boundary time i.e., $f_{1}(s) \neq f_{2}(s)$.}.

Again we have the following set of Schwarian equations of motion after  imposing (\ref{dilatonbc}) in (\ref{dilaton solution}) 
\begin{align}\label{Sch eom1}
   & \frac{a_{1}}{16\pi G_{N}}\frac{d(Sch\{f_{1}(\tau),\tau\})}{d\tau} = (f'_{1}(\tau))^{2}[T_{L}(f_{1}(\tau))-\bar{T}_{L}(f_{1}(\tau))] + \mathcal{O}(\epsilon^{2}) \; \text{and} \\
 &   \frac{a_{2}}{16\pi G_{N}}\frac{d(Sch\{f_{2}(\tau'),\tau'\})}{d\tau'} = (f'_{2}(\tau'))^{2}[T_{R}(f_{2}(\tau'))-\bar{T}_{R}(f_{2}(\tau'))] + \mathcal{O}(\epsilon^{2}),
\end{align}
where the subscript $L,R$ refer to stress tensors of matter CFTs in two AdS spacetimes. 
If we take again $\epsilon \rightarrow 0$ limit such that $\tau,\tau' \rightarrow s$, then in the proper boundary time frame $(\omega,\omega')$, (\ref{Sch eom1}) reduces to the following:
\begin{align}\label{totalenergy cons}
     \frac{a_{1}}{16\pi G_{N}}\frac{d(Sch\{f_{1}(s),s\})}{ds} - (T_{L}(\omega)-\bar{T}_{L}(\bar{\omega})) = 
   \frac{a_{2}}{16\pi G_{N}}\frac{d(Sch\{f_{2}(s),s\})}{ds} - (T_{R}(\omega')-\bar{T}_{R}(\bar{\omega'}))=0
\end{align}
We will now couple two CFTs as follows.  Consider the CFT actions as the following:
\begin{align}
    \sum_{i=1}^{2}I^{i}_{CFT} = -\int_{M_{1}}\mathcal{L}_{1} - \int_{M_{2}}\mathcal{L}_{2} - \int_{\partial M_{1}\cap \partial M_{2}} \mathcal{L}_{int}
\end{align}
Here $\mathcal{L}_{1}$, $\mathcal{L}_{2}$ are CFTs defined on $\theta <0$ and $\theta' >0$ respectively for all $s$. $\mathcal{L}_{int}$ is the defect term defined on $\theta=\theta'=0$ which specifies an existence of conformal interface. Note that, we just use this Lagrangian notation to simplify our argument which is not necessary though. In fact, we will not assume any specific form of the Lagrangian. Now consider a generic infinitesimal conformal transformation: $\omega_{\mu} \rightarrow \omega_{\mu}+\epsilon_{\mu}$ and $\omega'_{\mu} \rightarrow \omega'_{\mu}+\epsilon_{\mu}$, where $\omega_{\mu}=:(s,\theta)$ and $\omega'_{\mu}:=(s,\theta')$. The variation of the action is given by \cite{Meineri:2019ycm},\cite{Papadopoulos:2023kyd}:
\begin{align}
    \delta_{\epsilon}(\sum_{i=1}^{2}I^{i}_{CFT} )= -\int_{M^{1}}T_{L}^{\mu\nu}\nabla_{\mu}\epsilon_{\nu} - \int_{M^{2}}T_{R}^{\mu\nu}\nabla'_{\mu}\epsilon_{\nu} -\int_{\theta=\theta'=0} ds D^{\mu}\epsilon_{\mu}
\end{align}
The deformation of the defect is denoted as $D^{\mu}$ whose form is not known or rather not required. After integrating by parts along the line of \cite{Papadopoulos:2023kyd}, one can finally obtain
\begin{align}
   \delta_{\epsilon}(\sum_{i=1}^{2}I^{i}_{CFT} )= -\int_{\theta=\theta'=0} ds (D^{\mu}+(T_{L}^{\mu\theta}-T_{R}^{\mu\theta'}))\epsilon_{\mu} 
\end{align}
Here in between the steps, we have used the conservation of stress energy tensor in $M^{1,2}$: $\nabla_{\mu}T_{L}^{\mu\nu} = \nabla'_{\mu}T_{R}^{\mu\nu} = 0$. 
Next we will use the fact that we are using conformal preserving interface i.e. the interface Lagrangian is invariant under particular set of conformal symmetries. Since the interface at $\theta=\theta'=0$ respects time translation symmetry($\epsilon_{\mu} = \epsilon\delta^{s}_{\mu}$), we have 
 $D^{\mu}\epsilon_{\mu}=0$ in this case. Hence varying the full action $I_{tot}$ with respect to the time translation $s\rightarrow s+\epsilon$, we get
 \begin{align}
     \delta_{\epsilon}(\sum_{i=1}^{2}I^{i}_{CFT}) = - \int ds \epsilon[T_{L}^{s\theta}-T_{R}^{s\theta'}]
 \end{align}
Hence on the on-shell configuration, we have
\begin{align}\label{onshell1}
 \int ds \left[  T_{L}(\omega)-\bar{T}_{L}(\bar{\omega}) - T_{R}(\omega')+\bar{T}_{R}(\bar{\omega'})\right] = 0  
\end{align}
Similarly we shall also consider the defect Lagrangian is invariant under scaling of $s$. For this, $\epsilon_{\mu} = \epsilon\delta^{s}_{\mu}s$ such that $D^{\mu}\epsilon_{\mu}=0$. Similarly for this case $s \rightarrow s+\epsilon s$, the on-shell variation of the CFT action gives :
\begin{align}\label{onshell2}
    \int ds s \left[ T_{L}(\omega)-\bar{T}_{L}(\bar{\omega})- T_{R}(\omega')+\bar{T}_{R}(\bar{\omega'})\right] = 0 
\end{align}
Combining (\ref{onshell1}) and (\ref{onshell2}), one can show \cite{Nakayama:2012ed} the general solution of those above equations is the integrand vanishes itself. This implies:
\begin{align}\label{gluing1}
    T_{L}(\omega)-\bar{T}_{L}(\bar{\omega}) =  T_{R}(\omega')-\bar{T}_{R}(\bar{\omega'})|_{\theta=\theta' =0}
\end{align}
This gluing condition is the statement of matter energy conservation due to the conformal interface which glue two CFTs. 

If we combine (\ref{gluing1}) with the statement of total energy conservation or Schwarzian equation of motion (\ref{totalenergy cons}), we end up with
\begin{align}\label{mass gluing}
  \frac{a_{1}}{16\pi G_{N}}\frac{d(Sch\{f_{1}(s),s\})}{ds} =   \frac{a_{2}}{16\pi G_{N}}\frac{d(Sch\{f_{2}(s),s\})}{ds}
\end{align}
We can think of this as the condition of conformal gluing of two JT gravity theories. The solution of the above condition is
\begin{align}\label{cond onshell}
    a_{1}Sch\{f_{1}(s),s\} = a_{2}\{f_{2}(s),s\} + \tilde{a}
\end{align}
where $c$ is an arbitrary $s$ independent constant. Putting (\ref{cond onshell}) back into (\ref{Itot2}), we get the on-shell total action as:
\begin{align}\label{Itotonshell}
    I_{tot}^{on-shell} = \sum_{i=1}^{2} I_{CFT}^{i}+ \frac{\tilde{a}}{16\pi G_{N}}\int ds
\end{align}
Thus the total classical on-shell action has only CFT degrees of freedom and the corresponding gluing between them in two asymptotically AdS$_{2}$ geometries specified by $f_{1}$ and $f_{2}$. 
This suggests, \textit{if we want to glue two JT gravity theories by conformal coupling of matter CFTs, the time flow generated by the CFT Hamiltonians should be aligned with the proper boundary time of each. Hence the on-shell or classical action we get as in (\ref{Itotonshell}) after gluing two theories of JT gravity with conformal coupling, is nothing but the conformal gluing of two deformed CFTs, where the boundary condition knows about the Schwarzians.}


Combining (\ref{1.diff of Ham}) and (\ref{totalenergy cons}) with the assumption $T_{L,R}(\infty)-\bar{T}_{L,R}(\infty)=0$, we would immediately have the following relation
\begin{align}
    \frac{a_{1}}{16\pi G_{N}}\frac{d(Sch\{f_{1}(s),s\})}{ds}  + \frac{\partial H_{L}}{\partial s} = \frac{a_{2}}{16\pi G_{N}}\frac{d(Sch\{f_{2}(s),s\})}{ds} + \frac{\partial H_{R}}{\partial s} =0
\end{align}
This is another form of writing the Schwarzian equation of motions as total energy conservation in each sides. 

In the context of standard notion of \textit{semiclassical JT gravity}, one can promote the CFT stress tensor $T$ to it's expectation value $\langle T\rangle$ in the preferred state of interest \cite{Mertens:2022irh}, by \textit{fixing} $f(s)$ and thereby fixing the Schwarzian $Sch\{f(s),s\}$. In other words, the relation (\ref{totalenergy cons}) will be changed as the following:  
\begin{align}\label{semiclassical gluing}
    \frac{a_{1}}{16\pi G_{N}}\frac{d(Sch\{f_{1}(s),s\})}{ds} - (\langle T_{L}\rangle-\langle\bar{T}_{L}\rangle)|_{\theta=0} = 
   \frac{a_{2}}{16\pi G_{N}}\frac{d(Sch\{f_{2}(s),s\})}{ds} - (\langle T_{R}\rangle-\langle\bar{T}_{R}\rangle)|_{\theta'=0}=0
\end{align}
However this is strictly true for $c>>1$. Note that, if we now choose $Sch\{f(s),s\}$ is constant, we know the solutions corresponding to AdS$_{2}$ black hole and Poincare patches and CFT on these patches satisfy $\langle T\rangle -\langle \bar{T}\rangle =0$. Thus the gluing condition (\ref{semiclassical gluing}) is automatically satisfied. Hence for these special metric solutions with $c\sim \mathcal{O}(1)$ the semiclassical gluing condition (\ref{semiclassical gluing}) can still be satisfied. In the last section, we identify the semiclassical vacuum of these patches with the vacuum of the strip $|0\rangle_{S}$ and the deformed CFT admits the semiclassical description of CFT with $c\sim \mathcal{O}(1)$ coupled to JT gravity within these patches in $G_{N}\rightarrow 0$ limit. From the JT/1D duality \cite{Sachdev:1992fk},\cite{K},\cite{Saad:2019lba} \footnote{To be precise, we assume a dual description of quantum mechanical system which describes JT coupled to CFT in low energy limit}, we can couple two 1D dual quantum mechanical system which describe gluing of black hole and Poincare AdS$_{2}$ in the semiclassical limit, which we study in the next section to compute entanglement entropy holographically. Hence the crux point of the analysis is, \textit{we can couple two 1D dual theories to glue two theories of JT gravity with conformal matter using (\ref{totalenergy cons}), (\ref{gluing1}) and (\ref{mass gluing}), such that semiclassically  (\ref{semiclassical gluing}) must be satisfied to glue two classical JT gravity in AdS$_{2}$ background with matter CFTs. For central charge $c\sim \mathcal{O}(1)$ the gluing condition in the semiclassical limit (\ref{semiclassical gluing}) is trivially satisfied for solutions with constant Schwarzian.}\\

Before ending this section, we note that a \textit{problem} occurs if we want to construct a conformal coupling of JT gravity with non-gravitational AdS$_{2}$ bath following the previous analysis. This is also the main set-up of recent semiclassical resolution of black hole information paradox \cite{Penington:2019npb}-\cite{Mahajan:2025gfh} where a blackhole is attached to non-gravitational bath to make the black hole evaporating, where the CFT in the bath captures the Hawking radiation coming out of the black hole. In this set-up, one can get a Page curve of Hawking radiation by computing bath entanglement entropy using the holographic QES prescription \cite{Engelhardt:2014gca}. In our previous analysis, We can try to glue one JT gravity with matter CFT sector in an AdS$_{2}$ black hole, which is conformally coupled to the other CFT living in AdS$_{2}$ Poincare bath with no gravity in semiclassical limit. We assume that both the CFTs are conformally coupled using (\ref{gluing1}) at the  interface, but the gravity is being switched off in the Poincare bath. For this case, we have only one Schwarzian term in the action (\ref{Itotonshell}) instead of difference of two Schwarzians. Similarly, the conservation of total energy condition as in (\ref{totalenergy cons}) will also be modified and we will end up with $T_{L}-\bar{T}_{L}=T_{R}-\bar{T}_{R} = 0$ at $\theta=\theta'=0$. In other words, if we want to couple a gravitating system with a non-gravitational one using the previous procedure, the only solution of conformal gluing condition is that the interface must be factorizing. Note that, in contrast to the previous discussion, this is true  before satisfying (\ref{semiclassical gluing}). 

This problem has been observed \cite{Geng:2020qvw}-\cite{Geng:2021hlu} in the context of higher dimensional gravity theory, where the authors claimed the breaking of energy momentum conservation of the boundary CFT introduces mass to the graviton. Very recently \cite{Antonini:2025sur}, it is claimed that this type of violation at the boundary should not affect the local bulk physics much in the context of holography. However, in the context of JT gravity, such a violation would lead to a change of background geometry which is governed by (\ref{totalenergy cons}). 

\section{Towards a Page curve from gluing of JT gravities: A tale of QES with no island}\label{sec4}
Euclidean gravitational path integrals act as a \textit{magic box}, yielding numerous semiclassical predictions that match extensive tests. These include novel semiclassical saddles responsible for non‑trivial quantum extremal surface (QES) formation\cite{Almheiri:2019qdq}. Such saddles have been validated in both lower- and higher-dimensional AdS, including non‑black‑hole backgrounds \cite{Chen:2020uac}-\cite{Krishnan:2020oun}. However, most of studies include a coupling to non-gravitational bath which leads to \textit{violation} of energy momentum conservation of the boundary theory as we discussed in the last section. In this section, we compute the entanglement entropy of dual 1D quantum systems holographically in a dynamical setting of JT gravity \textit{without} attaching a bath. We do so by conformally gluing an AdS$_2$ black hole to Poincaré AdS$_{2}$, following the method from the previous section in JT gravity coupled to CFT framework. Using the inputs of modular quantization in the semiclassical limit($G_{N}\rightarrow 0$), we demonstrate a non‑trivial QES solution that reproduces a Page‑curve‑like behavior, restoring unitarity of the fine‑grained entropy. Crucially, this analysis does \textit{not} require an \textit{island} in the QES solution.

\subsection{Review of trivial and non-trivial QES}
QES prescription provides\cite{Engelhardt:2014gca} a semiclassical way of computing fine-grained entanglement entropy of a dual quantum system, which in principle can take care of all order quantum correction coming from gravity coupled to matter sector. This prescription suggests, the entanglement entropy of quantum state $\rho_{R}$ for some subregion $R$, can be computed as
\begin{align}
    S(\rho_{R}) =  min\left(ext_{X}\left[\frac{A(X)}{4G}+S_{semi-classical}(R\cup X)\right]\right),
\end{align}
where $X$ is a co-dimension 2 surface in the holographic bulk dual, $X\cup R$ is the spatial region of a slice extended from $X$ to $R$,  $A$ refers to the area and $S_{semi-classical}$ is the fine grained entropy of the quantum matter in classical gravity background. The QES is obtained by extremizing the generalized entropy $S_{gen}(X) \equiv (\frac{A}{4G}+S_{semi-classical})$ over $X$. If there exists more than one extremum, one must choose the minimum of them. Even it was also suggested\cite{Penington:2019npb} that the same formula can be used as a gravitational fine-grained entropy formula to obtain entropy of a state $\rho_{R}$ in a full quantum description. In that case, by definition $R$ could be thought of as a region in gravitating part. This fact is explicitly used to compute entropy of Hawking radiation to obtain a unitary Page curve. In all those dynamical situations \cite{Penington:2019npb}-\cite{Almheiri:2020cfm}, it is shown that at late time the minimum QES can be located inside the interior or just outside the exterior of a black hole in the semi-classical description, such that a disconnected part of the gravity region extended inside the interior, called the \textit{island}, can be formed in the entanglement wedge of the radiation region. However, in the absence of a manifestly smooth interior, one would expect to find a minimum just on or before the horizon. To test this in our setting, we will purely work in 2d JT gravity coupled to CFT as described in section (\ref{sec3}). The 2d version of the QES formula can be read as 
\begin{align}
    S(\rho_{0}) =  min\left(ext_{X}\left[\frac{A( X)}{4G_{N}}+S_{semi-classical}^{CFT}(\Sigma_{ X})\right]\right).
\end{align}
Here the $0$ subscript in $\rho_{0}$ suggests 0-dimensional region or a point which belongs to the spatial slice of 1d quantum mechanical theory at the boundary of AdS$_{2}$. $\Sigma_{X}$ denotes the spatial slice from $X$ to the boundary point. In the context of JT gravity, $A(\partial X)$ is replaced by the value of dilaton at the same point $\phi( X)$. Note that, here we use the nearly AdS$_{2}$/nearly CFT$_{1}$ duality in JT gravity, where the nearly CFT$_{1}$ refers to 1d quantum mechanical model or SYK like random matrix ensembles, whose low energy limit corresponds to the Schwarzian theory coupled to CFT. 

The simplest solution of the extremization comes from `no QES' contribution. This is generally called `\textit{trivial QES}' \cite{Almheiri:2020cfm}. In 2d, $X=\{0\}$ will automatically extremize this and the $S_{semi-classical}$ contribution of the matter entropy only exists at a single point\footnote{Here we use the identification of the area term with the value of dilaton  purely comes from dimensional reduction}. This should be zero by definition if the state of the 1d theory is pure. Hence for a single 1d theory in a pure state, the vanishing entanglement entropy can be computed semiclassically just by using trivial QES, which is by default the only minima. When the trivial QES is non-zero, one might need to look for other solution of QES by varying $X$ and then find the minima. Any other non-vanishing QES solution is known as `\textit{non-trivial QES}'. These type of situation may arise when there are more than one dual 1d theory exists, either they are in entangled state or they could be interacting with each other. Before going through the case study, we define the procedure of computing $S_{CFT}(\Sigma_{X})$.

\underline{\textit{CFT entanglement entropy in AdS$_{2}$:}} Since semiclassical contribution to bulk entanglement entropy coming from CFT entanglement entropy in AdS$_{2}$, we need to follow a rule for that. In \cite{Penington:2019npb},\cite{Almheiri:2019psf}, a formal way is followed to compute CFT entanglement entropy on a curved background by the virtue of Weyl transformation. 

However, in this context we will use inputs from SL(2,$\mathbb{R}$) deformed CFTs to obtain the entanglement entropy in semiclassical limit of JT gravity coupled to CFT in AdS$_{2}$ background. In our notation, CFT in proper AdS frame is defined as  $(\omega,\bar{\omega})$ plane or $(s,\theta)$ plane. We can conformally map CFT in $\omega$ plane to deformed CFT in $UHP$ or $z$ plane. Hence the rule is to map $(\omega,\bar{\omega}) \rightarrow(z,\bar{z})$ to compute twist operator two point function in the vacuum of deformed CFT on strip. From the semiclassical description, we have identified the vacuum of strip and the vacuum of AdS$_{2}$ patches: black hole, Poincare and global. Since we are computing this semiclassically, the stress energy tensor must be replaced by the expectation value in the respective semiclassical state we want to compute in. This implies, in $UHP$ we must have $T(z) \rightarrow \langle T(z)\rangle = 0$. To put it altogether, we follow this:
\begin{align}\label{twist corr}
    &S^{CFT}_{semi-classical}(\Sigma_{X}(s,\theta_{X})) = \lim_{n\rightarrow 1}\frac{1}{1-n}\log(\langle \tau_{n}(\omega,\bar{\omega})\rangle) , \; \text{where} \\ &\langle \tau_{n}(\omega,\bar{\omega})\rangle =\left(\frac{\partial z}{\partial \omega}\right)^{h_{n}} \left(\frac{\partial \bar{z}}{\partial \bar{\omega}}\right)^{h_{n}}\langle\tau_{n}(z,\bar{z})\rangle_{UHP}
\end{align}
However for the case of AdS$_{2}$ blackhole, we have used another near-horizon boundary condition at $\theta=\Lambda$(IR cut-off) in the UV frame, coming purely from modular quantization as an UV input. In the semiclassical description with $G_{N}\rightarrow 0$ for non holographic CFTs, we should take $c_{eff} \rightarrow \infty$ limit such that it should include $\Lambda \rightarrow \infty$, maintaining the relation (\ref{other effective c}). In the same limit, the leading finite term in $\Lambda$ would provide the semiclassical answer. Here we take the bulk matter CFT to be strictly non-holographic with finite and fixed $c$.

Hence to summarize the set-up we use the following rule:
\begin{itemize}
    \item We use nearly-AdS$_{2}$/nearly-CFT$_{1}$ set-up of JT gravity coupled to CFT with finite central charge $c\sim \mathcal{O}(1)$, to compute entanglement entropy of nearly CFT$_{1}$ system holographically using QES prescription. 
    \item To compute the the semi-classical contribution of matter CFT in a fixed classical background we have to promote $T$ by $\langle T\rangle$ in the respective vacuum state in standard semiclassical prescription of JT gravity. Even though this is true for $c>>1$, we have argued for $c\sim\mathcal{O}(1)$ there are special solutions with constant Schwarzians where (\ref{semiclassical gluing}) can still be satisfied trivially.
    \item Since the bulk CFT with $c\sim\mathcal{O}(1)$ in the fixed UV frame is defined by deformed CFT on a strip in a prescribed semiclassical limit $\Lambda\rightarrow\infty$, we will use this input in computing the matter entanglement entropy part. We also discussed in the last section that this $G_{N}\rightarrow 0$ limit is insensitive to the asymptotic boundary condition of CFT as it does not incorporate backreaction on metric as well for finite $c\sim\mathcal{O}(1)$.
\end{itemize}

\subsection{Case study of non-dynamical QES} We begin with QES computation of entanglement entropy(EE) involving dual 1d quantum mechanical systems in some state.

\textbf{(i) One 1d in a pure state:} For a single 1d system prepared in a pure state, EE must be zero. In our examples, this could be either one-sided AdS$_{2}$ blackhole or Poincare. Here the boundary region reduces to a point. As we already mentioned, this is guaranteed from trivial QES which gives $S_{CFT}=0$. Hence we do not need to check any non-trivial QES for such cases.

\begin{figure}
    \centering
   \begin{tikzpicture}[scale=.5]
    
         \draw[ thin](-5,-5)--(5,5);
         \draw[ thin,blue](-5,1)--(-1,1);
         \draw node at (-7,1.2){$(s,\theta=0)$} ;
         \draw node at (.1,1.4) {$(s,\theta=\Lambda)$};
       \draw[ thin](5,-5)--(-5,5);
        \draw[ thin](-5,5)--(-5,-5);
       \draw[ thin](5,5)--(5,-5);
    
   \end{tikzpicture}
    \caption{Dual to two 1D system in TFD: two-sided BH}
    \label{tfd draw}
\end{figure}
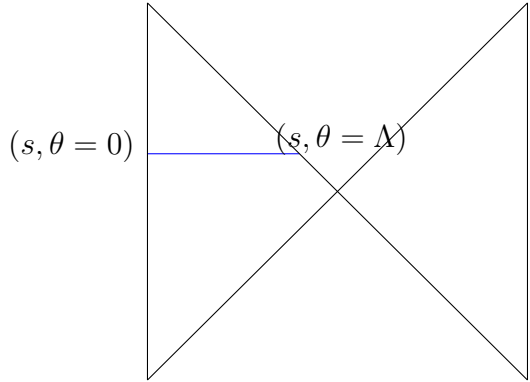

\textbf{(ii) Two 1d systems in a TFD state:} We can also consider two non-interacting 1d quantum mechanical systems prepared in a TFD state. From the holographic dictionary, the semiclassical dual should be a two-sided eternal AdS$_{2}$ black hole. The question we want to ask is \textit{how much one 1d system is entangled with the other 1d?} Holographically the answer should be $S_{BH}$. From the fig(\ref{tfd draw}), the trivial QES corresponds to the region $\Sigma_{0}=\{s,\theta:(0,\Lambda)\}$. Using (\ref{twist corr}) and the map $z=\sqrt{d}\tan[\mu(s+i\theta)]$ we get,
\begin{align}
    S^{trivial} \sim  \frac{c}{6}\log(\sinh(2\mu\Lambda)) \sim \frac{c}{3}\mu\Lambda = \frac{\mu a}{4G_{N}}
\end{align}

Here we used $\Lambda \rightarrow \infty$ to follow our semiclassical analysis of CFT in UV frame. The last equality follows from (\ref{other effective c}). We also note, this map coincide with that discussed in appendix (\ref{app1}), once we take $\mu=\sqrt{d}/2$. Hence we arrive at $S^{trivial} = S^{\text{AdS}_{2}}_{BH}$ with temperature $\mu=\sqrt{d}/2$. Since this is a non zero answer, we also need to compute non-trivial QES to check if there exists any other minima. To compute so, we consider a finite interval $(0,b)$ which ends at $b$ inside the bulk of one wedge. Hence the interval slice becomes $\Sigma_{b}=(s,\theta:(0,b))$. The dilaton value at point $b$ becomes $\mu a\coth(2\mu b)$. Hence the non-trivial QES is
\begin{align}\label{bh non trivial qes}
    S^{non-trivial} = ext_{b}\left(\frac{\mu a\coth(2\mu b)}{4G_{2}}+ \frac{c}{6}\log(\sinh(2\mu b))\right)
\end{align}
The solution to the extremization will be
\begin{align}\label{ee bh}
    \frac{c G_{N}}{2\mu a} \sinh(4\mu b^{*}) = 1
\end{align}
In the semiclassical analysis, we must have $G_{N}\rightarrow 0$. From (\ref{other effective c}), we identify $a \propto c$. Hence (\ref{ee bh}) suggests $\mu b^{*} \rightarrow \infty$. In other way, the solution of non-trivial QES is placed near the horizon. Once we put this solution back to (\ref{bh non trivial qes}), we will end up with
\begin{align}
    S^{non-trivial} \sim S_{BH}+ \frac{c}{3}\mu\Lambda > S^{trivial}
\end{align}
 Minimizing the solutions, we have 
 \begin{align}
      S(\rho_{L}^{TFD}) = S^{trivial}  \sim \frac{c}{3}\mu\Lambda = \frac{\mu a}{4G_{N}} = S_{BH} .
 \end{align}

\textbf{(iii) Two coupled (same) 1d systems:} Interesting set-up will be involved by coupling two 1d systems. In JT gravity set-up, the coupling is generated by an existence of conformal interface where conformal boundary condition is imposed. The simplest case is to study coupling of two same AdS$_{2}$ spacetime in JT gravity coupled to CFT via transparent boundary condition in conformal matter. We consider two same CFTs($L$ and $R$) with same central charge $c$, interact at the boundary via transparent boundary condition $T_{L}=T_{R}$. We denote $L(z,\bar{z})$ by $(s,\theta)$ and $R(z',\bar{z}')$ by $(s,\theta')$, where $\theta:(-\infty,0),$ and $\theta':(0,\infty)$. Hence we have
\begin{align}
    T_{L}(z) = T_{R}(z')|_{\theta=\theta'=0} .
\end{align}
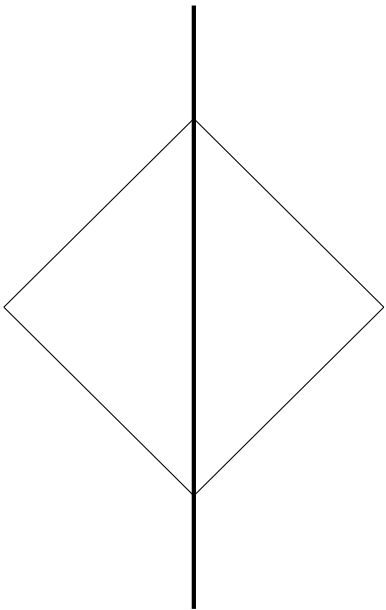
\begin{figure}
    \centering
   \begin{tikzpicture}[scale=.5]
        
       \draw[ultra thick](0,-8)--(0,8);
      \draw[ thin](0,5)--(-5,0);
      \draw[ thin](-5,0)--(0,-5);
      \draw[ thin](0,-5)--(5,0);
      \draw[ thin](5,0)--(0,5);
     
   \end{tikzpicture}
    \caption{Dual to gluing of two same 1D systems}
    \label{same 1d}
\end{figure}

In this set-up, the trivial QES should correspond to a zero interval or a point at $\theta=\theta' =0$ (see the fig(\ref{same 1d})). We can compute it as follows:
\begin{align}
    S^{trivial} = \lim_{\epsilon\rightarrow 0}S_{CFT}(\epsilon,-\epsilon)|_{\forall \; (\text{constant}\; s)} = 0
\end{align}
Hence schematically this would suggest writing $|i\rangle_{LR} =|i\rangle_{L}\otimes |i\rangle_{R}$ . However this means the interface must be factorizing. 
The holographic result suggests, that we can not couple them. This can be understood from our discussion in the previous section. From (\ref{Itot2}), we can see that the two same Schwarzian term will cancel each other. Hence, the CFT sectors will be decoupled into two isolated BCFTs.

\begin{figure}
    \centering
   \begin{tikzpicture}[scale=.5]
         \draw[ thin](-5,-5)--(5,5);
       \draw[ thin](5,-5)--(-5,5);
        \draw[ultra thick](-5,5)--(-5,-5);
       \draw[ultra thick](5,5)--(5,-5);
        \draw[ thin](5,5)--(10,0);
        \draw[ thin](10,0)--(5,-5);
        \draw[ thin](-5,5)--(-10,0);
        \draw[ thin](-10,0)--(-5,-5);
      
   \end{tikzpicture}
    \caption{Dual to two coupled 1D systems in TFD. Four triangles from left to right are labeled by $L'-L-R-R'$}
    \label{gluing 1d}
\end{figure}
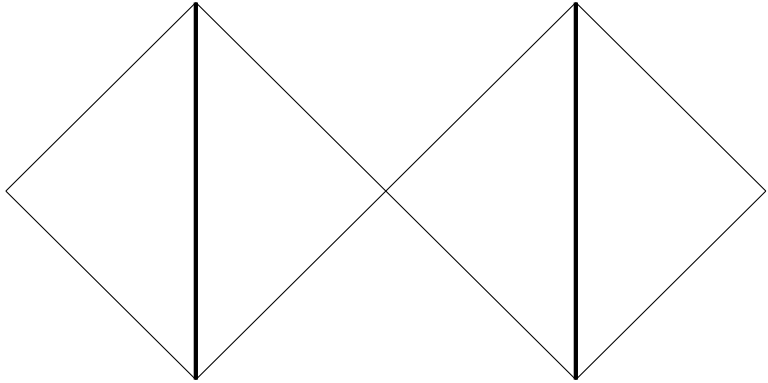

\textbf{(iv) Pair of coupled (same) 1d systems in a TFD state:}
Motivated by \cite{Almheiri:2019yqk}, we would like to study a joined set-up in a TFD state as in fig(\ref{gluing 1d}).

In particular, we consider two 1d coupled to 1d systems, say $LL'$ and $RR'$, in a TFD state. Each pairs( $LL'$ and $RR'$) will not couple to the other. We construct a time-dependent TFD state between each pair as follows:
\begin{align}
    |TFD\rangle_{LL'RR'} = e^{i(H_{L}+H_{R}+H_{L'}+H_{R'})s}\sum_{E}e^{-4\beta E} |E\rangle_{LL'}|E\rangle_{RR'} 
\end{align}
Here $H_{L}=H_{L'}=H_{R'}=H_{R}$, $\beta=\frac{1}{\mu}$ and $|E\rangle_{LL'}=|E\rangle_{L}|E\rangle_{L'}$, $|E\rangle_{RR'}=|E\rangle_{R}|E\rangle_{R'}$\footnote{As we have seen from the previous set-up.}. Here the dual two-sided blackhole has forward directed time $s$. However, one can easily see for such state,
\begin{align}
    \rho_{LL'}=\rho_{RR'} = \rho_{LR'}= \rho_{LR}=\rho_{L'R'} =\rho_{L'R} = \text{time independent}.
\end{align}
If we compute $S_{LR}$, the trivial QES would correspond to the region $\Sigma_{0}=\{-s,\theta_{L}:(0,\Lambda)\}\cup\{ s,\theta_{R}:(0,-\Lambda)\} $, for which we have
\begin{align}
    S_{LR}^{trivial} \sim  \frac{2c}{3}\mu\Lambda = 2S_{BH}
\end{align}
Similarly to the previous examples, one can also check $S^{non-trivial} > S^{trivial}$ and hence $S_{LR}=S^{trivial}_{LR}$. Note that, this is true once we use the input from modular quantization with a near-horizon cut-off and taking $\Lambda\rightarrow\infty$ limit. However, if we instead use the eternal black hole has a \textit{smooth interior}, the region for computing trivial QES would correspond to $\Sigma_{0}=\{(-s,\theta_{L}=0)\cup(s,\theta_{R}=0)\} $ with no cut-off at the horizon. For this case $S^{trivial}_{LR} \sim \frac{c}{3}\mu s$ at large $s$. However this is \textit{inconsistent} with the time-independent set-up. We can also check the correctness of the answer we got by computing $S_{LL'},S_{LR'}$. The trivial surface corresponding to $S_{LL'}$ would be $\Sigma_{0}=\{s,\theta_{L}:(\Lambda,0)\}\cup \{s,\theta_{L'}:(0,-\Lambda)\}$. This is because at $\theta=0$ there should be a factorizing interface as we argued previously. Hence the corresponding EE becomes:
\begin{align}
    S^{trivial}_{LL'} \sim \frac{2c}{3}\mu\Lambda = 2S_{BH}
\end{align}
Similarly one can also check $S_{LR'}=S_{L'R}=2S_{BH}$. For that, we should use factorization of twist two point function into two twist one point function in BCFT set-up as both $L$ and $R'$ (similarly $L'$ and $R$) are infinite distance away from each other. In each cases, trivial QES dominates and it shows the same time independent result of $2S_{BH}$. 

\subsection{Dynamical case: Two coupled 1d systems with different Hamiltonians}

From all of the previous non-dynamical examples, the key point we note that trivial QES term would actually be dominated in the entanglement entropy of the full quantum description in 1d. 
However, this is not the full story, since all these set-up provide uninteresting non-dynamical background. Now we will try to construct a dynamical set-up where non-trivial QES would play an important role.

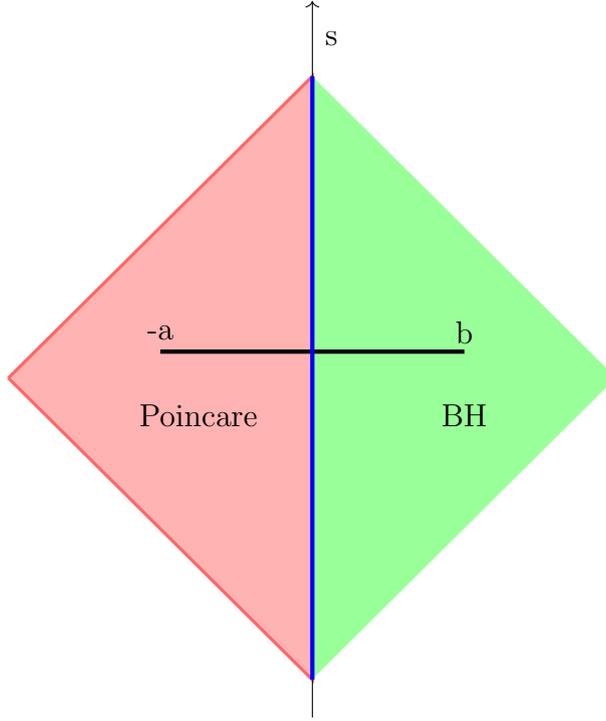
\begin{figure}
    \centering
   \begin{tikzpicture}[scale=.5]
    \filldraw[color=red!60, fill=red!30, very thick](-8,0)--(0,8)--(0,-8)--(-8,0);
       \filldraw[color=green!6, fill=green!40 ](8,0)--(0,-8)--(0,8);
        \draw[ultra thick](-4,.7)--(4,.7);
         \draw node at (-4,1.2) {-a};
         \draw node at (4,1.2) {b};
        \draw [thin, ->] (0,-9) -- (0,10) node at (.5,9) {s};
        \draw[ultra thick, blue](0,-8)--(0,8);
        \draw node at (4,-1) {BH};
        \draw node at (-3,-1) {Poincare};
   \end{tikzpicture}
    \caption{Black hole-Poincare gluing }
    \label{BH P gluing}
\end{figure}

For this purpose, we will couple two CFTs in two different AdS$_{2}$ background in JT gravity framework. In particular, we choose one to be the one-sided AdS$_{2}$ black hole and the other to be the AdS$_{2}$ Poincare. From equation (\ref{semiclassical gluing}), we know that we can consistently couple these two spacetimes (see fig(\ref{BH P gluing})) semiclassically. 

To begin with, we consider the Poincare patch ($\omega',\bar{\omega'}$) having coordinates $[s,\theta':(0,-\infty)]$ and the black hole patch($\omega,\bar{\omega}$) with coordinates $[s,\theta:(0,\infty)]$. They are coupled at $\theta=\theta'=0$. The trivial QES corresponding to this case, is the CFT entanglement entropy of the region $\theta=\theta'=0$. However we will have two possibilities of the trivial QES depending on two different limit of approaching $\theta=\theta'=0$ :
\begin{align}
    S^{trivial}=\lim_{\epsilon\rightarrow 0}S_{CFT}(\{(s,\theta=\epsilon)\cup (s,\theta'=-\epsilon)\} \; or, \; \lim_{\epsilon\rightarrow 0}S_{CFT}\{(s,\theta=\epsilon)\cup (s,\theta=0)\}
\end{align}
In 2d deformed Hamiltonian language, we can write them in terms of boundary twist field $\tau_{b}$ correlation functions as
\begin{align}
    &S^{trivial-I} = \lim_{n\rightarrow 1}\frac{1}{1-n}\log(\langle U_{heating}^{\dagger}\tau_{b}(0)U_{heating}U^{\dagger}_{pt}\tau'_{b}(0)U_{pt}\rangle), \\ &S^{trivial-II} = \lim_{n\rightarrow 1}\frac{1}{1-n}\log(\langle U_{heating}^{\dagger}\tau_{b}(0)\tau_{b}(0)U_{heating}\rangle) \; or, \; \lim_{n\rightarrow 1}\frac{1}{1-n}\log(\langle U_{pt}^{\dagger}\tau_{b}(0)\tau_{b}(0)U_{pt}\rangle).  
\end{align}
Here $U_{heating}$ refers to evolution operator for heating phase or modular Hamiltonian and $U_{pt}$ refers to that for phase transition Hamiltonian. 
Note that the last possibility corresponds to  taking $\theta'=-\epsilon$ and then  $\epsilon \rightarrow 0$. This last case would give zero entropy contribution or $S^{trivial-II}=0$. However the first case becomes time dependent as can be seen from the non-vanishing overlap of evolution operators $U_{heating}U^{\dag}_{pt}$ . Using the map $z=\sqrt{d}\tan[\mu(s+i\theta)]$ and $z'=\frac{-1}{\beta(s+i\theta')}$ we can map to $UHP$ and $LHP$ respectively and after joining them we have a full complex plane with an interface at $Im(z)=Im(z')=0$. Using twist correlation function under this map we get,
\begin{align}
S^{trivial-I}= \frac{c}{3} \log \left[ s\sinh\mu s-\cosh\mu s\right] \end{align}
In the limit of $\mu s >>1$, $S^{trivial-I} \sim \frac{c}{3}\mu s +\mathcal{O}(\log s)$. Here to find the time dependent trivial QES-I, we actually assume a transparent interface. In other words, these two different results correspond to two different boundary conditions. To impose transparent boundary condition, we must have $T_{L}(\omega) =T_{R}(\omega')$. However, as we have seen $\langle T_{L}(\omega)\rangle = \frac{c\mu^{2}}{12}$ and $\langle T_{R}(\omega')\rangle = 0$, the transparent boundary condition is not satisfied at $\theta=\theta'=0$. On the other hand, in the joined $(UHP(z)\cup LHP(z'))$ complex plane, we may still use transparent boundary condition, since $\langle T_{L}(z)\rangle =\langle T_{R}(z')\rangle = 0$ at $Im(z)=Im(z')=0$. Hence mapping to the complex plane, one can use transparent boundary condition to compute EE by which we obtained the $S^{trivial-I}$.  However, we are getting a \textit{wrong} result from trivial QES for transparent boundary condition which has everlasting growth in time. This, by default violates unitarity, since we expect the fine-grained entropy must be upper bounded by coarse grained entropy of the joined system or by the $S_{BH}$. For that matter, we would need to find a solution of non-trivial QES. We will comment on the \textit{correct} or direct way of obtaining the EE later, at the end part of this section.

Before computing the non-trivial QES, we would like to emphasize two comments on two trivial QES corresponding to two different boundary conditions. 
\begin{itemize}
    \item It is obvious that we can not impose both  reflecting boundary condition and transparent boundary condition at the boundary interface simultaneously. This fact is also reflected in getting two different answers of $S^{trivial}$. If we only consider reflecting condition, this corresponds to $S^{trivial-II}=0$ and we do not have interesting dynamical setting anymore. However, if we give up the Dirichlet boundary condition and only use the transparent one, we will have time dependent answer as reflected in the \textit{wrong} result of $S^{trivial-I}$.

    \item  From the conformal gluing of JT gravity set-up, we can use the transparent boundary condition in the full quantum picture of dual QM systems as we explained earlier. However, the \textit{wrong} CFT answer is a reflection of wrong CFT set-up. To be precise, we have used transparent interface in complex plane at $Im(z)=Im(z')=0$ and map it back to $\theta=\theta'=0$ to compute the entanglement entropy corresponding to trivial QES. At $\theta=\theta'=0$ it becomes non-conformal as we mentioned. In the literature, non-conformal boundary condition has been dealt in the `moving mirror'\cite{Akal:2021foz} and `moving interface' set-up \cite{Biswas:2024xut}, which was dubbed as `radiative boundary condition'. It was completely justified, since the dynamical mirror or interface explicitly breaks the conformal invariance. Hence a conformal mapping from a static boundary to a dynamical boundary manifests a breaking of conformal boundary condition. However, we are using non-conformal boundary condition in the static line $\theta=\theta'=0$. Thus the trivial QES provides a wrong CFT answer. To correct it, we need to find dynamical interface satisfying $z(s)=z'(s)$, which \textit{do not} have a local curve solution. 
\end{itemize}
Hence the question now we will address is whether in the JT/1D duality, non-trivial QES could rescue unitarity. 

To obtain the non-trivial QES, here we need to find solution of extremization in both AdS spacetimes. In the context of \textit{gravitating bath}, this type of extremization has been used \cite{Geng:2020fxl}. The modified QES proposal would be
\begin{align}
    S^{non-trivial} &=min\left[ext_{a,b}\left(S_{gen}(-a,b)\right)\right]\nonumber \\
    &\equiv min\left[ext_{a,b}\left( \frac{\phi_{Poincare}(-a)}{4G_{N}}+\frac{\phi_{BH}(b)}{4G_{N}}+S_{CFT}\{(s,-a\}\cup\{s,b)\}\right)\right]
\end{align}
Here the points $(b,-a)$ are indicated in fig(\ref{BH P gluing}). Hence to find the QES, we have to solve the equations:
\begin{align}\label{extrem}
    \partial_{a}S_{gen}(-a,b)|_{a=a^{*},b=\text{fixed}}=0 \; \text{and} \; \partial_{b}S_{gen}(-a,b)|_{b=b^{*},a=\text{fixed}}=0
\end{align}
Here the $S_{gen}$ will be
    \begin{equation}\label{generalized}
\begin{split}
S_{gen}(-a,b)= \frac{1}{4G_{N}}\left[\frac{ \mu\phi_{r}}{\tanh[2\mu b]} -\frac{\phi'_{r}}{a}\right] + \frac{c}{6} \log \left[ (s-a)\sinh[\mu(s+b)]-\cosh[\mu(s+b)]\right] \\  + \frac{c}{6} \log \left[ (s+a)\sinh[\mu(s-b)]-\cosh[\mu(s-b)]\right]
\end{split}
\end{equation}
Here we denote $\phi_{r}$ as the dilaton boundary value, which we described as `$a$' in the rest of the note. We have also used the dilaton value of Poincare as $\phi_{Poincare}(-a) = -\frac{\phi'_{r}}{a}$. 
From the first equation of (\ref{extrem}), we have
\begin{align}
    4G_{N}ca^{3}+6\phi'_{r}a^{2}\sim 6\phi'_{r}s^{2}|_{a=a^{*}} \; \text{for} \; s>>\frac{1}{\mu}
\end{align}
In the semiclassical limit of $G_{N}\rightarrow 0$ and identifying $c \propto \phi_{r} \propto \phi_{r}' \sim \mathcal{O}(1)$\footnote{The first proportionality comes from (\ref{other effective c}), whereas the second one is a demand, consistent with the transparent boundary condition.}, we get
\begin{align}
    a^{*} \sim s \; \text{at} \; s>>\frac{1}{\mu}.
\end{align}
Hence this indicates, at infinite time limit, $a$ will approach to the Poincare Horizon. Similarly, from the second equation of (\ref{extrem}), we will end up with
\begin{align}
    b^{*} \sim \frac{s}{3} \; \text{at} \; s>>\frac{1}{\mu}.
\end{align}
Hence again at large $s$ limit, the QES solution will tend to the black hole horizon. However note that, the solution is still outside the horizon even at infinite time limit. This is in parallel to the finding in \cite{Almheiri:2019yqk},\cite{Bousso:2023kdj}. If we just naively put these extremal solutions back to (\ref{generalized}), we will still have linear in time growth as $S_{gen}\sim S_{BH}+\frac{c}{3}\mu s+\mathcal{O}(\log s)$. However, in our picture, we have a near-horizon cut-off placed at $\theta=\Lambda$. Once the extremal surface reaches the cut-off $b^{*}=\Lambda$, the $S_{CFT}$ part will become vanishing \footnote{Since this includes the full spatial slice of the full black hole plus Poincare wedge.}. This will in turn gives
\begin{align}
    S^{non-trivial} \sim \frac{\mu\phi_{r}}{4G} = S_{BH}
\end{align}
Taking minimum of trivial and non-trivial QES will give the correct entanglement entropy at late time $s>>\frac{1}{\mu}$:
\begin{align}
    min\left(S^{trivial-I},S^{non-trivial}\right)|_{s>>\frac{1}{\mu}} = S_{BH}
\end{align}
This would provide a similar picture to \textit{Page curve}, where the Page time $s_{Page}$ is given as
\begin{align}
  s_{Page} \sim \frac{3}{c\mu}S_{BH} \propto \Lambda
\end{align}
Hence the Page time will be finite for a finite yet large cut-off $\Lambda$. In $\Lambda \rightarrow \infty$ the Page time also becomes infinite. We suspect, this calculation could mimic a collapsing star and creation of a black hole similar to \cite{Almheiri:2019yqk}. The crux point of this calculation is that, \textit{in the absence of a smooth interior or a disconnected island, one can still able to find a `Page curve'\footnote{The relation to actual Page curve for Hawking radiation is still not clear from this picture.} from QES calculation, where the stretched horizon cut-off plays the key role to restore the unitarity along the line \cite{Mathur:2009hf},\cite{Almheiri:2012rt} which predicted an existence of firewall in evaporating black holes.}

To conclude this section, we will summarize the key points:
\begin{itemize}
    \item \textit{For most of the non-dynamical cases, the entanglement entropy for states in 1d quantum system can be computed from trivial QES in proper UV frame of JT gravity or by direct CFT computation in deformed CFT set up in $c_{eff}\rightarrow \infty$ with finite $c$.}

    \item \textit{In our simple setting, we can still make an interesting dynamical set-up where non-trivial QES could play a role to manifest unitarity. In doing so, the stretched horizon cut-off would dictate a finite `Page time'. However the wrong trivial QES result would correspond to a wrong CFT set-up which we expect to be corrected by using dynamical non conformal interfaces.}

\end{itemize}
\section{Discussion}\label{sec5}
In this note, we have proposed an alternative UV completion of pure JT gravity (away from extremality) and that of a CFT (with central charge $c \sim \mathcal{O}(1)$) coupled to JT gravity on one-sided AdS$_{2}$ black hole background, in terms of certain SL(2,$R$) deformed CFTs on strip. In particular, following quantization of modular Hamiltonian on a strip as a deformed holographic CFT Hamiltonian with effective central charge $c_{eff}=c\Lambda$; the limit $c_{eff} \rightarrow \infty$ with a fixed $\Lambda$ and $c \rightarrow \infty$, provides a Hilbert space description of pure JT gravity in AdS$_{2}^{b}$ black hole. The dual CFT Hilbert space is defined by the holographic CFT made out of only stress tensor sector, in one-sided AdS$_{2}$ black hole with a stretched horizon placed at a distance $\Lambda$. On the other hand, taking $c_{eff} \rightarrow \infty$ limit with $\Lambda \rightarrow \infty$ while keeping $c$ finite, would describe the Hilbert space of a (non-holographic) CFT coupled to JT gravity in classical limit. In this note, we provide zeroth order checks on these two claims. 

However from our proposals, it is not entirely clear whether pure quantum JT gravity itself is well-defined or not. Strictly speaking, in our holographic identification, pure JT quantum gravity is related to the question of constructing CFT dual of pure Einstein gravity in 3D with all orders in $G_{N}$. It still remains a question of debates \cite{Witten:2007kt}-\cite{Castro:2011zq}, whether one can have a consistent CFT$_{2}$ dual of pure gravity in AdS$_{3}$, made out of only vacuum sector of certain CFT. However, there are evidences in literature \cite{Mertens:2017mtv},\cite{Mertens:2018fds} where JT gravity and Schwarzian theory can be embedded in Liouville theory as an expected dual of pure Einstein gravity. Also holographic large $c$ deformed CFTs(in heating phase) shows maximal Lyapunov behavior as noted in \cite{Das:2022jrr}.  Nevertheless, we must emphasize that the full holographic deformed CFT Hilbert space\footnote{Which contains both matter and stress tensor sectors and must be well-defined in standard AdS/CFT} on cylinder describes the quantum gravity Hilbert space of large BTZ black hole. On the other hand, for the case of JT with conformal matter, to study the backreaction of CFT($c>>1$) at finite $cG_{N}$, one should deform the boundary condition on strip. From the results of last section, we have already gathered hints that CFT entanglement entropy computation on a dynamical setting can be correctly done in a set-up, where the boundary condition on the strip should be deformed along a curve, which reproduces non-vanishing momentum flux at the asymptotic boundary in UV frame. It would be an interesting future direction to study deformed CFT on strip with dynamical boundary where we can impose non-conformal boundary condition. In this way, we can track the back-reaction problem of JT gravity purely from deformed CFT set-up. 

One natural follow-up question would be to find an extension, which can incorporate the topological part in JT gravity picture. We expect to have some additional conserved charge like U(1) in the Hilbert space of deformed CFT\footnote{See \cite{Banerjee:2024zqb}, where similar kind of Floquet CFT has been used in a different context.}, which could further mimic the additional entropy count of JT gravity black holes due to the presence of extremal sector. Studying such deformed CFT with additional charge sector in higher genus surface might shed some light on \textit{wormhole} amplitudes of JT gravity in path integral description. We would like to get back to this interesting direction in near future.

We now conclude this note by discussing another future direction with speculative remarks. Serious readers may skip this part.
\subsection{On the other scaling limit from the bulk} 
Our proposal on UV completion of pure JT gravity is motivated from AdS/CFT with a prescribed limit of $c_{eff}\rightarrow \infty$, in the Hilbert space constructed out of modular quantization of certain holographic CFT$_{2}$. The interesting question remains on how to understand another scaling limit from the bulk side, which is defined by $c_{eff} \rightarrow \infty$ such that $\Lambda\rightarrow \infty$ keeping $c$ fixed and $\mathcal{O}(1)$. We have described such limit as the classical limit of JT coupled to non-holographic CFT. 
Hence an interesting question to ask is whether such limit arises in any possible bulk description. A priori, without knowing the full details of the quantum gravity in bulk, one might expect the appearance of a coupling constant $G_{3,eff}\equiv \frac{G_{3}}{\Lambda}$ in certain effective description of bulk quantum gravity theory in BTZ. This is already appeared in taking $G_{3} \rightarrow 0$ with finite $\Lambda$ as JT term upon dimensional reduction of Einstein gravity in large BTZ or approximately planar BTZ with boundary cut-off spacetime
\begin{align}\label{bulk decoupling 1}
    S_{(G_{3}\rightarrow0,\Lambda \; \text{fixed})}^{b} = \frac{\Lambda}{8\pi G_{3}}\int_{\text{AdS}_{2}^{b}} \sqrt{-h}\;\tilde{\phi}(R+2) 
\end{align}
However the similar action might also be appeared in the other limit by taking $\Lambda \rightarrow \infty$ and $G_{3}$ finite:
\begin{align}\label{bulk decoupling 2}
   S_{(G_{3}\;\text{finite},\Lambda \rightarrow \infty)}^{b} =\left[ \underbrace{\frac{\Lambda}{8\pi G_{3}}\int_{\text{AdS}_{2}^{b}} \sqrt{-h}\;\tilde{\phi}(R+2)}_{\equiv W_{3}}\right]_{\text{finite}\; G_{3} \text{ background}} 
\end{align}
Usually one should not expect to have a classical gravity term in finite $G_{3}$ description, where the effect of background fluctuation takes over. Only the appearance of $\Lambda$ as the overall coefficient in the JT action enforces it to restore in such putative $\Lambda \rightarrow \infty$ limit. However we can \textit{not} treat it as a classical term by any means. Interestingly, in \cite{Solodukhin:1998tc}, it was observed that under a change of variables $\tilde{\phi}= 2\pi q G_{3,eff}\Phi_{h}\phi$ and $h_{ab}=e^{\frac{2}{q\Phi_{h}}\phi}\bar{h}_{ab}$, one can rewrite the JT gravity action as a \textit{deformed Liouville} action in 2D
\begin{align}\label{W3}
  W_{3} = \int_{\bar{AdS}_{2}^{b}} \sqrt{-\bar{h}}\left( \frac{1}{2}(\nabla\phi)^{2}+\frac{1}{4}q\Phi_{h}\phi R + U_{3}(\phi)\right)  \; \text{where} \; U_{3}(\phi) \equiv \frac{1}{2}q\Phi_{h}\phi e^{\frac{2}{q\Phi_{h}}\phi}.
\end{align}
Here, $q,\Phi_{h}$ are arbitrary constants. This action differs from the usual Liouville CFT action due to the different potential term $U_{3}(\phi)$. Hence this describes a \textit{non-conformal} theory having non-vanishing stress tensor component $T_{+-}$\cite{Solodukhin:1998tc}. The horizon value of classical dilaton field $\phi$ is related to $\Phi_{h}$ as $\Phi_{h}=\frac{\sqrt{d}}{2\pi qG_{3,eff}\phi_{h}}$.

Being ignorant of the quantum gravity literature, We now \textit{imagine} a description of certain UV theory in AdS-Rindler, could be possibly described by certain 2D theory lives in the worldsheet metric of AdS$^{b}_{2}$. This theory end on a stretched horizon brane located at $r=\sqrt{d}+\epsilon$. Further we assume, in $\epsilon\rightarrow 0$ or $\Lambda\rightarrow\infty$ limit, the worldsheet theory of field $\phi$ should be described by (\ref{W3}). This is not truly a worldsheet theory being a non-conformal one. However, we will now argue, this approaches to a conformal theory in the same limit.

If one now change the variable of the AdS$_{2}^{b}$ metric in terms of $\omega^{b}=s+\theta^{b}$, $\bar{\omega}^{b} = s-\theta^{b}$, where $d\theta^{b} \equiv \frac{dr}{r^{2}-d}$, such that $\theta^{b} = \frac{1}{2\sqrt{d}}\ln\left(\frac{r+\sqrt{d}}{r-\sqrt{d}}\right)$, it was shown in  \cite{Solodukhin:1998tc}, the classical equation of motion by varying $\bar{h}_{ab}$ and $\phi$, one can find the following:
\begin{align}
    \Box_{\omega^{b}}\phi = 0; \; T_{\omega^{b}\bar{\omega}^{b}}=0, \; \text{when} \; \theta^{b} \rightarrow \infty
\end{align}
Here, $\theta^{b} \rightarrow \infty$ dictates taking \textit{near horizon} limit. In \cite{Das:2024vqe}, it was argued one can identify
\begin{align}
    \Lambda = \frac{1}{2\sqrt{d}}\ln(\frac{2\sqrt{d}}{\epsilon}) \sim \theta^{b}(\epsilon),
\end{align}
 where $\epsilon$ is the stretched horizon cut-off in the AdS-Rindler defined as: $r=\sqrt{d}+\epsilon$. Hence, in the vicinity of the horizon, we can treat $\epsilon$ as the dynamical parameter and taking $\theta^{b}\rightarrow \infty$ reduces to $\epsilon \rightarrow 0$ or $\Lambda \rightarrow \infty$ and in that limit a \textit{classical conformal symmetry } emerges satisfying equation of motions of \textit{free massless scalar fields} $\phi$ with a vanishing stress tensor in the AdS$_{2}^{b}$($s,\theta_{b})$ which is conformally flat metric. In \cite{Solodukhin:1998tc}, it was further argued that exploiting this emergent near horizon conformal symmetry, one can derive a classical \textit{Virasoro algebra} in the vicinity of the horizon. However, in the emergent near horizon CFT, the central charge $c$ is defined as $c=3\pi q^{2}\Phi_{h}^{2}$. It might be possible to choose  arbitrary $q,\Phi_{h}$ such a way that $c =1$.  Hence we can rewrite (\ref{bulk decoupling 2}) as
\begin{align}\label{reduced bulk S}
    S_{(G_{3}\;\text{finite},\Lambda \rightarrow \infty)}^{b} =\left[ (\text{classical emergent CFT$_{c=1}$})^{\text{AdS}_{2}^{b}(\Lambda)}_{\Lambda\rightarrow \infty}\right]_{\text{finite}\; G_{3}\; \text{background}} 
\end{align}
We are again emphasizing that, the deformed Liouville theory of the scalar field $\phi$ defined in (\ref{W3}), is \textit{not} a CFT. The classical conformal symmetry corresponding to free massless scalar field with a classical Virasoro algebra emerges only in the near horizon limit by taking $\theta^{b}\rightarrow \infty$. Hence the quantum theory defined by (\ref{reduced bulk S}), should match with $\mathcal{H}_{c=1,\Lambda\rightarrow\infty}$ which is further described by a quantization of finite $c$ CFT on a AdS$_{2}$ black hole background with a stretched horizon approaching the horizon at $\Lambda \rightarrow \infty$. For further supporting evidence, we have also seen the appearance of \textit{infinite degenerate vacuum} sectors made by \textit{conformal primaries} at the horizon. It is also well known to have infinite number of primaries as vertex operators in the Hilbert space of free bosonic CFTs. Also, it is recently noted \cite{Burman:2023kko},\cite{Burman:2024egy} that, free massless scalar fields\footnote{Those studies include 3D massless scalars on the stretched horizon background, for which a fictitious cut-off on angular momentum modes has been appeared. This problem goes away, once free massless scalar theory on induced 2D plane would be considered, which does not affect any of the other claims of those papers.} in the presence of a stretched horizon in BTZ background, show certain conformal signature in the near horizon limit; which includes Cardy like density of state with an effective central charge\footnote{See also \cite{Krishnan:2024sle} for higher dimensional analogues.}. In the same description with same limit also describes an emergent Hartle Hawking correlators \footnote{See also \cite{Banerjee:2024dpl}-\cite{Banerjee:2024ivh}, for alternative takes on this in similar background with massive scalar fields$-$probably with some other motivation which does not resonate with the current discussion.} as a signature of low energy limit in BTZ background, which makes a translation to taking classical limit of JT coupled to CFT case. Hence from all of these observations, we can conjecture the following:
\begin{align}
    (\text{Modular quantization of CFT}_{ c=1})^{\text{AdS}_{2}(\Lambda)}_{\Lambda\rightarrow\infty} = (\text{Quantization of emergent near horizon CFT})^{\text{AdS}^{b}_{2}(\Lambda)}_{\Lambda\rightarrow\infty}
\end{align}
 We hope to return with more evidences of this conjecture from certain explicit theory of quantum gravity in AdS$_{3}$. We note that, the appearance of near-horizon CFT in 2d is universal in all higher dimensional spherically symmetric \textit{non-extremal} black-holes \cite{Solodukhin:1998tc} as well as arbitrary Rindler regions satisfying Einstein equation\cite{Banks:2021jwj}. \\

\section{Acknowledgement}
We extend our heartfelt gratitude to Bobby Ezhuthachan for collaboration and for his steadfast support throughout this project. His insightful critiques—often plain “no, it’s not possible”—and thought-provoking discussions greatly shaped our work.
Our thanks also go to Mir Afrasiar, whose casual suggestion during a friendly conversation sparked one core problem we have tackled here.
We gratefully acknowledge Parthajit Biswas, Diptarka Das, Bidyut Dey, Nilay Kundu, Koushik Ray, Rahul Roy, Krishnendu Sengupta, as well as ChatGPT and Meta AI, for numerous related discussions—both formal and informal—and for their direct and indirect help over the course of the research.
The research work by SD is supposed to be supported by the DST–INSPIRE Faculty Fellowship, which remains pending over 11 months due to some frustrating procedural delays.  He thanks his family members for their unconditional support during these frustrating days.
We also thank many cafes, restaurants, and many beautiful people of beautiful Kolkata for providing an inspiring and energizing environment conducive to research. The research work by AD  is supported in part by the National Key Research and Development Program
 of China under Grant No. 2020YFC2201504.

\appendix
\section{Modular quantization on strip}\label{app1}
Here we briefly review the modular quantization on a strip, which is a straightforward generalization of the same on cylinder as introduced and further studied in \cite{Tada:2019rls},\cite{Das:2024mlx}. Here we take the form of the Hamiltonian to be the same or even simpler, but defined on the strip,
\begin{align}
    H=\beta(L_{1}+L_{-1}+\bar{L}_{1}+\bar{L}_{-1}), \; \text{where} \; d\equiv -4\beta^{2}<0
\end{align}
Here $L_{n}=\frac{1}{2\pi i}\int_{C}dz z^{n+1}T(z)$ and $\bar{L}_{n}=\frac{1}{2\pi i}\int_{\bar{C}}d\bar{z} \bar{z}^{n+1}\bar{T}(\bar{z})$, with the condition $T(z)=\bar{T}(\bar{z})$ at $z=\bar{z}$. $C$ and $\bar{C}$ describes half circle contour on UHP. Note that, the above Hamiltonian preserve the symmetries of the UHP. An easy way to check this is by taking space-time generators of the Hamiltonian $\beta(l_{1}+l_{-1}+\bar{l}_{1}+\bar{l}_{-1})$ and acting it on the real line:
\begin{align}
    \beta(z^{2}\partial_{z}+\partial_{z}+\bar{z}^{2}\partial_{\bar{z}}^{2}+\partial_{\bar{z}})(z-\bar{z})|_{z=\bar{z}}=0
\end{align}
Following \cite{Ishibashi:2016bey},\cite{Tada:2019rls}, we can rewrite $H = \mathcal{L}_{0}+\bar{\mathcal{L}}_{0}$, which defines a new scaling operator in a new frame of quantization $(s,\theta)$. 
Here the flow in the spatial direction $\theta$ is given by $i(\mathcal{L}_{0}-\bar{\mathcal{L}}_{0})$. 
The next step is to find the time($s$) and space($\theta$) contours in the UHP($z,\bar{z}$). If we denote the frame by $\omega \equiv s+i\theta$, we get \cite{Das:2024mlx}
    \begin{align}\label{t,s curve}
   \omega= s+i\theta = \frac{1}{i\sqrt{d}}\ln\frac{z-z_{+}}{z-z_{-}} \; \text{and} \; \bar{\omega}=s-i\theta = -\frac{1}{i\sqrt{d}}\ln\frac{\bar{z}-z_{-}}{\bar{z}-z_{+}}; \; \text{where} \;  z_{\pm} = \frac{\pm i\sqrt{d}}{2\beta}
\end{align}
Here $z=x+iy$ with $y\geq 0$. Also from now on, we denote $|d|$ by $d$. From the above result, one can see $z=\frac{\sqrt{d}}{2\beta}\tan(\frac{\sqrt{d}}{2}\omega)$. Here we use slightly different notation from (\ref{deformed Ham1}), which can be recovered once we scale $\sqrt{d} \rightarrow 2\beta\sqrt{d}$ in the new definition. From (\ref{t,s curve}), we get constant $s$ and constant $\theta$ curves of the contour as the following:
\begin{align}\label{const t s curve}
    &(x-z^{R}_{*}+z^{I}_{*}\cot \sqrt{d}s)^{2}+y^{2} = (z^{I}_{*})^{2}(\csc\sqrt{d}s)^{2} \\
   & (x-z^{R}_{*})^{2} + (y-z^{I}_{*}\coth\sqrt{d}\theta )^{2} = (z^{I}_{*})^{2}(\csch\sqrt{d}\theta)^{2}
\end{align}
    
\begin{figure}
    \centering
   \begin{tikzpicture}[scale=.5]
      \draw[ thin](5,8)--(5,9);\draw[ thin](5,8)--(6,8); \draw[brown] node at ( 5.5,8.5) {Z};
        \draw[ thick,->](-8,0)--(8,0);
         \draw[ thick](0,-8)--(0,8);
    
        \draw [thin, ->] (0,-9) -- (0,10) ;
        \draw node at (.8,5.4) {$z_{+}$};
         \draw node at (-4,5.4) {constant $\theta$ curve};
          \draw node at (7,1) {constant $s$ curve};
         \draw node at (-.6,5.4) {$\epsilon$};
        \filldraw [red] (0,5) circle (3pt);
        
        \draw[dashed] (0,5) circle (.4cm);
         \draw(0,5)circle(1);
          \draw(0,5)circle(1.5);
           \draw(0,5)circle(2);
        \draw[color=green] (5,0) to [bend left=-35] (0,5);
       \draw[color=red]  (4,0) to [bend left=-35] (0,5);
       \draw[color=purple]  (3,0) to   [bend left=-35] (0,5);
        
       \draw[thick,->,color=purple]  (4,3) to   [bend left=-15] (3,4);
        
      \draw[thick,->,color=blue]  (-1,7.2) to   [bend left=-15] (-2,6.5);
   \end{tikzpicture}
    \caption{Contours of modular quantization on UHP($z$). The red arrow dictates flow in $\theta$ direction, while the blue arrow indicates time($s$) direction. $z_{+}$ is the fixed point and the smallest dashed circle with around it, indicates fixed point cut-off with radius $\epsilon$.}
    \label{contour}
\end{figure}
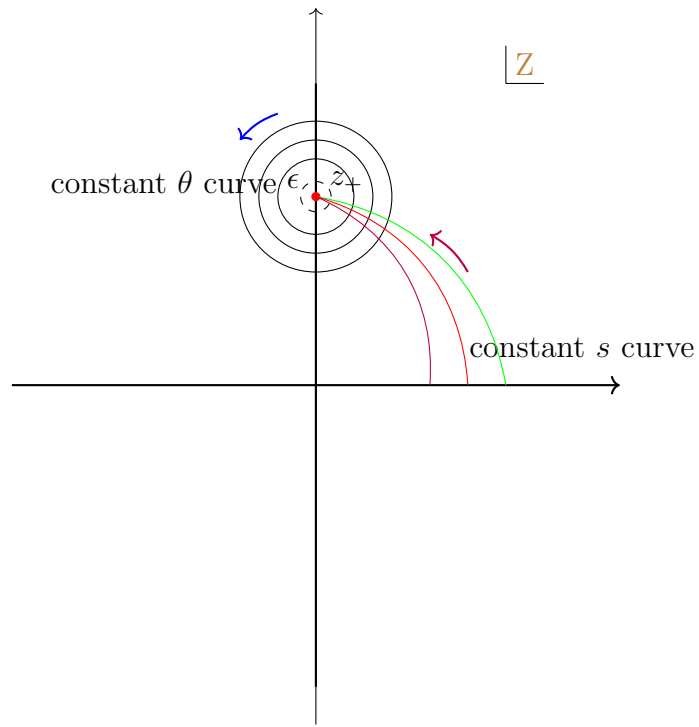

Here we defined $z_{\pm}=z^{R}_{*}\pm iz^{I}_{*}$, with $z^{R}_{*}=0$ for our set-up. From the constant $s$ curve, we can see the existence of fixed point(which corresponds to $z=z_{+}$ on the UHP), where the curve should end\cite{Tada:2019rls},\cite{Das:2024mlx}. This prevents the $\theta$ curves to form a complete semi-circle as in figure(\ref{contour}). However, constant $\theta$ curve will be circular and it can be shrunk to the fixed point at $\theta=\infty$. Hence the Euclidean $s$ generates contractible time circles. The range of $\theta:(0,\infty)$ on the space of quantization. The plot of the contours is described in fig(\ref{contour}).

In this constant $s$ contour, one can further define \cite{Tada:2019rls},\cite{Das:2024mlx} eigenmodes of the Hamiltonian as the following:
\begin{align}\label{Vir1}
   & \mathcal{L}_{k} = \frac{\beta}{2\pi i} \int_{C'}dz (z-z_{+})^{-\frac{ik}{\sqrt{d}}+1}(z-z_{-})^{\frac{ik}{\sqrt{d}}+1} T(z) \\
     &\mathcal{\bar{L}}_{k} = \frac{\beta}{2\pi i} \int_{\bar{C'}}d\bar{z} (\bar{z}-z_{-})^{\frac{ik}{\sqrt{d}}+1}(\bar{z}-z_{+})^{-\frac{ik}{\sqrt{d}}+1} \bar{T}(\bar{z})    
\end{align}
Here $C'$ and $\bar{C'}$ are the constant $s$ contours for $\theta:(\infty,0)$ and $\theta:(0,\infty)$ respectively. On the UHP, $C'$ corresponds to an arc extending from $z=z_{+}$ to $Im(z)=0$ and $\bar{C}'$ corresponds to $Im(\bar{z})=0$ to $\bar{z}=z_{-}$ in the opposite direction to $C'$. The UHP boundary condition $T(z)=\bar{T}(\bar{z})$ must be imposed at $z=\bar{z}$. When $k=0$, $\mathcal{L}_{0}+\bar{\mathcal{L}}_{0}$ reduces to $H$ with the same contours. 
To study the algebra generated by $\mathcal{L}_{k},\bar{\mathcal{L}_{k}}$, we deform the $\bar{C}'$ contour for $\bar{\mathcal{L}}_{k}$s to Lower Half Plane(LHP), where it extends from $\theta:(0,-\infty)$ or $Im(z)=0$ to $z=z_{-}$ for $Im(z)<0$. This is a standard unfolding trick in BCFT \cite{Cardy:2004hm}, where the Ward identity suggests that a bulk operator at $(z,\bar{z})$ in UHP transforms under any conformal transformation like two holomorphic operators placed at $z$ and $\bar{z}=z^{*}$ in the full complex plane \footnote{See Appendix D of \cite{Das:2019tga} for a review on this.}. Using this, we can define a new set of eigenmodes as the following:
\begin{align}
    \hat{\mathcal{L}}_{k} = \frac{\beta}{2\pi i} \int_{\hat{C}}dz (z-z_{+})^{-\frac{ik}{\sqrt{d}}+1}(z-z_{-})^{\frac{ik}{\sqrt{d}}+1} \hat{T}(z); \; \text{where} \;\Bigg( \begin{split}
        \hat{T}(z)=T(z); \; \text{when} \; Im(z)>0  \\
        \hat{T}(z)=\bar{T}(\bar{z}); \; \text{when} \; Im(z)<0 
    \end{split} 
    \Bigg)
\end{align}
Here $\hat{C}$ is the constant $s$ contour in the full complex plane with $\theta:(\infty,-\infty)$ or, $z:(z_{+},z_{-})$. Using this definition of modes, we write the following:
\begin{align}\label{algebra1}
   & [\hat{\mathcal{L}}_{k},\hat{\mathcal{L}}_{k'}] \nonumber\\
    &= \left(\frac{\beta}{2\pi i}\right)^{2} \int_{\hat{C}_{s_{1}}}dz\int_{\hat{C}_{s_{2}}}d\omega (z-z_{+})^{-\frac{ik}{\sqrt{d}}+1}(z-z_{-})^{\frac{ik}{\sqrt{d}}+1}  (\omega-z_{+})^{-\frac{ik}{\sqrt{d}}+1}(\omega-z_{-})^{\frac{ik}{\sqrt{d}}+1} [\hat{T}(z),\hat{T}(\omega)]
\end{align}
Here $\hat{C}_{s_{1}},\hat{C}_{s_{2}}$ are two contours of constant $s_{1},s_{2}$ such that $s_{1}>s_{2}$. In this way, we can define time ordering as:
\begin{align}
    \mathcal{T}(\hat{T}(z)\hat{T}(\omega)) = \Bigg\{ \begin{split}
        \hat{T}(z)\hat{T}(\omega), \; \text{when} \; |z|>|\omega| \\
        \hat{T}(\omega)\hat{T}(z), \; \text{when} \; |z|<|\omega|
    \end{split} \Bigg\}
\end{align}
From the contour diagram, we can obtain the following identity similar to the radial quantization:
\begin{align}\label{commutator ordering}
    \left[\int_{\hat{C}_{s_{1}}}dz a(z) \hat{T}(z),\int_{\hat{C}_{s_{2}}}d\omega b(\omega)\hat{T}(\omega)\right] = \int_{\hat{C}_{s_{2}}}d\omega b(\omega) \oint_{\omega}dz a(z)\hat{T}(z)\hat{T}(\omega)
\end{align}
T
Hence, using $\hat{T}\hat{T}$ OPE and  (\ref{commutator ordering}) in (\ref{algebra1}), we will obtain the following in a similar way described in \cite{Tada:2019rls}.
\begin{align}\label{algebra2}
    [\hat{\mathcal{L}}_{k},\hat{\mathcal{L}}_{k'}] = (k-k')\hat{\mathcal{L}}_{k+k'} + \frac{c}{12} \frac{(k^{3}+k d)}{2\pi i(k+k')}[f_{k+k'}(z_{+})-f_{k+k'}(z_{-})]
\end{align}
Here $f_{k}(z)$ is defined as follows
\begin{align}\label{f}
    f_{k}(z)= \left(\frac{z-z_{+}}{z-z_{-}}\right)^{-\frac{ik}{\sqrt{d}}}
\end{align}
As one can see $f_{k+k'}(z_{+})$ and $f_{k+k'}(z_{-})$ are divergent quantities, we need to put cut-off around $z=z_{\pm}$ as 
\begin{align}
    z_{c}=z_{\pm} + \epsilon e^{i\theta_{c}}; \; \text{such that} \; \theta_{c}=\Lambda=\frac{1}{\sqrt{d}}\ln\left(\frac{\sqrt{d}}{\beta\epsilon}\right)
\end{align}
Previously we have seen the need of a $\theta=\Lambda$ cut-off to define the Hamiltonian in $\omega$ frame. Here we will see the same cut-off leads to a well-defined Virasoro algebra. Hence, we redefine the modes as the following:
\begin{align}\label{redefined vir gen}
    \hat{\mathcal{L}}_{k}^{\epsilon} \equiv \frac{\beta}{2\pi i} \int_{\hat{C_{\epsilon}}}dz (z-z_{+})^{-\frac{ik}{\sqrt{d}}+1}(z-z_{-})^{\frac{ik}{\sqrt{d}}+1} \hat{T}(z)
\end{align}
where $\hat{C}_{\epsilon}$ defines the contour $z:(z_{+}+\epsilon e^{i\theta_{\epsilon}},z_{-}+\epsilon e^{i\theta_{\epsilon}})$ or $\theta:(\Lambda,-\Lambda)$. 
Inserting this in (\ref{algebra2}), we get
\begin{align}
     [\hat{\mathcal{L}^{\epsilon}_{k}},\hat{\mathcal{L}^{\epsilon}_{k'}}] = (k-k')\hat{\mathcal{L}}^{\epsilon}_{k+k'} + \frac{c}{12} \frac{(k^{3}+k d)e^{(k+k')t}}{\pi(k+k')}\sin\left(\frac{(k+k')}{\sqrt{d}}\ln\left(\frac{\sqrt{d}}{\beta\epsilon}\right)\right)
\end{align}
As reviewed in \cite{Das:2024mlx}, to have a closed algebra satisfying Jacobi identity, one must need to satisfy the following criterion for which the central extension term to be zero when $k+k' \neq 0$:
\begin{align}\label{normal modes}
    k = \frac{n\pi \sqrt{d}}{\ln\left(\frac{\sqrt{d}}{\beta\epsilon}\right)}  , \; n\in \frac{\mathbb{Z}}{2}
\end{align}
Note that, this exactly matches with the normal mode spectrum $\omega_{k}$ of free scalar(fermion) fields of AdS$_{2}$ blackholes in the presence of Dirichlet boundary condition at stretched horizon cut-off at $r=r_{H}+\epsilon$, where $r_{H}\sim \sqrt{d}$ \cite{Soni:2023fke}. This condition further reduces to a closed Virasoro algebra
\begin{align}
  [\hat{\mathcal{L}^{\epsilon}_{k}},\hat{\mathcal{L}^{\epsilon}_{k'}}] = (k-k')\hat{\mathcal{L}}^{\epsilon}_{k+k'} + \frac{c_{eff}}{12}(k^{3}+k d) \delta_{k+k'}  
\end{align}
Where we define $c_{eff} \equiv \frac{c}{\pi\sqrt{d}}\ln\left(\frac{\sqrt{d}}{\beta\epsilon}\right)$. In $c_{eff}\rightarrow \infty$ by taking $\epsilon \rightarrow 0$ limit (with finite $c$), a \textit{continuous Virasoro algebra} emerges with dirac delta function of continuum indices $k$ \cite{Das:2024mlx}.

We note that the structure of the Virasoro algebra is exactly same to the one holomorphic copy of the Virasoro algebra as constructed in \cite{Das:2024mlx} for modular quantization in cylinder. Hence we may also draw a parallel between the `emergent Virasoro algebra' in this modular quantization and emergent `near horizon classical conformal symmetry' \cite{Solodukhin:1998tc} as discussed and connected in \cite{Das:2024vqe}. 

We can rewrite the Virasoro modes in the $\omega$ plane as follows by choosing constant $s=0$ contour\footnote{We choose this contour for technical simplicity throuh out this note. Generalization to constant $s\neq0$ contour will be straightforward.}
\begin{align}
    \hat{\mathcal{L}}_{k}^{\Lambda} = \frac{1}{2\pi}\int^{\Lambda}_{-\Lambda}d\theta e^{ik\theta}\hat{T}(\theta) -\delta_{k,0}\frac{c\Lambda d}{12\pi}
\end{align}
Here we denote $\hat{\mathcal{L}}^{\Lambda}_{k}$ and $\hat{\mathcal{L}}^{\epsilon}_{k}$ in $\omega$ plane in the same footing. This suggests a mode expansion of $\hat{T}(\theta)$ as
\begin{align}
    \hat{T}(\theta) = \frac{2\pi^{2}}{\Lambda}\sum_{n=-\infty}^{\infty}e^{-ik(n)\theta}\hat{\mathcal{L}}_{k(n)}^{\Lambda}+\frac{c\pi d}{6}
\end{align}
Here $k(n) =\frac{n\pi}{\Lambda}$ as in (\ref{normal modes}) for $n \in \mathbb{Z}$.
On the other hand, we can also use the mode expansion of $T(\omega)$ and $\bar{T}(\bar{\omega})$ in terms of $\mathcal{L}_{k}^{\Lambda}$ and $\bar{\mathcal{L}}_{k}^{\Lambda}$ as 
\begin{align}
    T(\theta) = \frac{4\pi^{2}}{\Lambda}\sum_{n=-\infty}^{\infty}e^{-ik(n)\theta}\mathcal{L}^{\Lambda}_{k(n)}, \; \; \bar{T}(\theta) = \frac{4\pi^{2}}{\Lambda}\sum_{k=-\infty}^{\infty}e^{ik(n)\theta}\bar{\mathcal{L}}_{k(n)}^{\Lambda}
\end{align}
Hence, in $\omega$ plane, the boundary condition $T(\theta)-\bar{T}(\theta)|_{\theta=\Lambda}=0$ reduces to the following:
\begin{align}\label{fixedpoint bc}
    (-1)^{n}(\mathcal{L}_{k(n)}^{\Lambda}-\bar{\mathcal{L}}^{\Lambda}_{-k(n)})|B\rangle_{\Lambda} = 0
\end{align}
Here $|B\rangle_{\Lambda}$ is the conformal boundary state in the $\omega$ frame, localized at the cut-off. It will be useful to obtain $|B\rangle_{\Lambda}$ in terms of state in radial quantization on strip. Before obtaining that, we note two points. Firstly, we note that we have similar  form of (\ref{fixedpoint bc}) for imposing asymptotic boundary condition $(T(\theta)-\bar{T}(\theta))|_{\theta=0} =0$. The only difference is that the boundary state we get $|B'\rangle_{0}$, is localized on the asymptotic boundary $\theta=0$. Physically, the boundary condition at $\theta=0$ should not affect the boundary condition at $\theta=\infty$. The second point is to note that the boundary condition at $\theta=\infty$ is \textit{emergent} in the $\omega$ frame. It does not correspond to any boundary condition in $T(z)$ at $z=z_{+}$ in the UHP. By incorporating these two points, we first insert the mode expansion $T(z)=\sum_{m}(z-z_{+})^{-m-2}\tilde{L}_{m}$ inside (\ref{Vir1}) where the contour enclosed $z_{+}$. Here $\tilde{L}_{m}$ are modes in the usual radial quantization defined by $\tilde{L}_{m}=\oint_{z_{+}}dz (z-z_{+})^{m+1}T(z)$. Since we are interested to find the action of $\mathcal{L}^{\Lambda}_{k}$ on the state $|B\rangle_{\Lambda}$ as in (\ref{fixedpoint bc}), we can do this without loss of generality where $\tilde{L}_{m}$ acts on the states or operators\footnote{Both are same by the virtue of state-operator correspondence in radial quantization.} localized near the fixed points $z=z_{+}$. Hence we have
\begin{align}
   \mathcal{L}_{k} = \frac{\beta}{2\pi i}\int_{C'} dz (z-z_{+})^{-\frac{ik}{\sqrt{d}}+1}(z-z_{-})^{\frac{ik}{\sqrt{d}}+1}\sum_{m}(z-z_{+})^{-m-2}\tilde{L}_{m} 
\end{align}
Again choosing the $s=0$ contour, we obtain 
\begin{align}
     \mathcal{L}^{\Lambda}_{k} = -\frac{\sqrt{d}}{2\pi}\sum_{m}\left(\frac{i\sqrt{d}}{\beta}\right)^{-m+1}(-1)^{-\frac{ik}{\sqrt{d}}-m}\int^{\Lambda}_{0}d\theta \frac{e^{m\sqrt{d}\theta+ik\theta}}{(1-e^{-\sqrt{d}\theta})^{-m+2}}\tilde{L}_{m} 
     \end{align}
To make the above modes to be well-defined we may choose certain complex contour and find the following in $\Lambda \rightarrow \infty$ limit:
     \begin{align}
      \lim_{\Lambda \rightarrow \infty}\mathcal{L}^{\Lambda}_{k} =-\frac{\sqrt{d}}{2\pi}\sum_{m>1} \left(\frac{i\sqrt{d}}{\beta}\right)^{-m+1}(-1)^{-\frac{ik}{\sqrt{d}}-m}\frac{2\Gamma(m-1)\Gamma\left(-m-\frac{ik}{\sqrt{d}}\right)}{\sqrt{d}\Gamma\left(-1-\frac{ik}{\sqrt{d}}\right)}\tilde{L}_{m}
 \end{align}
 In a similar way, we can insert $\bar{T}(\bar{z})=\sum_{m}(\bar{z}-z_{-})^{-m-2}\bar{\tilde{L}}_{m}$ in the definition of $\bar{\mathcal{L}}^{\Lambda}_{k}$ and we obtain
\begin{align}
    \lim_{\Lambda \rightarrow \infty}\bar{\mathcal{L}}^{\Lambda}_{k} =-\frac{\sqrt{d}}{2\pi}\sum_{m>1} \left(\frac{i\sqrt{d}}{\beta}\right)^{-m+1}(-1)^{\frac{ik}{\sqrt{d}}-m}\frac{2\Gamma(m-1)\Gamma\left(-m+\frac{ik}{\sqrt{d}}\right)}{\sqrt{d}\Gamma\left(-1+\frac{ik}{\sqrt{d}}\right)}\bar{\tilde{L}}_{m}
\end{align}
Hence, we can immediately see
\begin{align}
    \lim_{\Lambda\rightarrow\infty}(\mathcal{L}^{\Lambda}_{k}-\bar{\mathcal{L}}_{-k}^{\Lambda})|B\rangle_{\Lambda} = -\frac{\sqrt{d}}{2\pi}\sum_{m>1} \left(\frac{i\sqrt{d}}{\beta}\right)^{-m+1}(-1)^{-\frac{ik}{\sqrt{d}}-m}\frac{2\Gamma(m-1)\Gamma\left(-m-\frac{ik}{\sqrt{d}}\right)}{\sqrt{d}\Gamma\left(-1-\frac{ik}{\sqrt{d}}\right)}(\tilde{L}_{m}-\bar{\tilde{L}}_{m})|B\rangle_{\Lambda}
\end{align}

Hence the near horizon boundary condition in $\omega$ plane (\ref{fixedpoint bc}) reduces to the following condition
\begin{align}\label{fixedpoint bc2}
 (\tilde{L}_{m}-\bar{\tilde{L}}_{m})|B\rangle_{\Lambda\rightarrow\infty} = 0 ,\; \forall m\geq 2
\end{align}
The solution of this is not a standard Ishibashi state, since this boundary condition is emergent purely in the $\omega$ plane as we mentioned earlier. The solution of (\ref{fixedpoint bc2}) will be any primary state localized on the fixed point or conformal vacuum. In particular, we can identify:
\begin{align}\label{shrinking cond}
    |B\rangle_{\Lambda\rightarrow\infty} = \mathcal{O}_{h,h}(z_{+},z_{-})|0\rangle \equiv |h,h\rangle_{new}
\end{align}
 Here $|0\rangle$ denotes the vacuum state (of radial quantization) which respects all global conformal symmetries. Interestingly, in \cite{Das:2024mlx}, this class of local primaries has been constructed at the fixed point that is annihilated by $\tilde{H}=\beta(\tilde{L}_{1}+\tilde{L}_{-1}+\bar{\tilde{L}}_{1}+\bar{\tilde{L}}_{-1})$. This was dubbed as `the other vacuum' of the quantization. In our case, to check whether $|h,h\rangle_{new}$ is the vacuum of $H$, we proceed by constructing a class of degenerate vacuum states $|0\rangle_{\Lambda}, |\tilde{0}\rangle_{\Lambda}$ of $H_{\Lambda}$:
 \begin{align}
H_{\Lambda}|0\rangle_{\Lambda}=H_{\Lambda}|\tilde{0}\rangle_{\Lambda} =     \hat{\mathcal{L}}_{0}^{\Lambda}|0\rangle_{\Lambda} = \hat{\mathcal{L}}_{0}^{\Lambda}|\tilde{0}\rangle_{\Lambda} =0; 
\; \text{such that} \;|0\rangle_{\Lambda\rightarrow\infty} = |0\rangle, |\tilde{0}\rangle_{\Lambda\rightarrow\infty}=|h,h\rangle_{new}
 \end{align}
 Here $H_{\Lambda} \equiv \hat{\mathcal{L}}_{0}^{\Lambda}$, such that $H_{\Lambda\rightarrow\infty}=H=\hat{\mathcal{L}}_{0}$. In the Lorentzian representation of the modular Virasoro algebra, we can always construct such state which satisfy $\hat{\mathcal{L}}_{k}^{\Lambda}|0\rangle_{\Lambda}=\hat{\mathcal{L}}_{k}^{\Lambda}|\tilde{0}\rangle_{\Lambda}=0$ for $k\geq 0$\cite{Das:2024mlx}. This is a trick of uncompactifying the Euclidean time by $s\rightarrow(-\infty,\infty)$, which is equivalent to obtain the Lorentzian representation of $\hat{\mathcal{L}}^{\Lambda}_{k}$ by taking $s\rightarrow is$ and $k \rightarrow ik$\footnote{This exactly matches with the Lorentzian modular Virasoro generators of a CFT vacuum appeared in \cite{Das:2020goe}, defined in a fine width Lorentzian strip $[z_{+},z_{-}]$.}. After doing this, one can define a vacuum by performing a standard path integral with the fixed point boundary condition placed at $s=-\infty$, which we eventually choose to be the location of the fixed point cut-off. From the appearance of a larger class of vacuum, one may suspect to use different boundary conditions at the cut-off. However, the degenerate vacuum sector is connected to a class of zero modes of the Hamiltonian $H_{\Lambda}$, which can be constructed by path integral with same boundary condition at the fixed point. The exact forms of $|0\rangle_{\Lambda}$ and $|\tilde{0}\rangle_{\Lambda}$ in terms of states constructed in radial quantization are not required for the subsequent analysis. The commutator of $\hat{\mathcal{L}}_{0}^{\epsilon}$ with any primary is given by:
 \begin{align}
     [\hat{\mathcal{L}}^{\epsilon}_{0},\mathcal{O}(z,\bar{z})] =\frac{\beta}{2\pi i}\oint_{z}d\omega (\omega-z_{+})(\omega-z_{-}) \hat{T}(\omega)\mathcal{O}(z,\bar{z}) \nonumber \\
  \end{align}
  Using $\hat{T}\mathcal{O}$ OPE, we can find the commutator.
  Let us now consider primaries of dimension $(h,h)$ located at $z=z_{+}+\epsilon$\footnote{Here we took $\theta_{\epsilon}=0$ which corresponds to taking $s=0$.}. The commutator of such primaries with $\hat{\mathcal{L}}_{0}^{\epsilon}$ is given by \cite{Das:2024mlx}:
  \begin{align}
      [\hat{\mathcal{L}}^{\epsilon}_{0},\mathcal{O}(z,\bar{z})]|_{z=z_{+}+\epsilon e^{i\theta_{\epsilon}}} &= 4\beta\epsilon h\mathcal{O}(z,\bar{z})|_{z=z_{+}+\epsilon e^{i\theta_{\epsilon}}}  
  \end{align}
  Hence
  \begin{align}
      \hat{\mathcal{L}}_{0}^{\epsilon} \mathcal{O}(z=z_{+}+\epsilon,\bar{z}=z_{-}+\epsilon)|0\rangle_{\epsilon} = 4\beta\epsilon h\mathcal{O}(z=z_{+}+\epsilon,\bar{z}=z_{-}+\epsilon)|0\rangle_{\epsilon}
  \end{align}
  Here we used the fact $\hat{\mathcal{L}}^{\epsilon}_{0}|0\rangle_{\epsilon} = 0$\footnote{We denote $|0\rangle_{\Lambda}=|0\rangle_{\epsilon}$ and $|\tilde{0}\rangle_{\Lambda}=|\tilde{0}\rangle_{\epsilon}$.}. Hence, taking $\epsilon\rightarrow 0$ limit yields
  \begin{align}
      \lim_{\epsilon\rightarrow 0} \hat{\mathcal{L}}_{0}^{\epsilon} \mathcal{O}(z=z_{+}+\epsilon,\bar{z}=z_{-}+\epsilon)|0\rangle_{\epsilon} \equiv \lim_{\epsilon\rightarrow 0} \hat{\mathcal{L}}_{0}^{\epsilon}|h,h\rangle_{new}^{\epsilon}= 0 =\hat{\mathcal{L}}_{0}|h,h\rangle_{new}
  \end{align}
  Hence from the above analysis, we can conclude that set of states $|h,h\rangle_{new}^{\epsilon}$ are the eigenstates of $\hat{\mathcal{L}}^{\epsilon}_{0}$ with eigenvalues $4\beta h\epsilon$. In $\epsilon\rightarrow 0$ limit, these states correspond to the vacuum $|h,h\rangle_{new}$ of $\hat{\mathcal{L}}_{0}$. Hence it is natural to expect that $|\tilde{0}\rangle_{\epsilon}$ and $|h,h\rangle^{\epsilon}_{new}$ are related. However the explicit relation is not required for the subsequent analysis. 
 One can also similarly construct $N_{lev}$ level \textit{modular descendant states} of the following form:
 \begin{align}
|\psi\rangle_{\epsilon} \equiv  \prod_{k}(\mathcal{\hat{L}}^{\epsilon}_{-k})^{N_{k}}|0\rangle_{\epsilon},\; \text{and} \; |\tilde{\psi}\rangle_{\epsilon} \equiv  \prod_{k}(\mathcal{\hat{L}}^{\epsilon}_{-k})^{N_{k}}|\tilde{0}\rangle_{\epsilon}\;\text{with} \; \sum_{k}kN_{k} = N_{lev}.
 \end{align}
 We note that, modular descendants acting on primaries $|h,h\rangle_{new}^{\epsilon}$, are not well defined in this prescribed Lorentzian representation of modular quantization\cite{Das:2024mlx}. Using the modular Virasoro algebra, we can see
 \begin{align}
     \hat{\mathcal{L}}_{0}^{\epsilon}|\psi\rangle_{\epsilon} = N_{lev}|\psi\rangle_{\epsilon}, \; \hat{\mathcal{L}}_{0}^{\epsilon}|\tilde{\psi}\rangle_{\epsilon} = N_{lev}|\tilde{\psi}\rangle_{\epsilon} \; \text{and} \; \hat{\mathcal{L}}_{0}|\psi\rangle_{\epsilon\rightarrow 0} = \hat{\mathcal{L}}_{0}|\tilde{\psi}\rangle_{\epsilon\rightarrow 0} = 0
 \end{align}
  From all of these observations, we can claim that we can construct the boundary state $|B\rangle_{\epsilon}$ out of the following basis states:
 \begin{align}\label{basis}
     |B\rangle_{\epsilon} = \{|0\rangle_{\epsilon},|\tilde{0}\rangle_{\epsilon,}|h,h\rangle_{new}^{\epsilon},|\psi\rangle_{\epsilon},|\tilde{\psi}\rangle_{\epsilon}\} ,\; \text{such that} \; |B\rangle_{\epsilon\rightarrow 0}=\{|0\rangle,|h,h\rangle_{new}\} 
 \end{align}
 The non-trivial aspect from all of these observations is that, \textit{in the highest weight Lorentzian representation of modular Virasoro algebra is defined by set of states $|B_{\epsilon}\rangle$ on the stretched horizon corresponding to the emergent boundary condition in the Hilbert space of modular quantization. These states correspond to the eigenstates of the Hamiltonian with cut-off. In $\epsilon \rightarrow 0$ or $\Lambda\rightarrow \infty$ limit, the boundary condition reduces to degenerate vacuum state of the quantization which is made out of zero mode of modular Hamiltonian.}

On the other hand, the other side of the half cylinder or at the boundary of $UHP$, there is another conformal boundary condition for which we can define the boundary states:
\begin{align}
 ( T(\omega)-\bar{T}(\bar{\omega}))|B'\rangle_{0}=  (T(z)-\bar{T}(\bar{z}))|B'\rangle_{0}= 0 \; \text{for} \; Im(z)=Im(\omega)=0.
\end{align}
The solution of this is the standard conformal boundary state or Cardy state $(L_{k}-\bar{L}_{-k})|B'\rangle_{0}=0$. For these states, we do not need to impose any shrinking limit by construction.

In order to evaluate the partition function of the annulus or half cylinder, we would now exploit open string-closed string duality as described in \cite{Cardy:2004hm}. Here the width of the annulus is $W= \sqrt{d}\Lambda =\\log\frac{2}{\epsilon}$\footnote{Here we used the fact that in our case $\sqrt{d}=2\beta$.}. In the open string picture, the time translation in the strip is generated by $H_{S}=\frac{1}{2}(\hat{\mathcal{L}_{0}^{\epsilon}}-\frac{c\Lambda d}{12\pi})$\footnote{Here we have a half factor in the Hamiltonian as in (\ref{redefined vir gen}), since $W$ contains a $2\beta$ factor\cite{Cardy:2016fqc}.} and hence the partition function is
\begin{align}\label{open string}
    Z_{\epsilon} = Tr_{\mathcal{H}_{\epsilon}}(e^{-\frac{4\pi^{2}}{W}H_{S}}) = \int dE \rho(E)e^{-\frac{2\pi^{2}}{W}(E-\frac{c\Lambda d}{12\pi})},
\end{align}
where $\rho(E)$ is the density of states in the truncated Hilbert space $\mathcal{H}_{\epsilon}$. In the closed string channel of the half-cylinder, one can compute the same partition function by evaluating the transition amplitude between two boundary state at two different boundaries with respect to the Hamiltonian $\tilde{H}=\tilde{L}_{0}+\bar{\tilde{L}}_{0}-\frac{c}{12}$ and the other modular parameter $W$.
\begin{align}
    Z_{\epsilon} = \;_ {0}\langle B'|e^{-W\tilde{H}}|B\rangle_{\epsilon} = \sum_{h,h',\Delta}d(h')C_{\epsilon}(h)\langle\langle h'|\Delta\rangle\langle\Delta|h\rangle\rangle_{\epsilon} e^{-W\Delta+\frac{c}{12}W}
\end{align}
Here we use $|B'\rangle_{0}=\sum_{h'}d(h')|h'\rangle\rangle$ and $|B\rangle_{\epsilon} = \sum_{h}C_{\epsilon}(h)|h\rangle\rangle_{\epsilon}$, where $|h'\rangle\rangle$ is standard Ishibashi state, whereas $|h\rangle\rangle_{\epsilon}$ is a basis state constructed out of linear combination of orthonormal states defined in (\ref{basis}). In the $c_{eff}\rightarrow \infty$ limit with $c\rightarrow\infty$ and $W$ fixed, the leading contribution of the sum comes from $\Delta=0$. Note that, this is true even for finite $c$ with $W \rightarrow \infty$. Hence for the limit $\frac{cW}{12}\rightarrow \infty$ we have,
\begin{align}\label{part st}
    \lim_{c_{eff}\rightarrow \infty}Z_{\epsilon} \sim e^{\frac{c}{12}W+\log(\sum_{h,h'}d(h)C^{\alpha}_{\epsilon\rightarrow 0}(h')\langle\langle h|0\rangle\langle 0|h'\rangle\rangle_{\epsilon})} \approx e^{\frac{c}{12}W}
\end{align}
From this, one could compute the thermal entropy with respect to inverse temperature $\frac{4\pi^{2}}{W}$ and get\cite{Das:2024mlx}
\begin{align}\label{entropy1}
    S|_{c_{eff}\rightarrow \infty} = \frac{c}{6}\log\frac{2}{\epsilon}+g_{a}+g_{b} = \frac{S^{(3)}_{AdS-Rindler}}{2} + (\text{boundary entropy}) \approx S^{(2)}_{BH}
\end{align}
Where $g_{a,b}$ are the boundary entropies\cite{Cardy:2016fqc}. Here we identify the the thermal entropy of this modular quantization with the AdS$_{2}$ black hole entropy and also with the entanglement entropy of BCFT \cite{Cardy:2016fqc}. Note that, here we took a different limit than taking $\epsilon \rightarrow 0$ which was carried out in \cite{Das:2024mlx} to compute thermal entropy as entanglement entropy or matching with AdS-Rindler entropy\cite{Das:2024vqe}. Hence, the leading term of the entropy is $\frac{c}{6}W=\frac{c\Lambda\sqrt{d}}{6}$, which is \textit{universal} in $c_{eff} \rightarrow \infty$ limit of modular quantization. Remarkably, this exactly matches with the black hole entropy in JT gravity which is reproduced purely from the classical action\cite{Maldacena:2016upp}:
\begin{align}
    S|_{c_{eff}\rightarrow\infty}=S_{BH}=4\pi^{2}CT=\frac{a\sqrt{d}}{8G_{2}}; \; \text{where} \; C=\frac{a}{16\pi G_{2}},\;T=\frac{\sqrt{d}}{2\pi}
\end{align}
\begin{itemize}
    \item \textbf{$S|_{c_{eff} \rightarrow \infty}$ with $c\rightarrow \infty$ and $\Lambda$ fixed:} Here we have identified $a=\frac{\Lambda}{\pi}$ and $c=\frac{3}{4\pi G_{2}}$, which is obtained from half dimensional reduction of JT gravity from AdS-Rindler. On the other hand, taking $a=1$ and $T=\frac{\Lambda\sqrt{d}}{2\pi^{2}}$ would give the same entropy, which corresponds to the half of the Bekenstein-Hawking entropy of a \textit{large BTZ black hole} with radius $r_{h}=\frac{\Lambda\sqrt{d}}{\pi}$ and mass $M_{BTZ}=\frac{c\Lambda^{2}d}{12\pi^{2}}$. This value of $a=1$ is obtained from the \textit{half} dimensional reduction of Einstein gravity in 3D BTZ background.
    \item \textbf{$S|_{c_{eff} \rightarrow \infty}$ with $\Lambda\rightarrow \infty$ and $c$ fixed:} From (\ref{other effective c}), we obtained $\frac{c\Lambda}{3}=\frac{a}{4G_{2}}$. Hence, $S|_{c_{eff}\rightarrow\infty}=\frac{c\Lambda\sqrt{d}}{6}=\frac{a\sqrt{d}}{8G_{2}}=S_{BH}$.
\end{itemize}

Using (\ref{open string}) and (\ref{part st}), we can write the density of state as a inverse Laplace transform of the partition function:
\begin{align}
    \rho(E) = \frac{1}{2\pi i}\int^{C+i\infty}_{C-i\infty}dW e^{\frac{c}{12}W+\frac{2\pi^{2}}{W}E}
\end{align}
Here $E>>\frac{c\Lambda d}{12\pi}$. In $c_{eff}\rightarrow \infty$, we can use the saddle point approximation, for which we would obtain an expression for asymptotically high energy density of state. We find the saddle to be $W^{*}= \sqrt{\frac{24\pi E}{c}} >>\sqrt{\Lambda}$. Using this we obtain
\begin{align}
   \rho(E)_{CFT} = e^{2\pi\sqrt{\frac{cE}{6}}}
\end{align}
This has the similar form with the high energy density of state corresponding to the Hilbert space of \textit{quantum} JT gravity with \textit{one-sided} black holes $\rho(E')_{JT}=e^{2\pi\sqrt{2CE'}}$ \cite{Stanford:2017thb},\cite{Kitaev:2018wpr}, where $C=\frac{a}{16\pi G_{2}}$. In fact these two coincide,
\begin{align}\label{jt deform equiv}
    \rho(E)_{CFT}=\rho(E')_{JT} = e^{2\pi\sqrt{\frac{a E'}{8\pi G_{2}}}}, \; \text{if} \; aE'=E \;.
\end{align}
Using $\rho(E)_{CFT}$, one could compute the \textit{microcanonical entropy} as $S_{micro}=\ln\rho(E)_{CFT}$. For $E'=\frac{c\Lambda d}{24\pi}$ with $a=\frac{\Lambda}{\pi}$ or $E'=\frac{c\Lambda^{2}d}{24\pi^{2}}=\frac{M_{BTZ}}{2}$ with $a=1$, we have $E=\frac{c\Lambda^{2}d}{24\pi^{2}}$ and correspondingly,
\begin{align}
    S_{micro}|_{E=\frac{c\Lambda^{2}d}{24\pi^{2}}=\frac{M_{BTZ}}{2}}= \frac{c\Lambda\sqrt{d}}{6}=S_{BH}
\end{align}
One can obtain the same expression for non-holographic CFT by taking same value of $E=\frac{c\Lambda^{2}}{24\pi^{2}}$ and using (\ref{other effective c}) . In this way, we can see that microcanonical and thermal entropy coincides with the one-sided AdS$_{2}^{b}$ black hole entropy, sourced from half-reduction of AdS-Rindler or large BTZ. Also the same coincides with the one-sided AdS$_{2}$ black hole entropy in the description of CFT coupled to classical JT in AdS$_{2}$ black hole. In both of these descriptions, the continuous density of state is \textit{emergent} in taking $c_{eff}\rightarrow \infty$ limit. 

\begin{figure}
    \centering
   \begin{tikzpicture}[scale=.5]
      \draw[ thin](5,8)--(5,9);\draw[ thin](5,8)--(6,8); \draw[brown] node at ( 5.5,8.5) {Z};
        \draw[ultra thick](-8,0)--(8,0);
 \draw [thin, ->] (0,-9) -- (0,10) ;
        \draw (0,0)-- (5,0) arc (0:90:5);
         \draw[dashed]  (0,5) arc (90:180:5);
      \draw[ultra thick, blue](0,-8)--(0,8);
   \end{tikzpicture}
    \caption{The other side of constant $s$ contour is denoted in the dashed line.}
    \label{other side}
\end{figure}
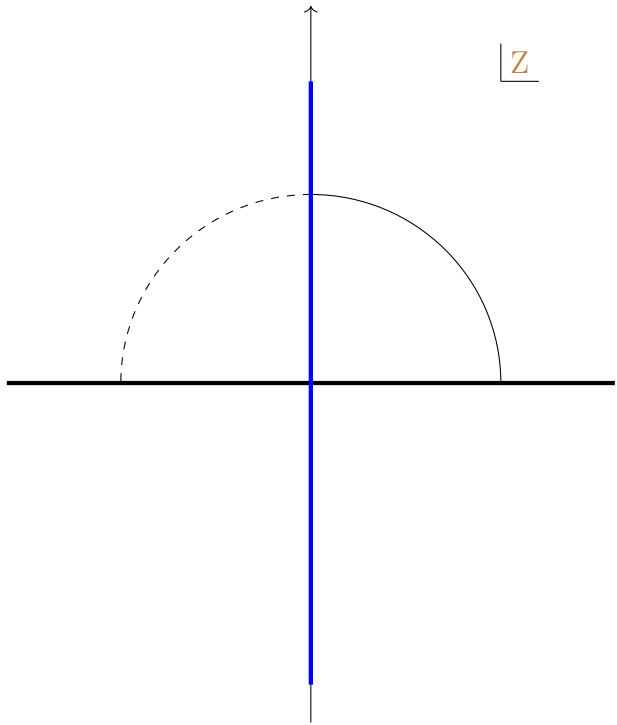

One immediate consequence from (\ref{entropy1}) is to interpret the thermal entropy as the entanglement entropy of the interval  $(\sqrt{d},0)$ in the half-plane $(0,\infty)$. However, the form of entanglement entropy in the half-line is universal for any CFT with any $c$. Hence, we should be careful here to give interpretation of entanglement entropy only in the limit $c_{eff}\rightarrow\infty$ with $\Lambda \rightarrow \infty$. According to the line of \cite{Das:2024vqe}, the vacuum of the strip $|0\rangle_{S}$ can be identified as a TFD state, where the other copy of the TFD must correspond to the other side of the quantization as shown in fig(\ref{other side}). 

\begin{align}\label{vacuum as tfd}
    |0\rangle_{S} \propto \lim_{c_{eff}\rightarrow\infty\; \text{with} \;\Lambda\rightarrow\infty}\sum_{\Delta_{B_{\epsilon}}}e^{-\frac{4\pi^{2}}{\log(2/\epsilon)}\Delta_{B_{\epsilon}}}|\Delta_{B_{\epsilon}}\rangle_{L}|\Delta_{B_{\epsilon}}\rangle_{R}
\end{align}
Here, $|\Delta_{B_{\epsilon}}\rangle_{L,R}$ refers to the eigenstates of $H_{L,R}=(\hat{\mathcal{L}}^{\epsilon}_{0})_{L,R}$ with eigenvalues $\Delta_{B_{\epsilon}}$. These states are orthonormal and constructed out of the basis states (\ref{basis}). In $\epsilon \rightarrow 0$, these states are constructed out of zero modes of the Hamiltonian and hence the identification with the conformal vacuum on strip $|0\rangle_{S}$ makes it self-consistent. We have also used the inverse temperature as $\frac{4\pi^{2}}{W}.$ Consequently, the full Hilbert space on the strip corresponding to both side of the interval ($L$ and $R$) generated by $H_{L}+H_{R}$ admits a factorization as\footnote{We must note that, this factorization is possible only in the limiting sense of $\epsilon$. This is related to the split property in QFT \cite{Fewster:2016mzz}. Strictly at $\epsilon=0$, we can not factorize the Hilbert space. This makes sense, since strictly $\epsilon=0$ makes the entropy divergent and the whole analysis will be ill-defined.}:
\begin{align}
    \mathcal{\tilde{H}}^{S(L,R)}_{c_{eff}\rightarrow\infty} = \mathcal{\tilde{H}}_{c_{eff}\rightarrow\infty}^{S(L)}\otimes\mathcal{\tilde{H}}_{c_{eff}\rightarrow\infty}^{S(R)}
\end{align}

On the other hand, to interpret the black hole entropy as thermal entropy in the other limit of $c_{eff}\rightarrow\infty$ with $c\rightarrow\infty$ and fixed $\Lambda$, we should have a TFD state $|HH\rangle_{\Lambda}$ constructed out of different Hilbert space of states:
\begin{align}\label{TFD1}
    |HH\rangle_{\Lambda} \propto \lim_{c_{eff}\rightarrow\infty\; \text{with} \;c\rightarrow\infty}\sum_{\tilde{\Delta}_{B_{\epsilon}}}e^{-\frac{4\pi^{2}}{\log(2/\epsilon)}\tilde{\Delta}_{B_{\epsilon}}}|\tilde{\Delta}_{B_{\epsilon}}\rangle_{L}|\tilde{\Delta}_{B_{\epsilon}}\rangle_{R}
\end{align}
Here due to matching this Hilbert space with pure JT gravity Hilbert space, we should consider states $|\tilde{\Delta}_{B_{\epsilon}}\rangle$ from holographic CFTs with vacuum and descendants. In particular, we should identify a subset of (\ref{basis}) as $|\tilde{\Delta}_{B_{\epsilon}}\rangle = \{|0\rangle_{\epsilon},|\psi\rangle_{\epsilon}\}$. Note that these basis states are constructed out of universal vacuum sector, that should exist in the Hilbert space of modular quantization for any CFTs. This is consistent with the other claim that JT with finite $c$ CFT can also be described by similar deformed CFT Hilbert space. In holographic CFTs, some extra constraints on spectrum should be imposed.   
Here we should remind that $|HH\rangle_{\Lambda}$ is constructed strictly at $c\rightarrow \infty$ limit of holographic CFT.  
 Similarly the full Hilbert space corresponding to two copies in this limit should be identified as:
\begin{align}
     \mathcal{H}^{S(L,R)}_{c_{eff}\rightarrow\infty} = \mathcal{H}_{c_{eff}\rightarrow\infty}^{S(L)}\otimes\mathcal{H}_{c_{eff}\rightarrow\infty}^{S(R)} 
\end{align}
Note that, this Hilbert space is not isomorphic to the Hilbert space of radial quantization on the strip. 

However, from the structural similarity of modular quantization and modular Virasoro algebra on the strip(one holomorphic copy) and the cylinder(two copies), we claim that\textit{ the Hilbert space 
$\mathcal{H}^{C(A)}_{c_{eff}\rightarrow\infty}$,  
subjected to modular quantization on the cylinder and corresponding to the subregion 
$A \equiv a \cup a^c$ 
of the ring (with the same Hamiltonian), is isomorphic to 
$\mathcal{H}^{S(a, a^c)}_{c_{eff}\rightarrow\infty}$.  
Moreover, the same conformal boundary conditions at the two fixed‐point cutoffs on the ring map to the two copies at the fixed point of two strips.}

\begin{align}
  \mathcal{H}^{C(a\cup a^{c})}_{c_{eff}\rightarrow \infty}   \cong \mathcal{H}^{S(a,a^{c})}_{c_{eff}\rightarrow \infty} = \mathcal{H}_{c_{eff}\rightarrow \infty}^{S(a)}\otimes\mathcal{H}_{c_{eff}\rightarrow \infty}^{S(a^{c})}.
\end{align}
We also note that, all the subsequent discussion will be valid for $\mathcal{\tilde{H}^{S}}$ or $\mathcal{\tilde{H}}^{C}$ which is obtained by $c_{eff} \rightarrow \infty$ with $\Lambda\rightarrow\infty$ and keeping $c$ fixed as we defined in the introduction.

This claim can be justified from the Partition function of modular quantization in $c_{eff}\rightarrow\infty$ limit. To be precise, the partition function corresponding to modular quantization on cylinder $Z^{C}_{\epsilon}$ can be written as \cite{Das:2024mlx}:
\begin{align}\label{factorization1}
    Z^{C}_{c_{eff} \rightarrow \infty} = \exp\left(\frac{c}{6}W\right) = \exp\left(\frac{c}{12}W+\frac{c}{12}W\right) = Z^{S}_{c_{eff}\rightarrow \infty}\times Z^{S}_{c_{eff}\rightarrow \infty}
\end{align}
Here $Z^{S}_{c_{eff}\rightarrow \infty}$ is the partition function of the modular quantization on strip as in (\ref{part st}). Since in the closed string picture, the thermal partition function is written as an amplitude between two fixed point boundary states, equivalently in the path integral picture we can write it as:
\begin{align}\label{factorization2}
 &Z^{C}_{c_{eff} \rightarrow \infty} = \int_{\phi(t=-W)=\phi\{B\}}^{\phi(t=W)=\phi\{B\}} D\phi\; e^{-\int^{W}_{-W}dt L} \nonumber \\
 &= \int D\phi(t=0) \int_{\phi(t=0)}^{\phi(t=W)=\phi\{B\}} D\phi\; e^{-\int^{W}_{0}dt L}\int_{\phi(t=-W)=\phi\{B\}}^{\phi(t=0)} D\phi\; e^{-\int^{0}_{-W}dt L}
\end{align}
Here, using the cutting principle of path integrals, we can write the path integral of the closed string action defined over the Euclidean time interval $(-W,W)$, over the fields with conformal boundary conditions at two fixed points $t=-W$ and $t=W$, as product of two path integrals with the actions defined for $(W,0)$ and $(0,-W)$ respectively and then integrating over all values of fields at $t=0$. However, if we choose the boundary condition at $t=0$ to be the Dirichlet one ($\phi(t=0)=0$), as a solution to the conformal boundary condition at the edges of the strip, the total partition function will be factorized into two copies of the strip. This choice is consistent with the discussion of JT gravity coupled to matter, where $\phi(t=0)$ should be identified with the boundary value of matter field and taking it to be zero corresponds to turn the boundary sources off. In this situation, the partition function of JT coupled to matter reduces to Schwarzian path integral and gives the same expression for density of state for pure JT quantum gravity \cite{Mertens:2022irh}. In this way, the choice of boundary condition will be consistent with (\ref{factorization1}). In other words, (\ref{factorization1}) suggests writing (\ref{factorization2}) with Dirichlet boundary condition at $t=0$ as the following
\begin{align}
&Z^{C}_{c_{eff} \rightarrow \infty}=\lim_{c_{eff}\rightarrow \infty} \int_{\phi(t=-W)=\phi\{B\}}^{\phi(t=W)=\phi\{B\}} D\phi\; e^{-\int^{W}_{-W}dt L} \nonumber \\
&\approx \int_{\phi(t=0)=0}^{\phi(t=W)=\phi\{B\}} D\phi\; e^{-\int^{W}_{0}dt L}\int_{\phi(t=-W)=\phi\{B\}}^{\phi(t=0)=0} D\phi\; e^{-\int^{0}_{-W}dt L} = (Z^{S}_{c_{eff}\rightarrow \infty})^{2}
\end{align}
 In the open string picture, (\ref{factorization1}) is same as writing:
\begin{align}\label{fact1}
   Tr_{\mathcal{H}^{C(a\cup a^{c})}_{c_{eff}\rightarrow\infty}}\left(e^{-\frac{2\pi^{2}}{W}H_{C}^{a\cup a^{c}}}\right) = Tr_{\mathcal{H}_{c_{eff}\rightarrow \infty}^{S(a)}}\left(e^{-\frac{4\pi^{2}}{W}H_{S}^{a}}\right) Tr_{\mathcal{H}_{c_{eff}\rightarrow \infty}^{S(a^{c})}}\left(e^{-\frac{4\pi^{2}}{W}H^{a^{c}}_{S}}\right)
\end{align}
Here we can identify $H_{C}^{a\cup a^{c}}=2(H_{S}^{a}+H_{S}^{a^{c}})$. Here $a:=\theta\in(\infty,0),\; a^{c}:=\theta\in(0,-\infty)$. The boundary terms at $\theta=0$ of the Hamiltonians will get canceled to each other. Since, $H_{S}^{a}$ and $H_{S}^{a^{c}}$ are made out of same Hamiltonian\footnote{They are same deformed Hamiltonian made out of same SL(2,$\mathbb{R}$) generators in UHP and LHP respectively.}, they will commute. Thus we can write
\begin{align}\label{fact2}
      Tr_{\mathcal{H}_{c_{eff}\rightarrow \infty}^{C(a\cup a^{c})}}\left(e^{-\frac{2\pi^{2}}{W}H_{C}^{a\cup a^{c}}}\right) =  Tr_{\mathcal{H}_{c_{eff}\rightarrow \infty}^{C(a\cup a^{c})}} \left(e^{-\frac{4\pi^{2}}{W}H_{S}^{a}} e^{-\frac{4\pi^{2}}{W}H_{S}^{a^{c}}}\right)
\end{align}
From (\ref{fact1}) and (\ref{fact2}), it is straightforward to argue
\begin{align}
    \mathcal{H}_{c_{eff}\rightarrow \infty}^{C(a\cup a^{c})}   \cong \mathcal{H}_{c_{eff}\rightarrow \infty}^{S(a)}\otimes\mathcal{H}_{c_{eff}\rightarrow \infty}^{S(a^{c})}
\end{align}
Physically, the crucial role behind this factorization is played by the modular quantization, in which the Lorentzian representation is built purely from the highest weight representation of modular Virasoro algebra. In the limit $c_{eff}\rightarrow\infty$, boundary condition at the edges of the strip \textit{does not} affect the near fixed point boundary condition.

This picture is reminiscent to our proposal that, \textit{pure JT gravity in one-sided AdS$_{2}$ black hole background is described by a holographic CFT$_{2}$ modular Hamiltonian on a strip. On the other hand, JT gravity on two sided AdS$_{2}$ black hole from the full reduction of AdS-Rindler is described by modular Hamiltonian of the holographic CFT on cylinder. Thus the Hilbert space of JT gravity on two sided AdS$_{2}$ black hole is factorized into that of two one sided AdS$_{2}$ black holes. The same discussion is also true for Hilbert space of non-holographic CFT in classical JT background with a different $c_{eff} \rightarrow \infty$ limit.}

\end{document}